\documentclass[aps,preprint]{revtex4}
\usepackage[dvipdfm]{graphicx}
\graphicspath{{figs/}}

\newcommand{\disregard}[1]{} 
\newcommand{\rs}[1]{_{\rm #1}} 
\newcommand{\rsu}[1]{^{\rm #1}} 
\newcommand{\bm}[1]{\mbox{\boldmath $#1$}} 
\newcommand{\dm}[1]{${\displaystyle #1 }$} 

\newcommand{\nosc}{N_{\rm osc}} 
\newcommand{\noscmax}{N_{\rm osc}^{\rm max}} 
\newcommand{\efree}[1]{\epsilon^{0}_{#1}} 
\newcommand{\cutoff}{f_{\rm c}} 
\newcommand{\fermiMicNeu}{\lambda_{\rm n}^{\rm mic}} 
\newcommand{\fermiMicPro}{\lambda_{\rm p}^{\rm mic}} 
\newcommand{\fermiMacNeu}{\lambda_{\rm n}^{\rm mac}} 
\newcommand{\fermiMacPro}{\lambda_{\rm p}^{\rm mac}} 

\newcommand{\spe}{\epsilon}  
\newcommand{\ecut}{\epsilon_{\rm cut}}
\newcommand{\rmax}{R_{\rm max}}
\newcommand{\ekin}{\epsilon_{\rm kin}} 
\newcommand{\ekmax}{\epsilon_{\rm kin}^{\rm max}}
\newcommand{\rhotf}{\rho_{\!\!\phantom{.}_{\mbox{\tiny TF}}}}
\newcommand{\gtf}{g_{\!\!\phantom{.}_{\mbox{\tiny TF}}}}
\newcommand{\gobtf}{g_{\!\!\phantom{.}_{\mbox{\tiny OB}}}}
\newcommand{\gammaobtf}{\Gamma_{\mbox{\tiny OB}}}
\newcommand{\rhoobtf}{\rho_{\!\!\phantom{.}_{\mbox{\tiny OB}}}}

\begin{document}

\title{
        Improved microscopic-macroscopic approach
       incorporating the effects of continuum states
}

\author{Naoki Tajima}
\affiliation{Department of Applied Physics, University of Fukui, 3-9-1 Bunkyo, Fukui 910-8507, Japan}
\author{Yoshifumi R. Shimizu}
\affiliation{Department of Physics, Graduate School of Science, Kyushu University, Fukuoka 812-8581, Japan}
\author{Satoshi Takahara}
\affiliation{Kyorin University, School of Medicine, Mitaka, Tokyo 181-8611, Japan}

\date{\today}

\begin{abstract}
The Woods-Saxon-Strutinsky method (the microscopic-macroscopic method)
combined with Kruppa's prescription
for positive energy levels, which is necessary to treat neutron rich
nuclei, is studied to clarify the reason for its success and to
propose improvements for its shortcomings.
The reason why the plateau condition is met for the Nilsson model but
not for the Woods-Saxon model is understood in a new interpretation of
the Strutinsky smoothing procedure as a low-pass filter.
Essential features of Kruppa's level density is extracted in terms of
the Thomas-Fermi approximation modified to describe spectra obtained
from diagonalization in truncated oscillator bases.
A method is proposed which weakens the dependence on the smoothing
width by applying the Strutinsky smoothing only to the deviations from
a reference level density.
The BCS equations are modified for the Kruppa's spectrum, which is
necessary to treat the pairing correlation properly in the presence of
continuum.
The potential depth is adjusted for the consistency between the microscopic
and macroscopic Fermi energies.
It is shown, with these improvements, that the microscopic-macroscopic
method is now capable to reliably calculate binding energies
of nuclei far from stability.
\end{abstract}

\maketitle

\section{INTRODUCTION} \label{sec:intro}

Understanding the properties of unstable nuclei is one of the most
interesting subjects of nuclear physics~\cite{DobNaz98}.
It is also important for astrophysics; for example, determination of
the precise position of neutron drip line is crucial for the r-process
nucleosynthesis~\cite{AGT07}.
A characteristic feature of unstable nuclei, among others, is
the weak binding of nucleons,
so that the proper treatment of continuum (scattering) states
is very important for the two basic ingredients of the nuclear structure,
the shell effect and the pairing correlation~\cite{DobNaz07}.
The most popular method of recent years
to treat this problem is the selfconsistent mean field theory,
especially the Hartree-Fock-Bogoliubov (HFB) theory~\cite{RS80},
with suitably chosen (density dependent) zero- or finite-range
effective interactions~\cite{BHR03}.
Such selfconsistent mean field models can reproduce
the very basic quantities like the nuclear mass rather well~\cite{LPT03},
and can be used to investigate the detailed deformation properties
of nucleus.
On the other hand, a non-selfconsistent semi-phenomenological method
of the Strutinsky shell correction
approach~\cite{Str67,Str68,NilStr,FunnyHills}, or often called the
microscopic-macroscopic method,
has been used for more than forty years in order to calculate nuclear masses,
deformations and fission paths.
In such an approach, the part of binding energy smoothly varying
as a function of nucleon (proton and neutron) number is 
represented by the liquid-drop or the droplet model with parameters
adjusted to reproduce the experimental binding energy,
on top of which is added the rapidly varying shell energy correction
evaluated by assuming some non-selfconsistent single-particle potential.

It is known that there is a close relationship between the two,
the selfconsistent mean field and the shell correction
approaches~\cite{Str74,BQ81}, but
the actual calculational procedures differ considerably
and their own merits are quite different.
The number of adjustable parameters is generally
fewer and the range of applicability is believed to be wider
in the selfconsistent mean field models, whereas
the shell correction approach requires much less computational power.
Thanks to recent progress of computer power,
the root mean square deviation between the calculated
and experimental masses, as an example,
in some of the selfconsistent mean field models~\cite{GCP09,GHG09}
is approaching the same level of accuracy
as that in the state of the art model of
the shell correction approach~\cite{MNM95} (or even better).
However, its ease of computation and its flexibility of choosing
the single-particle potential are still great merits
of the shell correction approach.
For example, the various effects of the single-particle
orbits can be more directly studied in the shell correction approach.
On the contrary, in selfconsistent mean field models,
a clear-cut picture is sometimes lost due to
the complicated selfconsistency between the nuclear mean field
and the effective interaction.

Although the qualities of the mass fit are similar in the two
approaches in stable nuclei, they often give quite different
predictions for very heavy nuclei and
unstable nuclei near the neutron drip line,
where no experimental date are available~\cite{LPT03,GCP09,GHG09}.
It should be noticed that the shell correction approach has
several difficulties for the calculation of unstable nuclei,
which are mainly related to the problem of unbound (continuum) states
characteristic to weakly bound systems.
The difficulties were carefully examined one by one in Ref.~\cite{NWD94}.

The first and the most crucial difficulty is that
the shell correction energy cannot be unambiguously determined
for the single-particle potential with finite depth,
which is indispensable for describing weakly bound states.
The energy of shell correction is defined as the difference between
the sum of single-particle energies up to the Fermi level and its smoothed part.
The conventional way of the smoothing procedure utilizes
the energy averaging of
the single-particle level density over the interval $\gamma$ of
the typical shell spacing,
$\gamma \approx \hbar \omega = 41/A^{1/3}$ MeV,
where $A$ is the mass number.
If the absolute energy of the Fermi level is smaller than the shell
spacing, the averaging inevitably involves the unbound states.
The continuum single-particle levels are usually discretized by using, e.g., 
the harmonic oscillator basis expansion,
but blind inclusion of them leads to divergent results as
the basis is enlarged
even in stable nuclei~\cite{BFN72,NWD94};
this is simply because the level density of continuum states
itself is a divergent quantity.
It is proposed that the level density above the threshold
should be replaced~\cite{Lin70}
with the so-called continuum level density~\cite{BU37,LandauLifshitz},
and the resultant smoothed energy is shown to be convergent~\cite{RB72}.

The evaluation of the continuum level density requires the energy derivative
of the phase shift, or of the scattering matrix in general,
so that the calculation is cumbersome for spherical nuclei and
difficult for deformed nuclei.
A breakthrough was given by Kruppa~\cite{Kru98},
who proposed a powerful practical prescription to calculate
the continuum level density by using the fact that it is written
as the difference between the level densities with and without
the mean field potential~\cite{TO75,Shl92}.
However, the problem remains; the so-called plateau condition~\cite{BP73},
which guarantees that the shell correction energy is independent of
the smoothing procedure, is not well satisfied generally~\cite{VKL98,VKN00}.
We reinvestigate the meaning of energy smoothing procedure and consider
a remedy to recover the plateau condition as much as possible employing
the Kruppa's prescription.

It is worth mentioning that there are different methods to calculate
the smoothed part.  One is to make averaging with respect to the particle
number, not to the single-particle energy, only by employing the bound
states~\cite{BKS72,SI75}. 
However, the resultant smoothed energy depends sensitively on how to
perform the averaging for nuclei near the drip
line~\cite{IS78,IS79a,Iva84,IS79b}, where there is no unoccupied bound
states and thus one has to tackle a difficult task to estimate
an average value at a point (a particle number) using data points
only on one side of that point (at smaller particle numbers).
It is also a problem that the smoothed part does
not necessarily behave like the liquid-drop model as a function of
deformation~\cite{Ton82}.
See Ref.~\cite{Pom04} for recent developments.
Another method is to apply the semiclassical Wigner-Kirkwood expansion
of the single-particle partition function~\cite{BR71,BP73,RS80,BB97}.
The relation to the Strutinsky shell correction method was
discussed~\cite{Jen73}, and the treatment of
realistic potentials with the spin-orbit term was developed~\cite{JBB75}.
This method was recommended in Ref.~\cite{NWD94,VKL98} to obtain
the smoothed energy unambiguously.
See Ref.~\cite{BVCprep09} for recent developments.
However,
to achieve the same accuracy as the conventional Strutinsky shell correction, 
one has to include up to the third order terms in $\hbar^2$.
The lowest term is nothing but the Thomas-Fermi energy.
The calculation is rather complicated
especially for the case without spherical symmetry.
It should be also noticed that the semiclassical level density diverges
at the threshold (or the barrier top in the case with the Coulomb potential),
which has non-negligible effects for drip line nuclei~\cite{VKL98}.
In this paper, we stick to the conventional energy smoothing procedure
and do not consider these other possibilities.

The second difficulty in the shell correction approach
with the continuum states included
is the treatment of the pairing correlation,
for which the simple BCS approximation is usually used with the ``diagonal''
(seniority) pairing force.  
The force strength is fixed according to
the model space employed by
the smoothed pairing gap method~\cite{Str68,FunnyHills}.
Since the pairing model space is taken to be within
about the major shell spacing above and below the Fermi level,
the same problem as that of
the smoothed energy arises for unstable nuclei, where the Fermi level
is so close to the threshold that the unbound states enter into the model space.
This is a serious problem because finite occupation probabilities
of unbound states lead to the formation of ``neutron gas'' surrounding nucleus.
A complete solution of this problem requires
the coordinate-space HFB method~\cite{DFT84}.
The pairing energy is also affected by the continuum states
in such an uncontrollable way that
it increases infinitely as more number of states are considered.
It is a major obstacle to the unambiguous prediction of the drip
line~\cite{NWD94}.  We extend the Kruppa's prescription to the
treatment of the pairing correlation and try to solve this problem.

The third difficulty in the the shell correction approach,
which is not particularly related to the unbound states,
is the inconsistency between the Fermi energy of the chosen
single-particle potential and that of the macroscopic part~\cite{Myers70};
this kind of problems do not appear in the selfconsistent mean field
approach, since the single-particle potential adjusts itself to give
the correct Fermi energy.
Though this problem is negligible in stable nuclei, 
it becomes severer near the particle threshold,
which easily shifts the drip line by about ten particle number.
In Ref.~\cite{NWD94}, parameters of a Woods-Saxon potential are adjusted
in accordance with the bulk nuclear asymmetry of the droplet model;
it is found that a fine tuning is necessary to obtain
the coincidence of the Fermi energy of the adjusted potential with
that of the macroscopic part.
In this paper we solve this problem with an automatic adjustment of
the potential depth in the Thomas-Fermi approximation.

The main purpose of the present work is to solve
the difficulties of the conventional microscopic-macroscopic approach.
We propose remedies to all the three difficulties mentioned above.
Although our remedies are not perfect ones, we believe that a combined use
of them gives much more reliable results for the shell correction
calculations of unstable nuclei near the drip lines.
This article is organized as follows.
In Sec.~\ref{sec:currentStatus}, the present status of the
shell correction method is reviewed and its difficulties are discussed
in details.
A new interpretation of the Strutinsky energy smoothing is also given
there.
In Sec.~\ref{sec:improvements},
the solutions to the difficulties are presented 
and the qualities of the improvements are examined in detail.  
Sec.~\ref{sec:conclusion} is devoted to the conclusion.

\section{The present state of the shell correction method}
\label{sec:currentStatus}

\subsection{The Woods-Saxon potential}
\label{sec:woodsSaxonPotential}

The Woods-Saxon potential is a finite-depth potential,
which has a continuum spectrum of unlocalized states
as well as a discrete spectrum of localized states.
Combined with a spin-orbit and proton's Coulomb potentials,
it resembles very well the potentials for a nucleon in atomic nuclei,
having a flat central part and a short-range tail.
The expression we employ is given by
\begin{equation} \label{eq:WoodsSaxonHamiltonian}
H\rs{WS} = \frac{\bm{p}^2}{2m} + V\rs{CE} + V\rs{SO} 
         + \frac{1}{2}\left(1-\tau_3\right) V\rs{CO},
\end{equation}
where the central part $V\rs{CE}$ and the spin-orbit part $V\rs{SO}$
are the standard ones~\cite{CDN87},
\begin{equation} \label{eq:centralPotential}
V\rs{CE} = V\rs{WS}(\bm{r};V\rs{0CE},\kappa\rs{CE},R\rs{0CE},a\rs{CE},\bm{\beta}),
\end{equation}
\begin{equation} \label{eq:SOpotential}
V\rs{SO} = \lambda\rs{SO}\left(\frac{\hbar}{2m\rs{red}c}\right)^2
\Bigl(\bm{\nabla}V\rs{WS}(\bm{r};V\rs{0CE},\kappa\rs{SO},R\rs{0SO},a\rs{SO},\bm{\beta})\Bigr)
\cdot (\bm{\sigma}\times\frac{1}{i}\bm{\nabla}),
\end{equation}
where $m\rs{red}=\frac{A-1}{A}m$ with $m$ being the nucleon mass,
$\tau_{3}$ the third component of the nucleon's isospin multiplied by two
(1 for neutrons and $-1$ for protons),
$\bm{\sigma}$ the Pauli matrix for the nucleon's spin
($\bm{s}=\frac{\hbar}{2}\bm{\sigma}$),
and the function $V\rs{WS}$ is defined by
\begin{equation} \label{eq:WSform}
V\rs{WS}(\bm{r};V_0,\kappa,R_0,a,\bm{\beta})=
-V_0\,\left[1\pm\kappa\frac{N-Z}{A}\right]
\frac{1}{1+\exp[D(\bm{r};R_0,\bm{\beta})/a]}.
\end{equation}
Here, $N$, $Z$, and $A$ are the neutron, proton, and mass numbers, respectively,
while $\pm$ in front of $\kappa$ means $+$ for proton and $-$ for neutron. 
$D(\bm{r};R_0,\bm{\beta})$ denotes the (perpendicular) distance 
(with minus sign if $\bm{r}$ is inside the surface)
between a given point $\bm{r}$ and the nuclear surface,
so that $D(\bm{r};R_0,\bm{\beta}=\bm{0})=r-R_0$ for spherical shape.
The surface is specified by the radius $R_0$
and the deformation parameters $\bm{\beta}\equiv(\beta_\lambda)$ as,
\begin{equation} \label{eq:surface}
R(\theta;R_0,\bm{\beta})=R_0\, c_v(\bm{\beta})\left[1+\sum_\lambda\beta_\lambda Y_{\lambda0}(\theta)\right],
\end{equation}
where the constant $c_v(\bm{\beta})$ takes care of
the conservation of the volume inside the surface against deformation
($c_v=1$ for $\bm{\beta}=\bm{0}$).
We consider only axially symmetric deformations
and take into account 
the quadrupole ($\beta_2$) and hexadecapole ($\beta_4$) ones in this paper.  
The Coulomb potential $V\rs{CO}$ acts only on protons,
and is created by electric charge $(Z-1)e$
distributed uniformly inside the nuclear surface
given by Eq.~(\ref{eq:surface}) with $R_0=R\rs{0CE}$.

The parameters $R\rs{0CE}$ ($R\rs{0SO}$) and $a\rs{CE}$ ($a\rs{SO}$) are
the radius and the surface diffuseness of the central (spin-orbit) potential.
For $N=Z$ nuclei,
the depth of the central potential is $V\rs{0CE}$, while
a dimensionless parameter $\lambda\rs{SO}$ specifies
the depth of the spin-orbit potential relative to the central potential.
The quantities $\kappa\rs{CE}$ and $\kappa\rs{SO}$ describe
the nuclear isospin dependence of the two potentials.
Note that the radii and diffusenesses of
the central and spin-orbit potentials are different in general,
but the shape of nuclear surfaces are taken to be the same, i.e.,
the common deformation parameters $\bm{\beta}$ are used in both of them.
The set of the values of these parameters mainly used to obtain
the results shown in this paper
is the universal parameter set of Ref.~\cite{CDN87}
(Note that the parameter $r_{0-\mbox{so}}(\mbox{P})=1.20$ in Table~1
of Ref.~\cite{CDN87} is a misprint and should be replaced to $1.320$,
see Ref.~\cite{DSW81}).
It should, however, be noted that we modify the depth of
the central potential
in order to be consistent with
the liquid-drop Fermi energy;
see Sec.~\ref{sec:potentialdepth} for details.

The Nilsson potential is a harmonic oscillator potential combined with
a spin-orbit and an $l^2$ terms. Since its depth (or height) is infinite,
its spectrum does not have a continuum part. See, e.g., Ref.~\cite{NR95}
for the equations to define the potential.
We employ the $N\rs{osc}$-dependent $ls$ and $l^2$ parameters of
Ref.~\cite{BR85}.

We use the anisotropic harmonic oscillator basis to diagonalize
these single-particle Hamiltonians.
The oscillator frequencies, $\omega_3$ and $\omega_\bot$,
along the symmetry axis and the perpendicular axis, respectively,
are determined by the two conditions; the volume conservation and
the condition that they are inversely
proportional to the root mean square length of each axis,
which is calculated
assuming the uniform sharp-cut density inside the nuclear surface
given by Eq.~(\ref{eq:surface}).
Namely the conditions are $\omega_3\omega_\bot^2=\omega_0^3$ and
$\omega_3/\omega_\bot
=\sqrt{\langle x^2 \rangle\rs{uni}/\langle z^2 \rangle\rs{uni}}\,$,
where $\omega_0$ is the frequency for spherical shape and
$\langle \;\rangle\rs{uni}$
denotes average value based on the uniform sharp-cut density.
The number of the basis states is specified by the total oscillator
quantum number $\nosc=n_\bot+n_3$ ($n_\bot =n_1+n_2$),
where $n_i$ $(i=1,2,3)$ is the number of oscillator quanta in the $i$-th axis.
In the following discussions, we use the standard harmonic oscillator energy,
$\hbar\omega=41/A^{1/3}$ MeV, and the Woods-Saxon potential
is diagonalized in the oscillator basis with a frequency $\omega_0=1.2\,\omega$.

\subsection{The shell correction method}
\label{sec:shellCorrectionMethod}

In the shell correction method, the total energy of a nucleus is assumed to be
decomposed as
\begin{equation} \label{eq:Etot}
E = E_{\rm mac} + \sum_{{\rm q}={\rm n},{\rm p}} \left( E\rs{sh}^{(\rm q)} + E\rs{pair}^{(\rm q)} \right),
\end{equation}
where
$E_{\rm mac}$ is the energy of a macroscopic model like the liquid-drop model
while $E^{\rm (q)}\rs{sh}$ and $E^{\rm (q)}\rs{pair}$ are the microscopic corrections.
Because the equations to define the contributions from neutrons (q=n) and protons (q=p) are
very similar, we show only the terms for neutrons in the rest of this paper.
For the sake of conciseness, we omit the superscript (n)
for the most part, i.e.,
$E\rs{sh}$ and $E\rs{pair}$ designate $E\rs{sh}^{(\rm n)}$
and  $E\rs{pair}^{(\rm n)}$, respectively.

The term $E\rs{sh}$ is the shell correction energy, which is defined by
\begin{equation} \label{eq:Esh}
E\rs{sh}=E\rs{s.p.}-\tilde{E}\rs{s.p.}.
\end{equation}
The first term on the right-hand side is
the sum of the single-particle energies of occupied levels,
\begin{equation} \label{eq:Esp}
E\rs{s.p.} = \sum_{i=1}^{N} \epsilon_{i},
\end{equation}
where $\epsilon_1 \le \epsilon_2 \le \cdots$ are the neutron single-particle
energies.  Since we are going to discuss about the Kruppa method
(see Sec.~\ref{sec:KruppaMethod}),
these levels are the results of diagonalizations of the single-particle 
Hamiltonian in a truncated harmonic oscillator basis and thus they are discrete
through negative and positive energies.

By introducing the (single-particle) level density,
\begin{equation} \label{eq:leveldens}
 g(\epsilon)=\sum_i \delta(\epsilon-\epsilon_i),
\end{equation}
the quantity $E\rs{s.p.}$ in Eq.~(\ref{eq:Esp}) can be written as an integral,
\begin{equation} \label{eq:EspIntg}
 E\rs{s.p.}=
 \int_{-\infty}^\lambda \epsilon g(\epsilon)d\epsilon,
\end{equation}
up to the Fermi energy $\lambda$.
Analogously, the second term on the right-hand side of Eq.~(\ref{eq:Esh})
is the integral of the product of the energy and a smoothed level density 
$\tilde{g}(\epsilon)$ over a semiinfinite energy interval
up to the corresponding Fermi energenergy $\tilde{\lambda}$,
\begin{equation} \label{eq:Esmoothed}
\tilde{E}\rs{s.p.} = \int_{-\infty}^{\tilde{\lambda}} \epsilon \tilde{g}(\epsilon) d\epsilon,
\end{equation}
with $\tilde{\lambda}$ determined to satisfy a constraint on the
number of particles,
\begin{equation} \label{eq:particleNumberEquation}
\int_{-\infty}^{\tilde{\lambda}} \tilde{g}(\epsilon) d\epsilon = N.
\end{equation}
The term $E\rs{pair}$ is the correction for the pairing energy,
which is defined by
\begin{equation} \label{eq:Epair}
E\rs{pair} = \left(E\rs{BCS} - E\rs{s.p.}\right) 
           - \left(\tilde{E}\rs{BCS} - \tilde{E}\rs{s.p.} \right).
\end{equation}
$E\rs{BCS}$ and $\tilde{E}\rs{BCS}$ are the energies of the BCS solutions of the
pairing Hamiltonian with discrete and smoothed level densities, respectively.
The terms in the first parentheses in the right-hand side represent 
the energy gain due to the pairing correlation.
The terms in the second parentheses 
are the part of the pairing energy gain smoothly changing as a function of
$N$ and $Z$, 
which should be subtracted since it is already included in $E_{\rm mac}$.
Explicit expressions for these quantities
are given in Secs.~\ref{sec:KruppaBCS} and~\ref{sec:pairingForceStrength}.

Using Eqs.~(\ref{eq:Esh}) and (\ref{eq:Epair}), one can simplify
Eq.~(\ref{eq:Etot}) as
\begin{equation} \label{eq:EtotB}
E = E_{\rm mac} + \sum_{\rm q=n,p} \left( E^{\rm (q)}\rs{BCS}
 - \tilde{E}^{\rm (q)}\rs{BCS} \right).
\end{equation}
However, from a physical point of view,
we discuss $E\rs{sh}$ and $E\rs{pair}$ separately.
It may be worth noticing that
one often uses simplified expressions for the smoothed part of the pairing energy
in many of existing calculations, e.g., Refs.~\cite{FunnyHills,MNM95},
assuming that the single-particle
levels are uniformly distributed with the smoothed level density
at the Fermi energy.  In such cases Eq.~(\ref{eq:EtotB}) does not hold exactly.
In this paper we calculate $\tilde{E}\rs{BCS}$ consistently
without such simplifications, as is discussed
in Secs.~\ref{sec:KruppaBCS} and~\ref{sec:pairingForceStrength}.

As for the energy of the macroscopic part, we use the liquid-drop model
of Ref.~\cite{MS67} in this paper;
see also Ref.~\cite{AFLR99} for its explicit form.


\subsection{The Strutinsky smoothing method as a low-pass filter}
\label{sec:strutinskySmoothing}

\begin{figure}[!ht] 
\includegraphics[width=75mm]{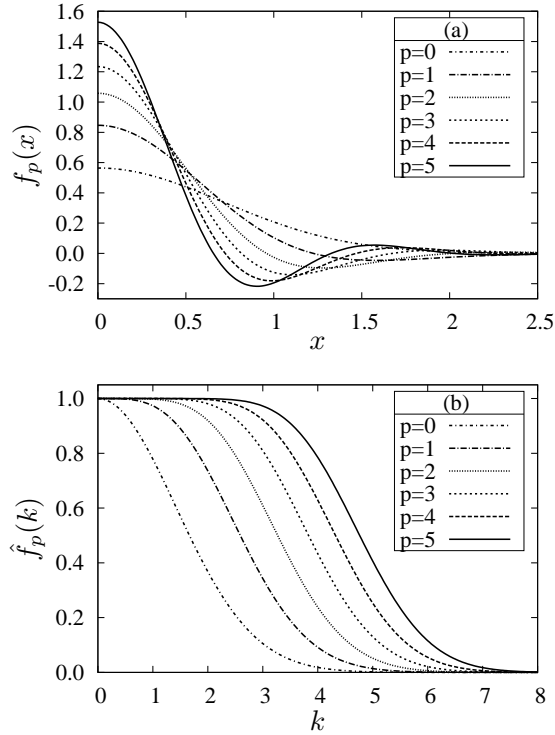}
\vspace*{-5mm}
\caption{The Strutinsky smoothing function in panel (a) and
its Fourier transform in (b).
The parameter $p$ is a half of the order of the polynomial part of the function.
}
\label{fig:filter}
\end{figure}

In the conventional Strutinsky smoothing method,
the smoothed level density is obtained by a convolution integral
with respect to the single-particle energy,
\begin{equation} \label{eq:smoothedLevelDensity}
\tilde{g}(\epsilon) = \frac{1}{\gamma} \int_{-\infty}^{\infty} g(\epsilon')
f_p \left( \frac{\epsilon-\epsilon'}{\gamma} \right) d \epsilon' ,
\end{equation}
where $f_p(x)$ is a smoothing function normalized as
\dm{\int_{-\infty}^{\infty} f_p(x) dx=1},
while $\gamma$ is the width parameter.
The smoothing function is chosen as
\begin{equation} \label{eq:f}
f_p (x) = \frac{1}{\sqrt{\pi}} e^{-x^2} L^{1/2}_{p}(x^2).
\end{equation}
Here, $L^{1/2}_{p}(x)$ is a polynomial of order $p$
(the generalized or associated Laguerre polynomial~\cite{AS65}), with which
the transformation~(\ref{eq:smoothedLevelDensity})
leaves $g(\epsilon)$ unchanged, i.e., $\tilde{g}(\epsilon)=g(\epsilon)$,
if $g(\epsilon)$ is a polynomial of order $2p$.
Note that the order of polynomial is denoted by ``$p$''
in, e.g., Refs.~\cite{NWD94,VKL98,VKN00},
so that the parameter $p$ in these references is $2p$ in this work.
Fig.~\ref{fig:filter}~(a) shows
the graphs of $f_p(x)$ for several values of $p$.

For a discrete level density,
\begin{equation} \label{eq:discreteLevelDensity}
g(\epsilon) = \sum_{i=1}^{M} \delta(\epsilon - \epsilon_i) ,
\end{equation}
the smoothed density is given by
\begin{equation} \label{eq:smoothedLevelDensityB}
\tilde{g}(\epsilon) = \frac{1}{\gamma} \sum_{i=1}^{M} f_p \left( \frac{\epsilon - \epsilon_i}{\gamma} \right),
\end{equation}
where $M$ is the number of single-particle levels included in the calculations.
Owing to the gaussian form factor in $f_p(x)$,
this transformation has a short-range character,
which is a large advantage
because it makes high positive energy levels
unnecessary to evaluate Eq.~(\ref{eq:Esmoothed})
since they hardly affect $\tilde{g}(\epsilon)$ at negative energies.

Let us unveil another aspect of this transformation.
The Fourier transform of a convolution of two functions is
proportional to the product of each function's Fourier transform.
Therefore, by denoting the Fourier transform of a function $F$ as $\hat{F}$ like
\begin{equation} \label{eq:fourierTransformation}
 \hat{F}(k)=\int_{-\infty}^\infty F(x)\,e^{-ikx}dx,
\end{equation}
one can rewrite Eq.~(\ref{eq:smoothedLevelDensity}) as
\begin{equation} \label{eq:FourierTransformOfConvolution}
\hat{\tilde{g}}(\tau) = \hat{f}_{p}(\gamma \tau) \, \hat{g}(\tau).
\end{equation}
(A similar expression in terms of the Laplace transformation
is given in Ref.~\cite{Jen73} in a different context.)
The ``wavenumber'' $\tau$ in Eq.~(\ref{eq:FourierTransformOfConvolution})
has a dimension of (energy)$^{-1}$
and may be regarded as a time variable (divided by $\hbar$).
Now, we show that the function $\hat{f}_{p}$ has a typical shape
of a low-pass filter.  The Laguerre polynomial can be expressed in terms of
the Hermite polynomials $H_{2l}$ as,
\begin{equation} \label{eq:LaguerreByHermite}
  L^{1/2}_{p} (x^2)=\sum_{l=0}^p C_l H_{2l}(x), \quad C_l=(-)^l(2^{2l}l!)^{-1}.
\end{equation}
By multiplying $e^{-x^2}$ to
the generating function of Hermite polynomials,
\begin{equation} \label{eq:generatinigFunctionOfHermitePolynomials}
  e^{-s^2+2xs}=\sum_{n=0}^\infty H_n(x)\,\frac{s^n}{n!},
\end{equation}
one obtains
\begin{equation} \label{eq:GFHPMultipliedByGaussian}
  e^{-(s-x)^2}=\sum_{n=0}^\infty H_n(x)\,e^{-x^2}\,\frac{s^n}{n!}.
\end{equation}
The Fourier transform of the left-hand side can be calculated easily as
\begin{equation} \label{eq:FourierTransformOfLHS}
 \int_{-\infty}^\infty \,e^{-(s-x)^2} \,e^{-ikx} dx
 =\sqrt{\pi}\,e^{-k^2/4}e^{-iks}
 = \sum_{n=0}^{\infty} \sqrt{\pi}(-ik)^n e^{-k^2/4} \frac{s^n}{n!},
\end{equation}
which means that the Fourier transform of $H_{n}(x) e^{-x^2}$ is
\begin{equation} \label{eq:FTofHermitePolynomialTimesGaussian}
 \int_{-\infty}^\infty \,H_{n}(x) e^{-x^2} dx =
 \sqrt{\pi}(-ik)^n e^{-k^2/4}.
\end{equation}
Using above results, 
one obtains the Fourier transform of the Strutinsky smoothing function as
\begin{equation} \label{eq:FTofStrutinskySmoothingFunction}
 \hat{f}_p(k) = \sum_{l=0}^p C_l (-ik)^{2l} e^{-k^2/4}
 = \left[\sum_{l=0}^p \frac{1}{l!} \left(\frac{k}{2}\right)^{2l}\right] e^{-(k/2)^2} .
\end{equation}
The term in the square brackets is
the Taylor series of $e^{(k/2)^2}$ truncated at order $2p$.
For $k \ll k_p\rsu{cut} \equiv 2\sqrt{p}$, 
the term is very close to $e^{(k/2)^2}$
and hence $\hat{f}_{p} \simeq 1$, i.e.,
the filter is almost perfectly transparent.
From this fact, one may give an alternative definition of
the polynomial part of the Strutinsky smoothing function:
It is a polynomial which minimizes the distortion of this
low-pass filter
in such a way that $\hat{f}_{p}^{(l)}(0)=0$ for $1 \le l \le 2p+1$.
For $k \gg k_p\rsu{cut}$,
the term in the square brackets is much smaller than $e^{(k/2)^2}$
and thus $\hat{f}_{p} \simeq 0$, i.e., the filter is almost
completely opaque.

\begin{table}[hbt] 
\caption{
Changes in the characteristics of the low-pass filter $\hat{f}_{p}$, 
i.e., the Fourier transform of the Strutinsky smoothing function,
versus the order $p$ of its polynomial part. 
The normalization is chosen as $\hat{f}_{p}(0)=1$.
$\hat{f}_{p}^{-1}$ denotes the inverse function of $\hat{f}_{p}$.
}
\label{tab:filter}
\begin{center}
\begin{tabular}{rrcc}
\hline
$p$ & $\phantom{w}\hat{f}_{p}^{-1}(0.5)$ &  $\phantom{w}\hat{f}_{p}^{-1}(0.5)/\sqrt{p}\phantom{w}$ & $\phantom{w}\hat{f}_{p}^{-1}(0.1)-\hat{f}_{p}^{-1}(0.9)$ \\
\hline
 0 & 1.665 & 1.665 & 2.386 \\
 1 & 2.591 & 2.591 & 2.486 \\
 2 & 3.271 & 2.313 & 2.514 \\
 3 & 3.833 & 2.213 & 2.528 \\
 4 & 4.322 & 2.161 & 2.535 \\
 5 & 4.762 & 2.130 & 2.540 \\
 10 & 6.533 & 2.066 & 2.551 \\
 20 & 9.092 & 2.033 & 2.557 \\
 50 &14.236 & 2.013 & 2.561 \\
100 &20.067 & 2.007 & 2.562 \\
\hline
\end{tabular}
\end{center}
\end{table} 

In Fig.~\ref{fig:filter}~(b), the Fourier transform $\hat{f}_{p}(k)$ of
the smoothing function is shown for several values of $p$.
One sees that they are almost a constant function near $k=0$
and decrease monotonically to zero.
They become a half of the maximum around 
$k \approx k_p\rsu{cut}=2\sqrt{p}$ (except $p$=0).
The length of the interval where the function drops
from 90\% to 10\% of the maximum ($\hat{f}_p(0)=1$) 
is $\sim 2.5$ and almost independent of $p$.
One can verify these features in Table~\ref{tab:filter}.
In this way, the usage of higher order polynomials lengthens
the width of the filter.
At the same time, it shortens the width of the smoothing function
in Fig.~\ref{fig:filter}~(a) in a complementary manner.

\begin{figure}[!ht] 
\includegraphics[width=75mm]{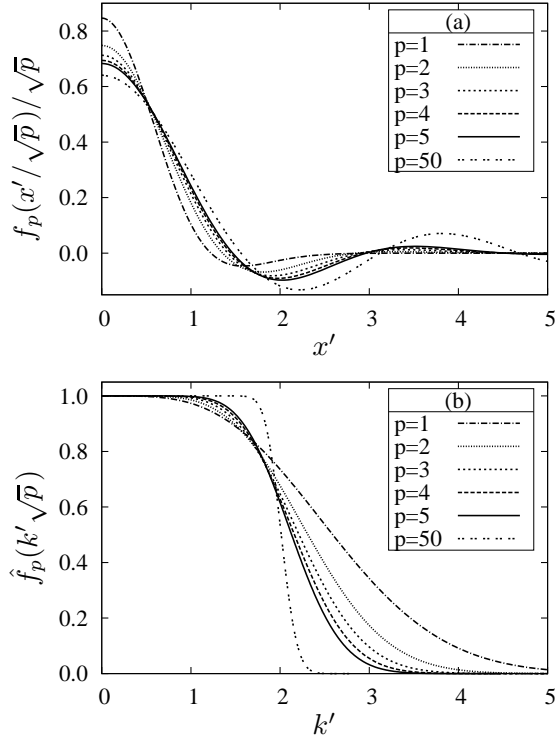}
\vspace*{-5mm}
\caption{Rescaled Strutinsky smoothing function in panel (a) and
its Fourier transform in (b).
}
\label{fig:scaled_filter}
\end{figure} 

Since the width of the filter of order $p$ is proportional to $\sqrt{p}$,
it is convenient to use variables scaled with $\sqrt{p}$, $k'=k/\sqrt{p}$.
In Fig.~\ref{fig:scaled_filter}~(b),
we show $\hat{f}_{p}(k'\sqrt{p}\,)$ versus $k'$
for several values of $p$.
As the order $p$ is increased, the cutoff becomes sharper
while the position of the cutoff
converges to $k'=2$ independent of $p$;
it approaches a step function $\theta\left( 2-|k'| \right)$
in the limit of $p \rightarrow \infty$.
Corresponding changes in the function $f_{p}$ can be found 
by using a dimensionless variable
 $x' = \sqrt{p}\,(\spe - \spe')/\gamma$
and a rescaled smoothing function $f_p(x'/\sqrt{p}\,)/\sqrt{p}$ 
to rewrite Eq.~(\ref{eq:smoothedLevelDensity}) as
\begin{equation} \label{eq:smoothedLevelDensityRewritten}
\tilde{g}(\epsilon) = \int_{-\infty}^{\infty}
g(\epsilon -\gamma x'/\sqrt{p}\,) f_p(x'/\sqrt{p}\,)\,dx'/\sqrt{p}.
\end{equation}
In Fig.~\ref{fig:scaled_filter}~(a), $f_p(x'/\sqrt{p}\,)/\sqrt{p}$ 
is shown as a function of $x'$ for several values of $p$.
Although the convergence is slow,
for very large values of $p$, 
$f_p(x'/\sqrt{p}\,)/\sqrt{p} \simeq (\sin 2x') /\pi x'$,
which can be obtained as the inverse Fourier transform of
 $\theta\left( 2-|k'| \right)$.
The curve for $p=50$ is quite close to this function
in the interval shown in the figure.
For larger $x'$, however, the rescaled smoothing function decreases
much faster than $x'^{-1}$ due to the Gaussian form factor.

Using the semiclassical periodic-orbit theory,
the quantum mechanical level density~(\ref{eq:discreteLevelDensity})
for a certain class of potentials can be represented
by a sum of contributions from classical periodic (closed) orbits,
the so-called trace formula~\cite{Gut67,BT76,GutText}.
It is discussed that the origin of the gross shell structure can be
well understood in terms of a few important short periodic orbits;
for example,
a beating pattern of the level density in the spherical billiard~\cite{BB72},
the shell structure in deformed nuclei~\cite{SMOD77},
and the supershells in metal clusters~\cite{NHM90}.
The smooth part of the energy corresponds to the gross shell structure,
to which only short periodic orbits contribute.
The low-pass filter expression~(\ref{eq:FourierTransformOfConvolution})
demonstrates clearly that the conventional Strutinsky smoothing
cuts off the contributions of long periodic orbits
with period (divided by $\hbar$)
$\tau =k/\gamma \gg \tau_p\rsu{cut}$ for the filter $\hat{f}_p(k)$,
where the cutoff period is
$\tau_p\rsu{cut} \equiv k_p\rsu{cut}/\gamma = 2\sqrt{p}/\gamma$.

If one changes $\gamma$ as $\gamma \propto \sqrt{p}$ for different choices of
the order $p$ in the smoothing function, the cutoff period
$\tau_p\rsu{cut}$ is independent of $p$, while the cutoff becomes sharper
for larger $p$ value as is clearly seen in Fig.~\ref{fig:scaled_filter}~(b).
In Sec.~\ref{sec:plateau} and Sec.~\ref{sec:improvedplateau},
we use this fact for discussions on the plateau condition, i.e., the stability
of the smoothed energy with respect to the smoothing width $\gamma$
and the order $p$ specifying the curvature correction polynomials.

\subsection{Kruppa's prescription for the positive energy levels}
\label{sec:KruppaMethod}

\begin{figure*}[!ht] 
\includegraphics[width=150mm]{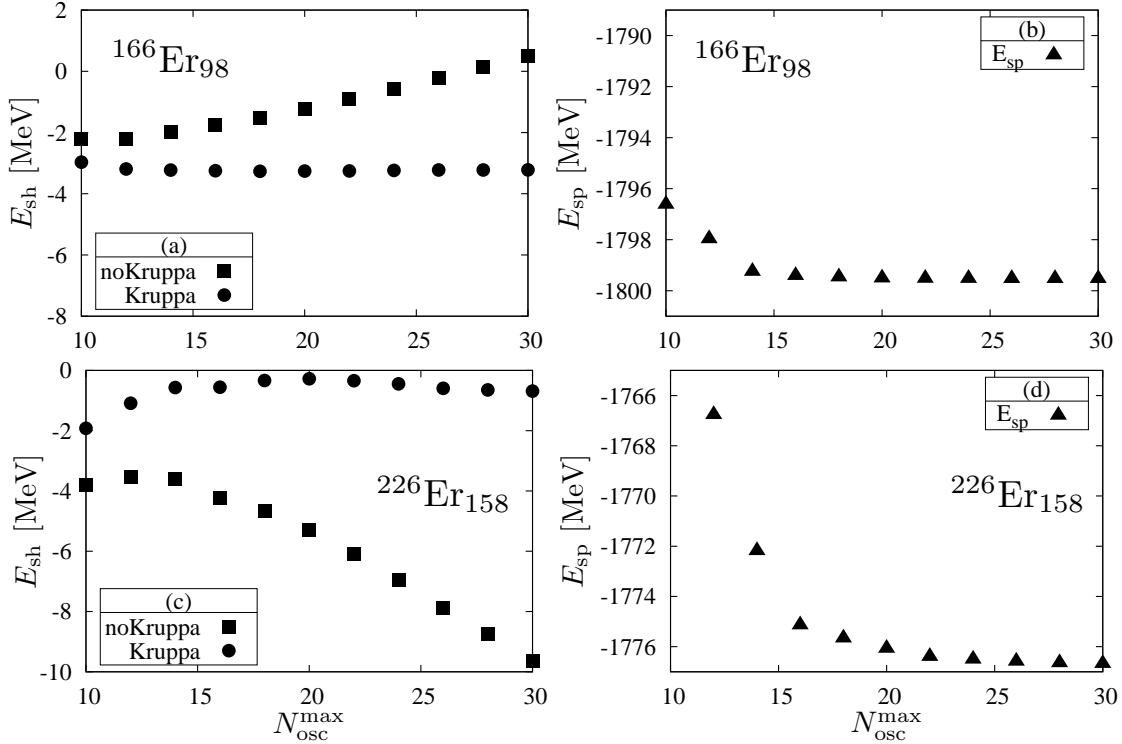}
\vspace*{-5mm}
\caption{ 
Neutron shell correction energies $E\rs{sh}$ ((a),(c)) and the sum of
single-particle energies $E\rs{s.p.}$ ((b),(d)) for $^{166}$Er ((a),(b)) and
$^{226}$Er ((c),(d)) as functions of the oscillator basis cutoff
$\noscmax$.  The values of $E\rs{sh}$ obtained with and without the
Kruppa's prescription are designated by filled circles and squares,
respectively.
The deformation parameters are $\beta_2=0.280$ ($0.255$)
and $\beta_4=0.005$ ($-0.033$) 
for $^{166}$Er ($^{226}$Er)
while the smoothing parameters are
$\gamma=1.2\, \hbar \omega$ and $p=3$.  
}
\label{fig:KruppaEsh}
\end{figure*}

For finite-depth potentials like the Woods-Saxon one,
positive energy levels appear as continuum states.
They also affect the energy of bound nuclei 
through Eqs.~(\ref{eq:Esmoothed}) and (\ref{eq:smoothedLevelDensityB}). 
Their contribution becomes
larger when $\tilde{\lambda}$ is closer to zero.
If one obtains the positive energy spectrum by diagonalizing
the Hamiltonian in a truncated oscillator basis,
the positive energy spectrum is not continuous but discrete. 
Thus one can calculate the summation in Eq.~(\ref{eq:smoothedLevelDensityB}) 
straightforwardly. However, the result depends strongly on
the size of the basis $M$.
In fact, the smoothed level density~(\ref{eq:smoothedLevelDensityB}) diverges
%
in the continuous limit and so does the smoothed energy~(\ref{eq:Esmoothed});
it is monotonically increasing or decreasing
as increasing the size of the basis~\cite{BFN72,NWD94}.
A practical way to avoid this problem is to restrict the size of the basis;
it is recommended in Ref.~\cite{BFN72,MNM95}
to take $\noscmax \approx 12$ for the harmonic oscillator basis.
However, in such small bases, negative energy levels 
may not be sufficiently accurate as we will see in the followings
(see, e.g., Fig.~\ref{fig:KruppaEsh}).

A way to circumvent the diverging single-particle level density $g(\epsilon)$
due to the particle continuum is to replace it
with the so-called continuum level density $g\rs{c}(\epsilon)$.
In the case of spherically symmetric potentials,
it is written as~\cite{BU37,LandauLifshitz},
\begin{equation} \label{eq:gphaseshift}
 g(\epsilon) \quad\Rightarrow\quad
 g\rs{c}(\epsilon)
 =\sum_{i:{\mbox{\scriptsize bound}}}\delta(\epsilon_i-\epsilon)
  +\frac{1}{\pi}\sum_{lj} (2j+1)\frac{d\delta_{lj}(\epsilon)}{d\epsilon},
\end{equation}
where $\delta_{lj}(\epsilon)$ is the scattering phase shift.
This expression was used for the calculation of shell correction
energy and it was found that the contributions of the particle continuum
(the second term on the right-hand side) through $\tilde{E}\rs{s.p.}$
are never negligible even in stable nuclei
for finite-depth potentials~\cite{Lin70,RB72}.

One can roughly regard the continuum level density as the difference
between the full and the free level densities~\cite{LandauLifshitz}.
Taking the energy derivative of the phase shift in Eq.~(\ref{eq:gphaseshift})
means calculating the level density from the number of states.
The number of states is actually
proportional to the phase of the radial oscillation of the wavefunction
because an increase in the phase by $\pi$ corresponds to the addition
of one radial node in the box boundary condition.  The phase shift is the
difference of the phases between full and free solutions.  Therefore,
the definition in terms of the phase shifts is actually
equal to taking the limit of infinite volume of the difference between
the full and free level densities in a finite volume cavity.

This can be shown more rigorously.
The generalization of Eq.~(\ref{eq:gphaseshift}) for non-spherically
symmetric potentials is given by~\cite{TO75,Kru98}
\begin{equation} \label{eq:gSmatrix}
 g\rs{c}(\epsilon)=\frac{1}{2\pi i}\,
 {\hat{\rm tr}}_{\mbox{$\epsilon$}}\left[
 S^\dagger(\epsilon)\frac{dS(\epsilon)}{d\epsilon}\right],
\end{equation}
where $S(\epsilon)$ is the on-shell S-matrix corresponding to the
single-particle Hamiltonian $H$ with energy $\epsilon$, and
${\hat{\rm tr}}_{\mbox{$\epsilon$}}$ means the restricted trace operation
with respect to the eigenstates with energy $\epsilon$. 
Note that Eq.~(\ref{eq:gSmatrix}) contains the contributions
from the bound states because they appear as poles of the S-matrix.
This quantity is related to the time-delay~\cite{GW64},
and shown to be identical to the trace of the difference between
the full and free single-particle Green's functions~\cite{TO75}.
In this way, the level density can be written as
\begin{equation} \label{eq:gGreenf}
 g\rs{c}(\epsilon)=\frac{1}{\pi}\mbox{Im}\left[
 {\rm tr}\,\frac{1}{H-\epsilon}-{\rm tr}\,\frac{1}{H_0-\epsilon}\right],
\end{equation}
where $H_0$ is the free Hamiltonian
(with the repulsive Coulomb potential for proton),
and ${\rm tr}$ here is the full trace operation.
This expression clearly tells that
both full and free level densities are divergent for positive
energies but their difference is finite.
It is used for investigation of the level density
in Ref.~\cite{Shl92} by using the Green's function technique~\cite{SB75}.

Inspired by Eq.~(\ref{eq:gGreenf}),
Kruppa has introduced a prescription~\cite{Kru98}, which is suitable to treat
the particle continuum by the diagonalization method with, e.g.,
the harmonic oscillator basis.
He has demonstrated that results with his prescription 
have much weaker dependence 
on the size of the basis and converge for enough large basis.
Let us call his prescription the Kruppa method.
This method changes the definition of $g(\epsilon)$ as
the difference of the single-particle level density
between the full Hamiltonian (including the potential) and
the free-particle Hamiltonian,
\begin{equation} \label{eq:KruppaDensity}
g(\epsilon)\quad \Rightarrow \quad
g\rsu{K}(\epsilon) = \sum_{i=1}^{M} \delta \left( \epsilon - \epsilon_i \right)
-  \sum_{j=1}^{M} \delta ( \epsilon - \efree{j} ) ,
\end{equation}
where
$\epsilon_i$ and $\efree{j}$ are the eigenvalues of the
full and the free Hamiltonians, respectively.
Here $M$ is the dimension of the basis commonly used in the two diagonalizations,
and $g\rsu{K}(\epsilon)\rightarrow g\rs{c}(\epsilon)$ as $M\rightarrow\infty$
(see Eq.~(\ref{eq:gGreenf})).
Note that, for one-body observables like the total single-particle energy
in Eq.~(\ref{eq:EspIntg}), the free energy terms
in Eq.~(\ref{eq:KruppaDensity}) do not contribute as long as $\lambda<0$,
i.e., when the Fermi energy does not exceed the particle threshold.
However, they contribute to the smoothed quantities.
Now the smoothed level density $\tilde{g}(\epsilon)$
should be obtained by applying the Strutinsky smoothing to this $g\rsu{K}(\epsilon)$:
\begin{equation} \label{eq:smoothedLevelDensityBK}
\tilde{g}(\epsilon)\quad \Rightarrow \quad
\tilde{g}\rsu{K}(\epsilon)
= \frac{1}{\gamma} \sum_{i=1}^{M} f_p \left( \frac{\epsilon - \epsilon_i}{\gamma} \right)-
\frac{1}{\gamma} \sum_{j=1}^{M} f_p \left( \frac{\epsilon - \epsilon^0_j}{\gamma} \right).
\end{equation}
The redefined level density $\tilde{g}\rsu{K}(\epsilon)$ converges
to $\tilde{g}\rs{c}(\epsilon)$ for sufficiently large basis sizes,
the reason of which is explained transparently in Sec.~\ref{sec:OBTF}.

The continuum level density was originally used to calculate the
second virial coefficient (related to the deviation of the equation of
state from that for the ideal gas) arising from the interaction
between gas particles~\cite{BU37,LandauLifshitz}.
For this purpose, one naturally has to separate the part corresponding
to the free motion of non-interacting particles from the integral over
the continuous spectrum.
Unlike this case, the reason to subtract the free spectrum is not so
obvious in the calculation of the shell correction.
At present, we do not know whether it can be derived rigorously from
a more basic theoretical framework.  Nevertheless, it certainly seems
to be the most reasonable prescription so far to obtain physically
meaningful results.

\begin{figure}[!ht] 
\includegraphics[width=75mm]{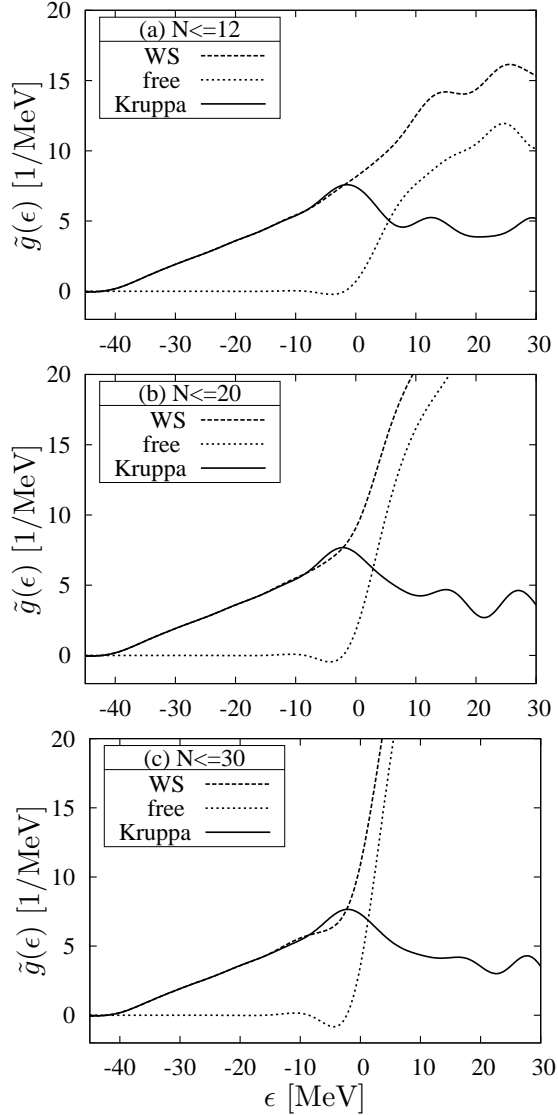}
\vspace*{-5mm}
\caption{
Smoothed level densities for the full and free Hamiltonians and their
difference (i.e., the Kruppa's level density) as functions of the
single-particle energy in MeV.
The smoothing parameters used are $\gamma=1.2\,\hbar\omega$ and $p=3$.
Panels (a), (b) and (c)
are for $\noscmax=12$, $20$ and $30$, respectively.
The nucleus is $^{164}$Er with deformation $\beta_2=0.27$ and $\beta_4=0.02$.
}
\label{fig:KruppaDensity}
\end{figure} 

In Fig.~\ref{fig:KruppaDensity} we show three kinds of level
densities, i.e., the full (with Woods-Saxon potential),
the free, and the Kruppa for $^{164}$Er.
They are the results of the Strutinsky smoothing
with $\gamma=1.2\,\hbar \omega$ and $p=3$.
The potential is deformed with $\beta_2=0.27$
and $\beta_4=0.02$.  The number of basis states $M$
in Eq.~(\ref{eq:KruppaDensity}) is specified
by the maximum number of the oscillator quanta $\noscmax$,
$M=\frac{1}{3}(\noscmax+1)(\noscmax+2)(\noscmax+3)$.
Comparing panels (a), (b) and (c),
one can see that positive energy part of the full and the free level densities
are increased
rapidly as $\noscmax$ is increased from 12 to 20 and to 30,
while the Kruppa's level density does not change essentially.
This clearly shows the fact that continuum parts of both
the full and free densities are divergent but their difference is convergent.
The energy range of the most influential part is $\spe \le \lambda$
for the smoothed single-particle energy $\tilde{E}\rs{s.p.}$
and $ \left\vert \spe - \lambda \right\vert \le \Lambda \sim \hbar \omega$
for the smoothed BCS energy $\tilde{E}\rs{BCS}$.
Though the difference in this part
between the calculated level density with $\noscmax=12$, 20 and 30
is much smaller than that in positive energy,
e.g., at $\spe \sim 10$MeV, 
it brings about significant differences to
the resulting nuclear properties,
especially to the pairing correlation (see Sec.~\ref{sec:resultKruppaBCS}).

All the smoothed quantities in the Kruppa method are obtained by 
replacing $\tilde{g}(\epsilon)$ with $\tilde{g}\rsu{K}(\epsilon)$.
The shell correction energy $E\rs{sh}$ by this prescription
is investigated in Ref.~\cite{VKN00}, and shown to be also convergent
when increasing the size of the basis.
Examples are depicted in Fig.~\ref{fig:KruppaEsh} as functions of the basis
cutoff $\noscmax$.  Without the Kruppa's prescription, $E\rs{sh}$
depends on the size of the basis even in a stable nuclei $^{166}$Er, and the dependence
is much stronger in a neutron rich nuclei $^{226}$Er.  The subtraction of the continuum
contributions reduces the dependence on the model space drastically
and the shell correction energy with the Kruppa method converges
in the large $\noscmax$ limit.
These examples clearly show that the Kruppa's prescription is indeed promising.
We extend it for other observables in Sec.~\ref{sec:ExtKruppa}.

It is also worth noting that
while the shell correction energy $E\rs{sh}$ almost
converges at $\noscmax\approx 12$, the sum of the single-particle
energies $E\rs{s.p.}$ itself does not;
especially for the unstable nuclei $^{226}$Er the single-particle energies
are not obtained very accurately when $\noscmax \le 20$.
From this fact one may compose a syllogism on the necessity of the Kruppa method.
(1) For large $\noscmax$, the Kruppa method is necessary to treat correctly the dense positive
energy spectrum. For small $\noscmax$, it is not necessary.
(2) One has to use large $\noscmax$ for sufficiently accurate bound-state energies.
(3) One needs the Kruppa method.

\subsection{Oscillator-basis Thomas-Fermi approximation for Kruppa's level density} 
\label{sec:OBTF}

One can roughly reproduce the shape of the Kruppa's level density 
in terms of a new variant of the Thomas-Fermi approximation
within the limited phase space corresponding to the truncated oscillator basis.
We call it the oscillator-basis Thomas-Fermi (OBTF) approximation
in this paper.
One can also demonstrate the independence of the results of the Kruppa method
from $\noscmax$ (if it is sufficiently large) in this approximation.

We study only spherically symmetric potentials without spin-orbit couplings.
Lifting these restrictions is possible but does not seem to be very meaningful,
because it turns out that the OBTF approximation
is not sufficiently accurate to be used as a replacement
of the smoothed energy in the realistic Strutinsky calculations.
This corresponds to the known fact that,
in order to obtain the Strutinsky smoothed energy,
one has to include up to the third order terms
in the semiclassical $\hbar^2$ expansion~\cite{Jen73,JBB75},
in which the Thomas-Fermi approximation is the lowest.

Hence we express the Hamiltonian for a nucleon as
\begin{equation} \label{eq:sphamiltonian}
H(p,r) = \frac{\bm{p}^2}{2m} + V(r), \;\;\; 
V(r)   = V\rs{CE}(r)+\frac{1}{2}\left(1-\tau_3\right) V\rs{CO}(r),
\end{equation}
where $V\rs{CE}(r)$ and $V\rs{CO}(r)$ are the central and Coulomb
potentials in Sec.\ref{sec:woodsSaxonPotential} with spherical shape,
i.e., with the deformation parameters $\bm{\beta}=\bm{0}$.
The states are assumed to be doubly degenerated for 
the two spin states $s_z=\pm \frac{1}{2}$. 
In the Thomas-Fermi approximation,
the number of particles in the potential well
for a given single-particle energy $\spe$ is given by
\begin{equation} \label{eq:gammae}
{\Gamma}(\spe)=4\pi\int_{0}^{\infty} \rhotf(r,\spe)r^2 dr
\end{equation}
where $\rhotf(r,\spe)$ is the particle density
at position $r$ for Fermi level $\spe$ expressed as
(using the Heaviside function  $\theta$),
\begin{equation} \label{eq:rhore}
\rhotf(r,\spe) = \frac{(2m)^{3/2}}{3\pi^2 \hbar^3}
\left\vert \spe - V(r) \right\vert^{3/2} \theta\left(\spe-V(r)\right).
\end{equation}
The level density is related to the number of particles ${\Gamma}(\spe)$ as
\begin{equation} \label{eq:gTilde}
\gtf(\spe) = \frac{d{\Gamma}(\spe)}{d\spe} =
4\pi \int_{0}^{\infty} \frac{d\rhotf(r,\spe)}{d\spe} r^2 dr.
\end{equation}
This level density diverges above the particle threshold, $\spe>0$,
for finite-depth potentials because of the infinite volume of the space.

The idea of OBTF is to extend the Thomas-Fermi approximation
in such a way that the phase space is limited within
a subspace spanned by a truncated harmonic oscillator (HO) basis,
which can be specified by the maximum kinetic energy
as a function of position as in the followings.  
By replacing $V(r)$ with the oscillator potential
$V\rs{HO}(r)=\frac{1}{2} m \omega^2 r^2$ in Eq.~(\ref{eq:rhore}), 
one obtains
\begin{equation} \label{eq:gammaec}
 {\Gamma}\rs{HO}(\spe)=\frac{1}{3}\left( \frac{\spe}{\hbar \omega}\right)^3,
\end{equation}
which is always finite.
A truncated oscillator basis is usually defined by the maximum
oscillator quantum number $\noscmax$. 
Equating the right-hand sides of Eqs.~(\ref{eq:gammaec}) to 
the number of states with $\nosc \le \noscmax$ leads to
the cutoff energy of the truncated basis, $\epsilon=\ecut(\noscmax)$,
\begin{equation}
\ecut = \hbar \omega \left[
(\noscmax+1)(\noscmax+2)(\noscmax+3)
\right]^{1/3}.
\end{equation}
We also define $\rmax$ by a condition $V\rs{HO}(\rmax)=\ecut$, i.e.,
\begin{equation}
  \rmax = \sqrt{\frac{2\ecut}{m\omega^2}},
\end{equation}
and the local maximum kinetic energy expressed as
\begin{eqnarray}
\ekmax(r) & = & \left(\ecut -
V\rs{HO}(r)\right) \theta\left(\ecut - V\rs{HO}(r)\right)
\nonumber \\
& = & \frac{1}{2}m\omega^2\left( \rmax^2-r^2\right)
\theta\left( \rmax-r\right). 
\label{eq:ekmax}
\end{eqnarray}
Now we define the level density in OBTF,
similarly to Eq.~(\ref{eq:gTilde}), as
\begin{equation}
\gobtf(\epsilon) = 
4 \pi \int_{0}^{\infty} \frac{d\rhoobtf(r,\spe)}{d\spe} r^2 dr,
\end{equation}
where $\rhoobtf(r,\epsilon)$ is equal to the right-hand side of 
Eq.~(\ref{eq:rhore}) with an additional restriction that the energy $\spe$
should be smaller than $\ekmax(r)+V(r)$.
Its derivative is given by (with the $\delta$-function
contribution from the Heaviside function having no effects),
\begin{equation} \label{eq:drhoBde}
\frac{d\rhoobtf(r,\spe)}{d\spe} = 
\frac{(2m)^{3/2}}{2\pi^2 \hbar^3}
\left\vert \spe - V(r) \right\vert^{1/2}
\theta\left( \spe - V(r) \right) \theta\left(\ekmax(r)+ V(r) -\spe \right).
\end{equation}
In this way, the finite level density $\gobtf(\spe)$ is obtained
for a given maximum oscillator quantum number $\noscmax$.

\begin{figure*}[htbp] 
\includegraphics[height=150mm,angle=270]{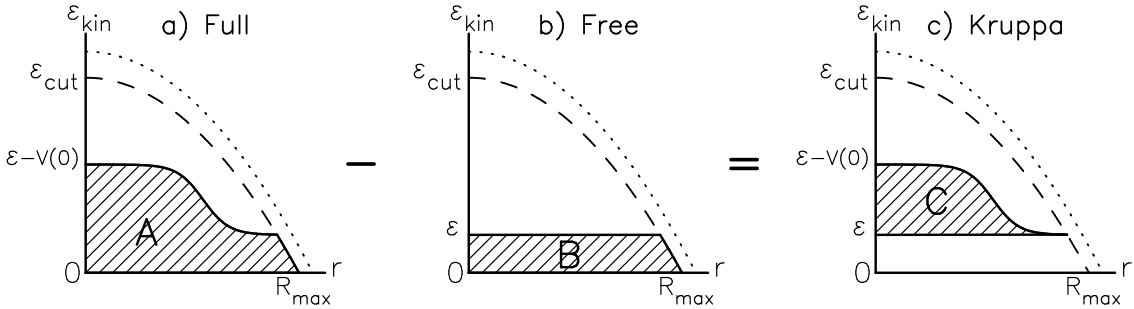}
\vspace*{1mm}
\caption{
A schematic figure to explain the convergence of the Kruppa's level
density in the oscillator-basis Thomas-Fermi approximation.  The
abscissa represents the radius $r$ from the center of the nucleus,
while the ordinate is the kinetic energy $\ekin=\spe-V(r)$ of a single nucleon.
Hatched area A, B, and C are the domain of integrations to obtain $\gammaobtf$,
$\gammaobtf^{0}$, and $\gammaobtf\rsu{K}$, respectively.
Parabolas drawn with dash and dot lines stand for
the maximum kinetic energy $\ekmax(\spe)$ in the oscillator basis
with $\noscmax$ and  ${\noscmax}'$ ($> \noscmax$), respectively.
}
\label{fig:OBTFsketch}
\end{figure*} 

The Kruppa's level density $\gobtf\rsu{K}(\spe)$ is defined as
$\gobtf\rsu{K}(\spe)=\gobtf(\spe)-\gobtf^{0}(\spe)$,
where $\gobtf^{0}(\spe)$ is the
free-particle level density expressed as
$\gobtf^0(\epsilon) = 4 \pi {\displaystyle \int}
\left[ d\rhoobtf^0(r,\epsilon)/d\epsilon \right] r^2 dr$
with $\rhoobtf^0(r,\epsilon)$ obtained by omitting
$V(r)$ in the right-hand side of Eq.~(\ref{eq:drhoBde}).
(This is for neutrons and changes necessary for protons are
described in the following paragraph).
It is readily shown that
\begin{equation} \label{eq:GammaOBTFasTwoDimIntegral}
\gammaobtf(\spe) 
= \int_{-\infty}^{\spe} \gobtf(\spe') d \spe'
= \frac{2(2m)^{3/2}}{\pi \hbar^3} \int\rs{A} \ekin^{1/2}\, r^2 d\ekin dr,
\end{equation}
where $\ekin=\spe'-V(r)$,
and the domain A of integration is  depicted in panel (a) of
Fig.~\ref{fig:OBTFsketch}. Changing the domain to B and C shown in panels~(b)
and (c) gives the similar expressions for 
\dm{\gammaobtf^{0}(\spe)=\int_{-\infty}^{\spe}\gobtf^{0}(\spe') d\spe'}
and
\dm{\gammaobtf\rsu{K}(\spe)=\int_{-\infty}^{\spe}\gobtf\rsu{K}(\spe') d\spe'},
respectively.
By enlarging the harmonic oscillator basis (i.e., by increasing
$\noscmax$ to ${\noscmax}' > \noscmax$), domains A and B are expanded
while domain C is left unchanged. The unchanged domain results in an
unchanged number of levels and thus an unchanged level density.  This
explains pictorially why the Kruppa's level density converges for
large $\noscmax$ above the particle threshold, $\spe>0$.

It is also possible to show that
$\gobtf\rsu{K}(\epsilon) \propto \epsilon^{-1/2}$
as $\epsilon\rightarrow\infty$
after the limit $\noscmax\rightarrow\infty$ is taken.
For an arbitrarily given $\epsilon>0$,
one can take sufficiently large $\noscmax$ to express
the region C as
$
  \left\{ \left( \ekin, r \right) \vert 
  \, 0 \le r < \infty, \; \epsilon \le \ekin \le \epsilon - V(r) \right\}
$
with an approximation that $V(r)=0$ for $r > R\rs{max}$ to obtain 
\begin{equation} \label{eq:gammaobtfkruppa}
 \gammaobtf\rsu{K}(\epsilon)\approx
  \frac{4(2m)^{3/2}}{3\pi \hbar^3} \int_0^{\infty}
 \left\{ \left[ \epsilon -V(r) \right]^{3/2} - \epsilon^{3/2} \right\} r^2 dr .
\end{equation}
Thus, for the level density
$g\rs{OB}\rsu{K}(\epsilon) = d \gammaobtf\rsu{K}(\epsilon)/d\epsilon$,
\begin{eqnarray}
  \epsilon^{1/2} g\rs{OB}\rsu{K}(\epsilon) & \approx &
- \frac{2(2m)^{3/2}}{\pi \hbar^3} \int_0^{\infty}
\frac{V(r) r^2 dr}{1+ \left[ 1- V(r)/\epsilon \right]^{1/2}}
\nonumber \\
& \rightarrow &
- \frac{(2m)^{3/2}}{\pi \hbar^3} \int_0^{\infty} V(r) r^2 dr 
\qquad (\epsilon \rightarrow \infty).
 \label{eq:cobk}
\end{eqnarray}
It can be confirmed that the following expression is
a very good approximation for the $\noscmax\rightarrow\infty$ limit
of the Kruppa level density
in the whole range of single-particle energy:
\begin{equation}\label{eq:KgOBapprox}
\gobtf\rsu{K}(\spe) \approx
\frac{2(2m)^{3/2}}{\pi\hbar^3}
 \int_0^\infty
 \left[\left(\spe-V(r)\right)^{1/2}\theta\left(\spe-V(r)\right)
 -\spe^{1/2}\theta\left(\spe\right)\right]\,r^2dr.
\end{equation}

For protons, 
one can repeat the same argument if one includes $V\rs{CO}(r)$
in the free Hamiltonian, because $V\rs{CO}(r)$ is not negligible even at
$r=R\rs{max}$ and $V(r)$ includes $V\rs{CO}(r)$.  In the same way,
it is readily seen that the Kruppa level density is convergent
as $\noscmax\rightarrow\infty$ and in a very good approximation,
\begin{equation}\label{eq:KgOBapproxProton}
\gobtf\rsu{K}(\spe) \approx
\frac{2(2m)^{3/2}}{\pi\hbar^3}
 \int_0^\infty
 \left[\left(\spe-V(r)\right)^{1/2}\theta\left(\spe-V(r)\right)
 -\left(\spe-V\rs{CO}(r)\right)^{1/2}\theta\left(\spe-V\rs{CO}(r)\right)
 \right]\,r^2dr.
\end{equation}

\begin{figure*}[htb] 
\includegraphics[width=150mm]{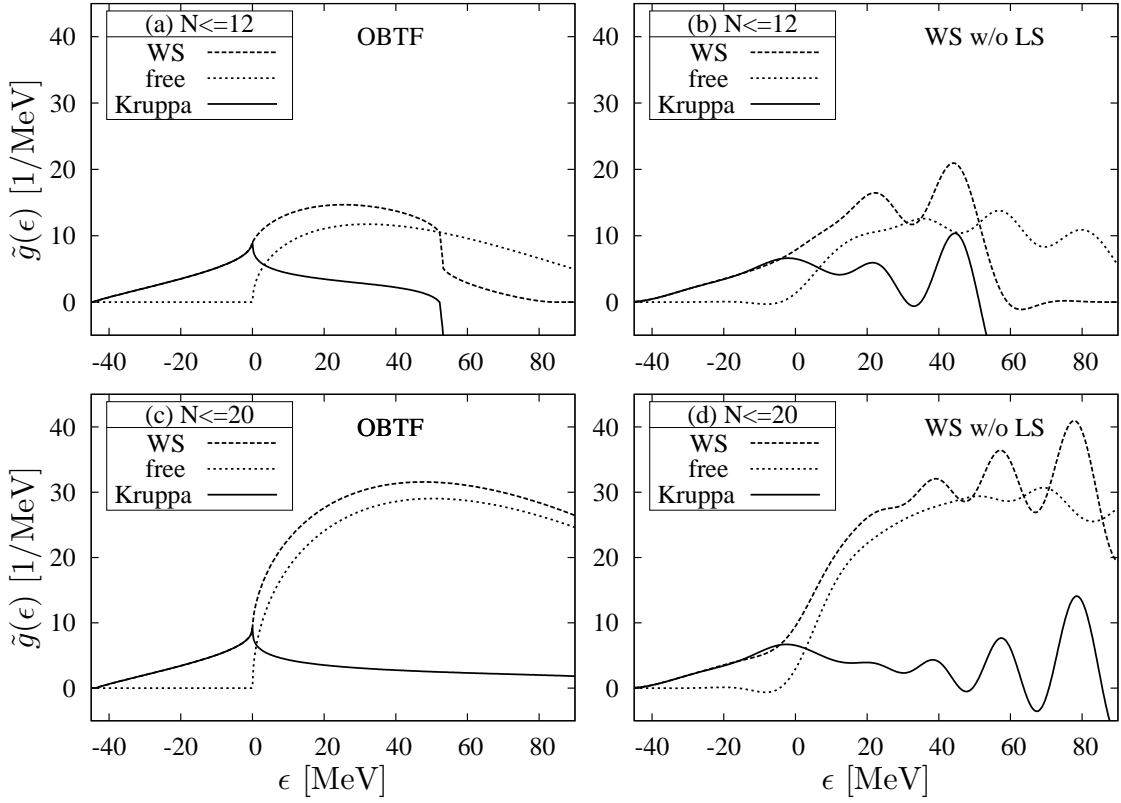}
\vspace*{-5mm}
\caption{
Neutron's level densities for the full and free Hamiltonians and their
differences obtained in the oscillator-basis Thomas-Fermi
approximation ((a),(c)) or with the Strutinsky smoothing method ((b),(d)).
The smoothing parameters are $\gamma=1.8\hbar\omega$ and $p=3$.
The oscillator basis has $\noscmax=12$ ((a),(b)) or  $\noscmax=20$ ((c),(d)).
The nucleus is $^{154}$Er.
The potential is spherical ($\beta_2=\beta_4=0$) and
the spin-orbit potential is turned off.
}
\label{fig:OBTFdensity}
\end{figure*} 

\begin{figure}[htb] 
\includegraphics[width=75mm]{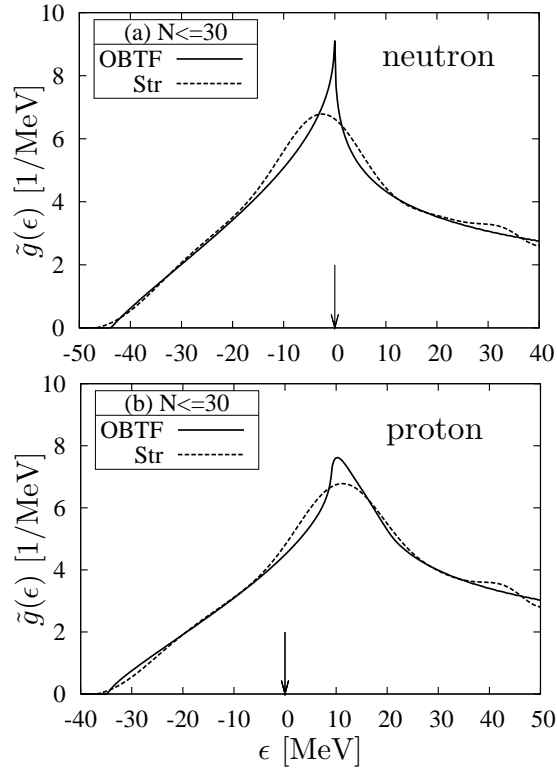}
\vspace*{-5mm}
\caption{
Kruppa Level densities obtained in the OBTF
approximation and with the Strutinsky method for $^{154}$Er;
panel (a) is for neutron and (b) for proton.
The arrows denote the particle threshold ($\spe=0$).
The same calculation as in Fig.~\ref{fig:OBTFdensity} is used except
that the oscillator basis has $\noscmax=30$.
}
\label{fig:OBTFKruppa}
\end{figure} 

In Figs.~\ref{fig:OBTFdensity} and~\ref{fig:OBTFKruppa},
the OBTF level density is compared
with the smoothed exact level density for a spherical nucleus $^{154}$Er.
Figure~\ref{fig:OBTFKruppa} includes the proton Kruppa densities.
The spin-orbit force is neglected and a larger smoothing parameter
$\gamma=1.8\hbar\omega$ is used with $p=3$
for this calculation to make the comparison more appropriate.
One can see that the OBTF is a fairly good approximation
for both $\noscmax=12$ and $\noscmax=20$.
The proton Kruppa level density is very similar to the neutron one
except that the single particle energy is shifted by the Coulomb barrier,
about 10 MeV in this nucleus, as is shown in Fig.~\ref{fig:OBTFKruppa}.
Threshold behaviors
of the OBTF neutron and proton level densities are slightly different,
which reflects the effect of the long-range Coulomb potential,
while such differences do not exist for the smoothed exact level densities.

Apart from oscillations at large $\spe$, the average behavior
of the continuum level density is well reproduced in the OBTF approximation
in Fig.~\ref{fig:OBTFdensity}.
It can also be clearly seen how subtracting the free level density, 
Eq.~(\ref{eq:KruppaDensity}), works to diminish the dependence on
$\noscmax$ in the Kruppa method.  However, as it is clearly shown
in Fig.~\ref{fig:OBTFKruppa}, the precise shape of
the smoothed level density cannot be obtained; especially,
the peak near the threshold ($\epsilon \approx 0$) shows cusp behavior
in the OBTF density, which is characteristic to the semiclassical approximation,
while the smoothed density looks like a broad peak.
In order to obtain more precise level density one has to go beyond
the Thomas-Fermi approximation.

The Kruppa OBTF level density $\gobtf\rsu{K}(\spe)$ becomes negative
in the range of single-particle energy, $\ecut+V(0) \le \spe \le \ecut$
(see Fig.~\ref{fig:OBTFdensity}~(a)), and it can be shown,
for a finite $\noscmax$, that the positive and negative contributions exactly
cancel out, \dm{\int_{-\infty}^\infty \gobtf\rsu{K}(\spe)d\spe=0}.
This behavior is also known~\cite{Shl92}
in the exact continuum level density $g\rs{c}(\epsilon)$ defined
by Eqs.~(\ref{eq:gphaseshift})$-$(\ref{eq:gGreenf}),
and reflects the Levinson's theorem~\cite{GW64}, 
i.e., \dm{\int_{-\infty}^\infty g\rs{c}(\spe)d\spe=0}.
In this way the OBTF Kruppa level density satisfies the desired property
of the continuum level density.

\subsection{Plateau condition}
\label{sec:plateau}

It would be preferable if $\tilde{E}\rs{s.p.}$ of Eq.~(\ref{eq:Esmoothed})
did not depend on the parameters concerning the smoothing of the level
density ($\gamma$ and $p$ in Eq.~(\ref{eq:smoothedLevelDensity}))
because their values can be chosen arbitrarily.
Since a perfect independence is unlikely to be satisfied,
one usually demands a weaker condition that
the dependence is very weak in a certain interval of $\gamma$
for a few values of $p$.
This is the meaning of the plateau condition in this paper.

For the Nilsson spectrum, a long plateau appears in most cases
(see, e.g., Ref.~\cite{NilStr,RB72}).
On the other hand,
for finite-depth potentials like the Woods-Saxon potential,
the situation is subtle.  If the oscillator basis is truncated at
$\noscmax \approx 10$ to $12$, reasonable plateau are
obtained in many cases~\cite{BFN72},
and $\noscmax=12$ is a recommend value
as a working prescription in Ref.~\cite{MNM95}.  
However, the model space defined by $\noscmax=12$ is not large enough
to calculate single-particle states accurately, especially for unstable nuclei
(see, e.g., Fig.~\ref{fig:KruppaEsh}~(d)), and this truncation is not justified.
The appearance of plateau obtained by the relatively small model space
with $\noscmax \approx 10$ to $12$ is accidental and
increasing $\noscmax$ drastically change the situation~\cite{BFN72,NWD94};
the shell correction energy depends strongly on the smoothing width $\gamma$.
This clearly indicates that a na{\"i}ve inclusion of
continuum states by the diagonalization method does not work.
Then, the continuum level density, Eq.~(\ref{eq:gphaseshift}), is used
to calculate the shell correction energy~\cite{Lin70,RB72}.
Although the dependence on $\gamma$ is weaker if the phase shift is
calculated up to enough high energies, no good plateau like in the case
of the Nilsson potential is obtained~\cite{VKL98}.

\begin{figure*}[htb] 
\includegraphics[width=150mm]{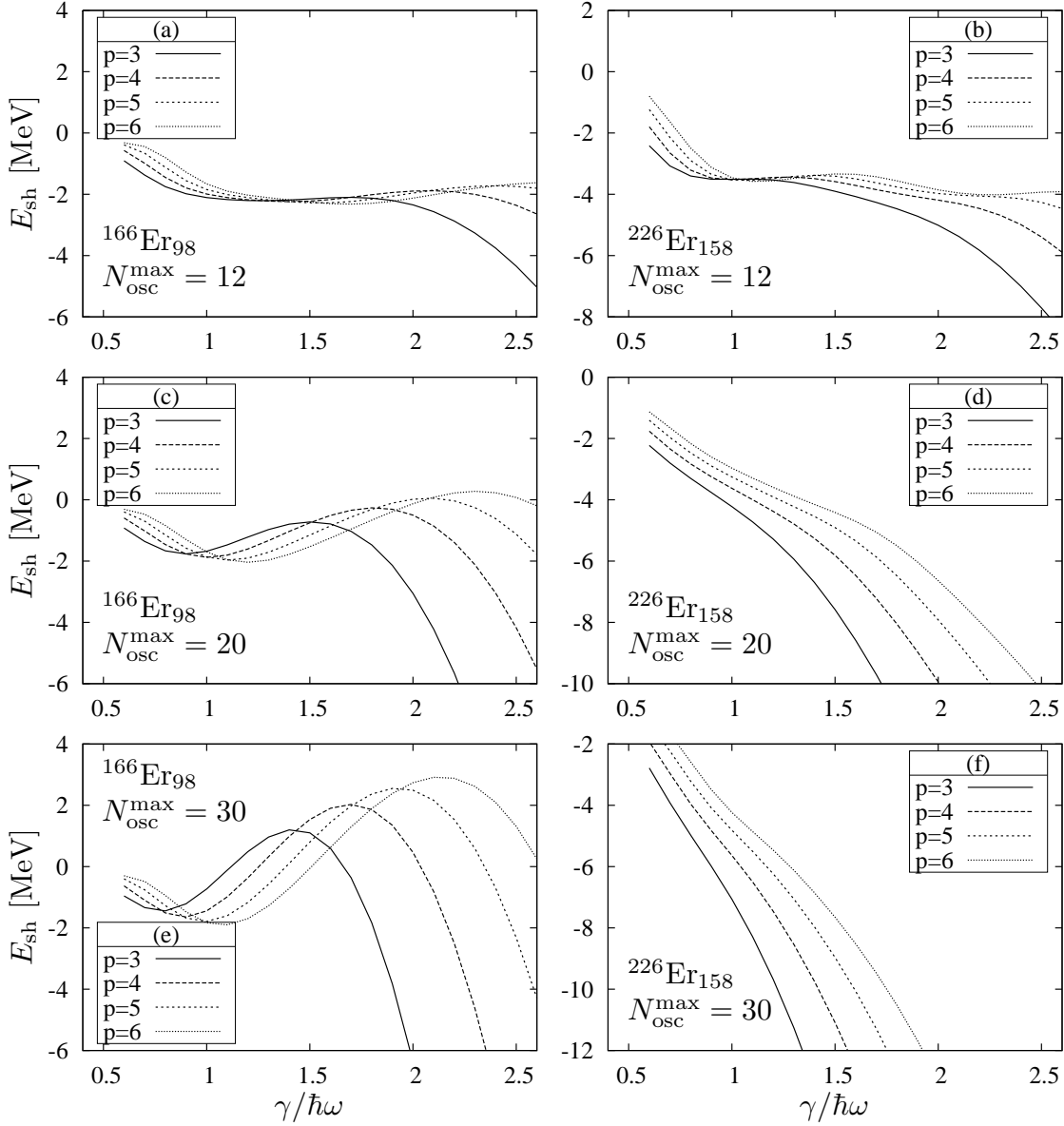}
\vspace*{-5mm}
\caption{
Neutron shell correction energies 
without using the Kruppa's prescription
as functions of the smoothing parameter $\gamma$ in units of $\hbar\omega$.
Each curve represents the result with different order $p=3$ to 6
of the smoothing function~(\ref{eq:f}).
The diagonalization basis is $\noscmax=12$ ((a),(b)), 
$\noscmax=20$ ((c),(d)), and $\noscmax=30$ ((e),(f)).
The nucleus is $^{166}$Er ((a),(c),(e)) and $^{226}$Er ((b),(d),(f)).
The deformation parameters are the same as in Fig.~\ref{fig:KruppaEsh}.
}
\label{fig:EshPlat1}
\end{figure*} 

\begin{figure*}[htb] 
\includegraphics[width=150mm]{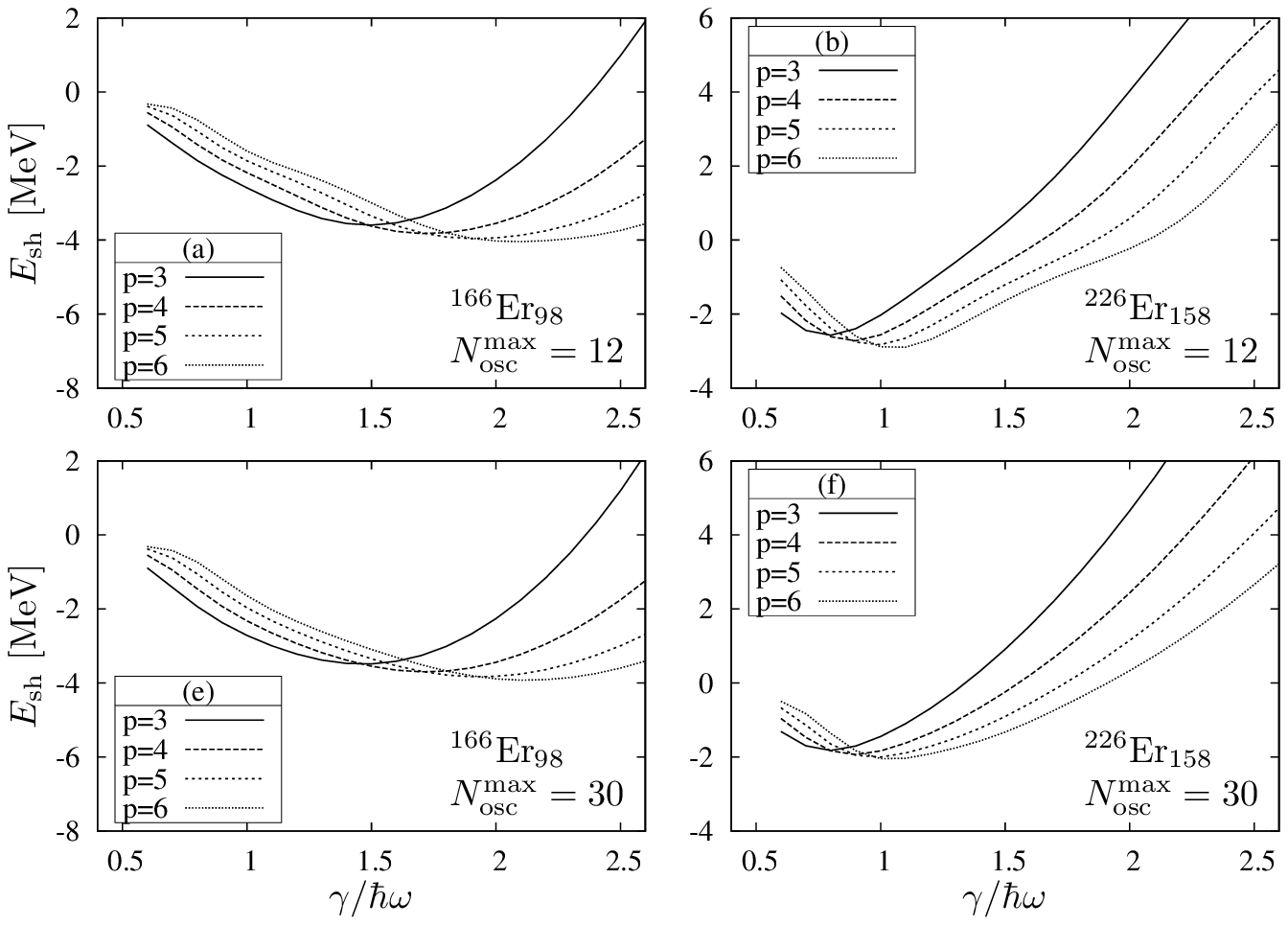}
\vspace*{-5mm}
\caption{
Same as Fig.~\ref{fig:EshPlat1} but with the Kruppa's prescription employed.
The results with $\noscmax=20$ are omitted since they are very similar to
the results with $\noscmax=30$.
}
\label{fig:EshPlat2}
\end{figure*} 

\begin{figure*}[htb] 
\includegraphics[width=150mm]{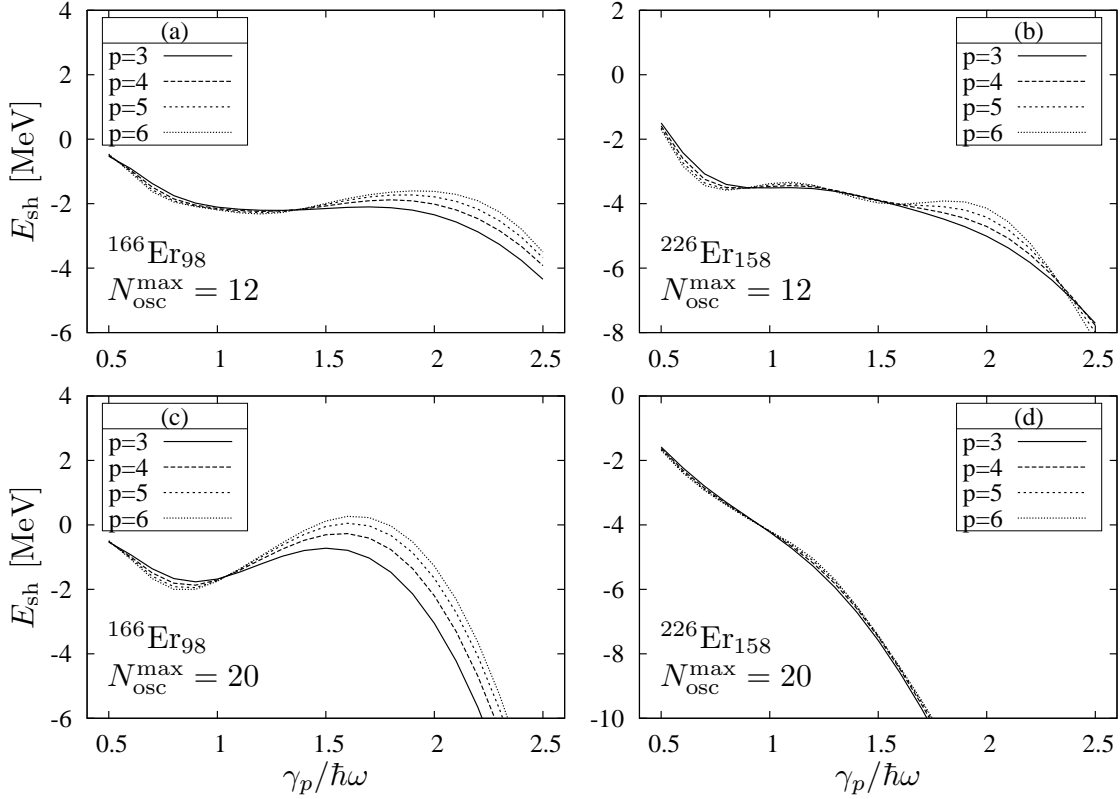}
\vspace*{-5mm}
\caption{
Same as Fig.~\ref{fig:EshPlat1} but plotted as functions of
the scaled smoothing width parameter $\gamma_p=\gamma/\sqrt{p/3}$.
The results with $\noscmax=30$ are omitted.
}
\label{fig:EshPlat3}
\end{figure*}

\begin{figure*}[htb] 
\includegraphics[width=150mm]{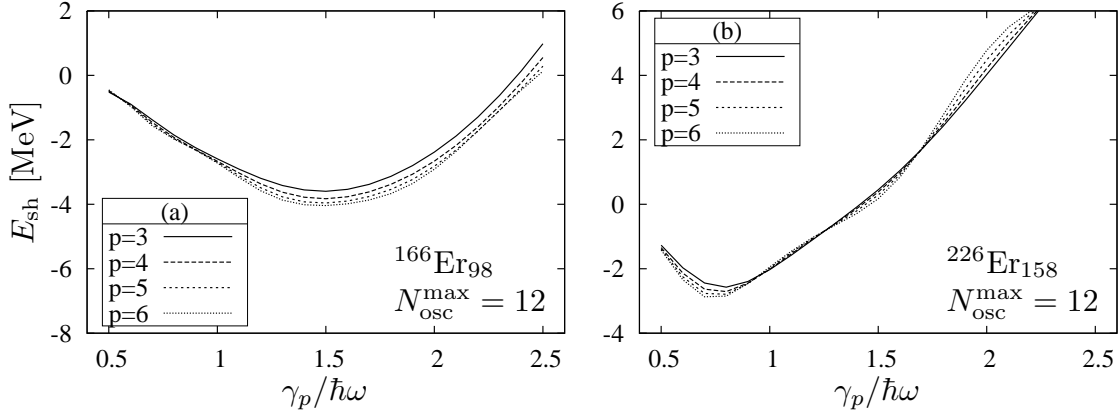}
\vspace*{-5mm}
\caption{
Same as Fig.~\ref{fig:EshPlat2} but plotted as functions of
the scaled smoothing width parameter $\gamma_p=\gamma/\sqrt{p/3}$.
Only the results of $\noscmax=12$ are shown
since the results with different $\noscmax$ look similar.
(See the results with $\noscmax=30$ in panels (a) and (c)
of Fig.~\ref{fig:refdnsEnergy}.)
}
\label{fig:EshPlat4}
\end{figure*}

We show examples in Figs.~\ref{fig:EshPlat1}~to~\ref{fig:EshPlat4}.
Figures~\ref{fig:EshPlat1} and~\ref{fig:EshPlat2} depict
the neutron shell correction energies $E\rs{sh}$
calculated with the standard Strutinsky smoothing method and
with the Kruppa-method, respectively, changing the basis size specified
by the maximum oscillator quantum number $\noscmax$.
The results for the stable and unstable nuclei,
$^{166}{\rm Er}$ and $^{226}{\rm Er}$,
are compared.
If $\noscmax=12$ is used for the basis size, a plateau-like behavior
is observed in a reasonably long range for $^{166}{\rm Er}$ and
in a shorter range for $^{226}{\rm Er}$.
But this is ``spurious'' because, using larger $\noscmax$, the shell correction
energy depends more strongly both on the smoothing width $\gamma$ and
the order $p$ of the curvature correction polynomial,
while the range of ``real'' plateau
in the case of the harmonic oscillator potential
grows as the basis size increases~\cite{BP73}.
The possible reason of this ``spurious'' plateau is that
the number of discretized continuum states with $\noscmax=12$ 
is just suitable for the smoothed level density to be approximated by
the lower order polynomial functions across the particle threshold
$\epsilon \approx 0$.
Increasing the basis size the curvature of the smoothed level density
changes suddenly at $\epsilon \approx 0$,
as is shown in Fig.~\ref{fig:KruppaDensity}, which no longer be
approximated by a simple polynomial; leading to the strong dependence
of $E\rs{sh}$ on $\gamma$ and $p$.
Therefore, it is difficult to obtain reliable shell correction energies
in the standard Strutinsky method.

In contrast, the Kruppa's prescription reduces the basis-size dependence
dramatically, as can be seen in Fig.~\ref{fig:EshPlat2}.
Compared with the standard method, where the plateau condition
is more and more unsatisfied as increasing the basis size,
the stability against the increase of $\noscmax$ is a very important
feature of the Kruppa method.
However, although there are almost degenerate local minima
with different order $p$'s, the plateau is not well established generally.
The situation is worse for unstable nucleus $^{226}$Er.
A possible improvement will be discussed
in Secs.~\ref{sec:refdens}~to~\ref{sec:improvedplateau}.

At first sight, the dependence of the shell correction energy on
the smoothing width $\gamma$ is quite different when the order $p$
of the smoothing function~(\ref{eq:f}) is changed.
Close inspection reveals, however, that the different curves
in each panel of Figs.~\ref{fig:EshPlat1} and~\ref{fig:EshPlat2}
are almost isomorphic if they are drawn as functions of
the $\sqrt{p}\,$--scaled width parameter,
\begin{equation} \label{eq:scaledgamma}
 \gamma_p \equiv \gamma/\sqrt{p/3},
\end{equation}
as shown in Figs.~\ref{fig:EshPlat3} and~\ref{fig:EshPlat4}.
Here we choose $\gamma_{p=3}=\gamma$ because $p=3$ is a standard choice
for the curvature correction polynomial.
The reason of this ``isomorphism'' between the results
with different order $p$'s can be understood
from the discussion in Sec.\ref{sec:strutinskySmoothing};
the range of the low-pass filter increases when employing the larger order $p$,
and if is used the variable scaled with $\sqrt{p}$ the cutoff ranges
are the same but the filter becomes sharper,
as is clearly shown in Fig.~\ref{fig:scaled_filter}~(b).
Therefore, the complete isomorphism means the shell correction energy
is independent of the sharpness of the filter.
We have found that, for calculation with larger $\noscmax$,
better isomorphism is generally obtained
by the Kruppa smoothing method than by the standard one;
compare Fig.~\ref{fig:EshPlat3} with Figs.~\ref{fig:EshPlat4}.
Even better isomorphism is obtained in the improved treatment
in Sec.~\ref{sec:refdens} (see Fig.~\ref{fig:refdnsEnergy}).
In the following discussions for the plateau condition,
we always use the Kruppa prescription and show the results
as functions of the $\sqrt{p}\,$--scaled
width parameter $\gamma_p$~(\ref{eq:scaledgamma}).

In the course of writing the present paper, we noticed that a similar
scaled smoothing width is used for investigating the plateau condition
in Ref.~\cite{SKVprep10}, where
the isomorphism of the smoothing width dependence
between different order $p$'s is not as good as in our case.
This is due to a different choice of basic smoothing function
that is not gaussian.  We believe that the Fourier transform
of the smoothing function will be useful for more detailed comparison
of our results with those of Ref.~\cite{SKVprep10}.

\subsection{The reason for no good plateaux}
\label{sec:noplateau}

A clue to find the origin of this difference between the Nilsson
(or the harmonic oscillator) potential
and the Woods-Saxon (or the finite-depth, in general) potential
is the fact that the Strutinsky smoothing is a low-pass filter
as discussed in Sec.~\ref{sec:strutinskySmoothing}.
In the Fourier transformed world, the smoothed level density
is simply the original density multiplied by the filter.
Therefore, the Fourier transform of the original level
density~(\ref{eq:discreteLevelDensity}),
or the Kruppa density~(\ref{eq:KruppaDensity}),
\begin{equation} \label{eq:gFourier}
 \hat{g}(\tau)=\sum_{i=1}^M \exp(-i\tau \epsilon_i),
 \qquad
 \hat{g}\rsu{K}(\tau)=
 \sum_{i=1}^M \exp(-i\tau \epsilon_i)
                    -\sum_{j=1}^M \exp(-i\tau \epsilon^0_j)
\end{equation}
should be investigated.

In this subsection, we employ the units
$\hbar \omega$ for the energy ($\epsilon$, $\gamma$, and $\sigma$)
and $(\hbar \omega)^{-1}$ for the Fourier transformed time
variable $\tau$, and regard $\spe$ and $\tau$ as if they were dimensionless.

\begin{figure*}[htb] 
\includegraphics[width=150mm]{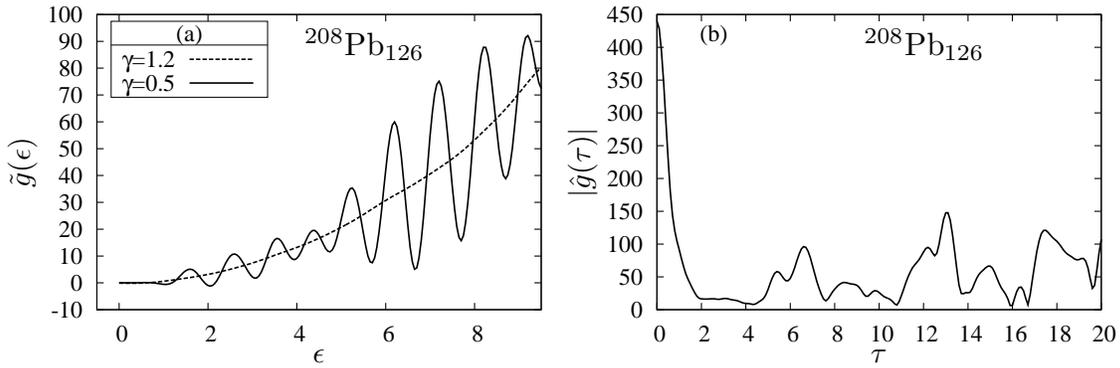}
\vspace*{-5mm}
\caption{
Neutron's level density of the Nilsson Hamiltonian in panel (a)
and the absolute value of its Fourier transform in (b) for $^{208}$Pb.
$\noscmax=9$ is used for the basis size.
In panel (a), smoothed level densities with $\gamma=1.2\, \hbar \omega$
and with $\gamma=0.5\, \hbar \omega$ ($p=3$ for both) are included,
while the Fourier transform in (b) has not been smoothed and calculated
directly by Eq.~(\ref{eq:gFourier}).
The units are $\hbar \omega$ for $\epsilon$,
$(\hbar \omega)^{-1}$ for $\tau$ and
$\tilde{g}$, and $\hat{g}$ is dimensionless.
}
\label{fig:nilssonLevelDensities}
\end{figure*} 

The level density $g(\epsilon)$
and its Fourier transform $\hat{g}(\tau)$ for the Nilsson potential
are shown in Fig.~\ref{fig:nilssonLevelDensities} for the spherical
nucleus $^{208}$Pb.
The single-particle states up to $\noscmax=9$
are included because the $ls$ and $l^2$ parameters are given
only for them~\cite{BR85}.
Fig.~\ref{fig:nilssonLevelDensities}~(a) shows
that the smoothed level density with $\gamma=1.2\,\hbar\omega$ is approximately
a quadratic function in $\spe$, while the major shell oscillation
is clearly seen in that with $\gamma=0.5\,\hbar\omega$.
This indicates that the semiclassical
property of the Nilsson spectra is essentially the same
as that of the HO potential; its Thomas-Fermi level density
is $\gtf\rsu{HO}(\spe)=\spe^2/(\hbar\omega)^3$, see Eq.~(\ref{eq:gammaec}).
Note that the so-called ``iso-stretching'' is done for the neutron
and proton frequencies in the Nilsson potential~\cite{NR95},
$\omega\rs{n}=(2N/A)^{1/3}\omega$ and $\omega\rs{p}=(2Z/A)^{1/3}\omega$,
so that the coefficient of $\spe^2$ is reduced by
a factor $208/(2\times 126)$ in Fig.~\ref{fig:nilssonLevelDensities}~(a).

Concerning the behavior of $\hat{g}(\tau)$
shown in Fig.~\ref{fig:nilssonLevelDensities}~(b),
one can see a very low density interval ($2 <  \tau < 5 $).
One may probably call its origin as the harmonicity of the potential.
If the cutoff period $\tau_p\rsu{cut} = 2\sqrt{p} \gamma^{-1}$
of the filter $\hat{f}_{p}(\tau \gamma)$ is in this interval,
the result of the filtering, $\hat{\tilde{g}}(\tau)$,
hardly depends on $\tau_p\rsu{cut}$ (see Sec.~\ref{sec:strutinskySmoothing}).
Namely, the
dependences on $\gamma$ and $p$ are weak and there appears a plateau.
This feature can be qualitatively understood by considering the case
of the anisotropic harmonic oscillator potential, for which
the spectra are equidistant and the sum in Eq.~(\ref{eq:gFourier})
with $M\rightarrow\infty$
can be evaluated as an infinite geometric series to be
\begin{equation} \label{eq:gHOFourier}
 \hat{g}\rs{HO}(\tau)=\left[
 (2i)^3 \sin\left(\frac{1}{2}\tau\hbar\omega_x\right)
        \sin\left(\frac{1}{2}\tau\hbar\omega_y\right)
        \sin\left(\frac{1}{2}\tau\hbar\omega_z\right)\right]^{-1},
\end{equation}
with $\omega_x\approx \omega_y \approx \omega_z\approx \omega$,
which has a long low density interval between $\tau=0$ and
$\tau=2\pi$ (in units of $(\hbar \omega)^{-1}$).

\begin{figure*}[htb] 
\includegraphics[width=150mm]{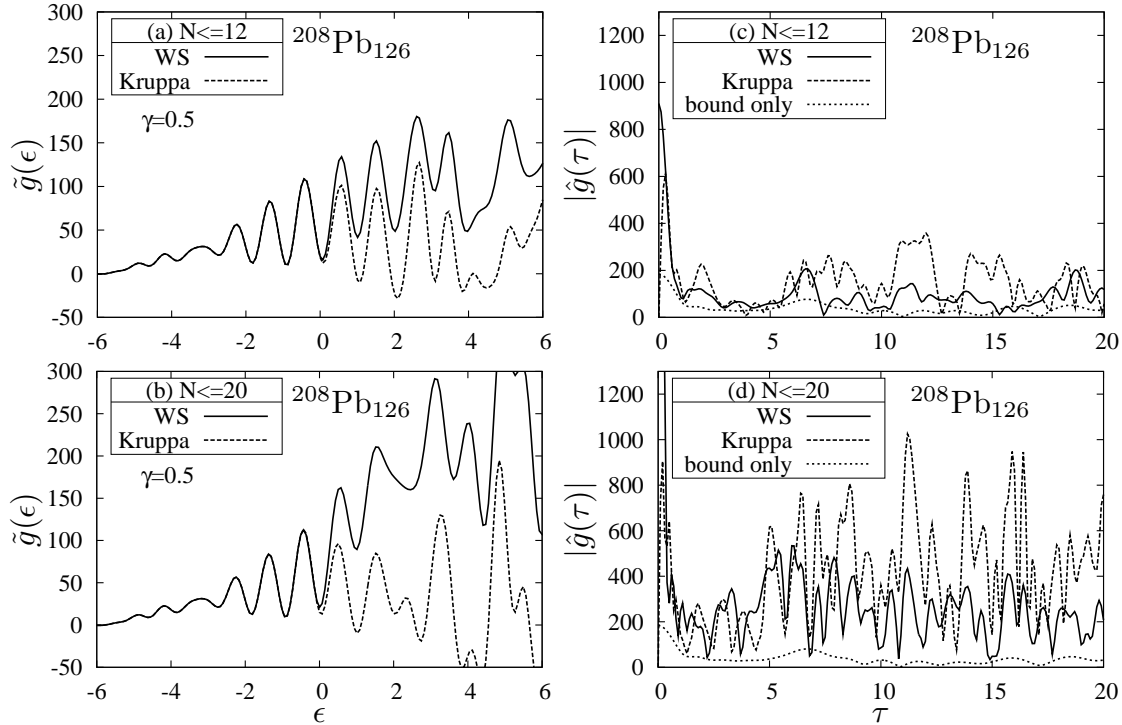}
\vspace*{-5mm}
\caption{
Neutron's level densities of the Woods-Saxon Hamiltonian
((a),(b)) and their Fourier transform ((c),(d)).  
The curves in panels (a) and (b) are the results of smoothing
with $\gamma=0.5\, \hbar \omega$ and $p=3$,
while those in (c) and (d) have not been smoothed.
The harmonic oscillator basis for the diagonalization
is $\noscmax=12$ in panel (a) and (c), and $\noscmax=20$ in (b) and (d).
The full and Kruppa spectra are shown with solid and dash curves, respectively.
In panels (c) and (d), the Fourier transform of only negative energy levels
are also shown with dot curves.
The units are the same as in Fig.~\ref{fig:nilssonLevelDensities}.
}
\label{fig:WoodsSaxonLevelDensities}
\end{figure*} 

In contrast to the case of Nilsson potential,
the Fourier transform of the Woods-Saxon level density
shown in Fig.~\ref{fig:WoodsSaxonLevelDensities}
does not have this low-amplitude region, although the smoothed
level density in the panels (a) and (b) clearly shows the similar
major shell oscillation to that in Fig.~\ref{fig:nilssonLevelDensities}~(a).
In Fig.~\ref{fig:WoodsSaxonLevelDensities} the Fourier transforms
of not only the Woods-Saxon spectra but of the Kruppa spectra and
of the restricted spectra within the bound states are also depicted.
The Fourier transform of the bound-states spectra
is smaller than that of the Woods-Saxon spectra,
but is still about a factor two to three larger
than that of the Nilsson spectra in the low
density interval ($2<\tau<5$) (note the difference of scale
in ordinates in Fig.~\ref{fig:nilssonLevelDensities}
and~\ref{fig:WoodsSaxonLevelDensities}).
Moreover, the Fourier transform of the Kruppa spectra is larger
than that of the Woods-Saxon spectra on average, and
increasing the basis size makes the situation worse.
This clearly shows that the Kruppa prescription does not help
to make a plateau in the shell correction energy,
which is already confirmed in the previous subsection.
The mechanism to develop a long plateau in the Nilsson spectrum
is not functioning in the Woods-Saxon spectrum.
This seems to be the very reason 
for the absence of plateau for the Woods-Saxon spectrum.

\begin{figure*}[htbp] 
\includegraphics[width=150mm]{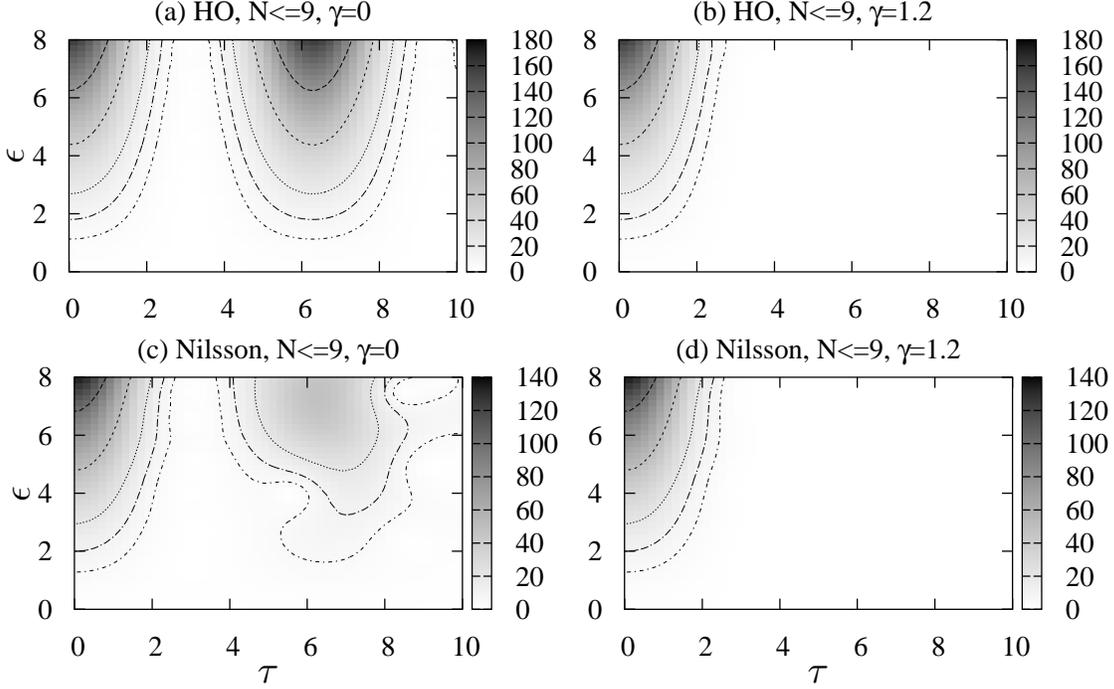}
\vspace*{-5mm}
\caption{
Absolute value of the short-time Fourier transform of the
level densities of the harmonic oscillator ((a),(b))
and the Nilsson ((c),(d)) Hamiltonians. 
The units are $\hbar \omega$ for $\epsilon$ and
$(\hbar \omega)^{-1}$ for $\tau$.  See text for explanations.
}
\label{fig:shortTimeFourierTransformResultHONIL}
\end{figure*}

The Woods-Saxon potential is different from the Nilsson potential
not only in the anharmonicity but also in the finite depth.
It seems interesting to investigate further the difference between
negative and positive parts of the spectrum.
In order to examine this point, the short-time Fourier transform
\cite{shortTimeFourierTransformation} seems
useful. It is defined by
\begin{equation} \label{eq:shortTimeFourierTransform}
\begin{array}{lll}
\hat{F}(k, x ; \sigma) &=&
{\displaystyle
 \int_{-\infty}^{\infty}
 F(x') w_\sigma(x'-x) e^{-ikx'} dx', }
 \vspace*{1mm}\\
 &=&
{\displaystyle
 \frac{1}{2\pi}e^{-ikx} \int_{-\infty}^{\infty}
 \hat{F}(k') \hat{w}_\sigma(k-k') e^{ik'x} dk',}\end{array}
\end{equation}
where for the window function $w_\sigma(\xi)$ and its Fourier transform
we employ
\begin{equation} \label{eq:windowFunction}
w_\sigma(\xi)=e^{-(\xi /\sigma)^2},\qquad
\hat{w}_\sigma(\kappa)=\sqrt{\pi}{\sigma}\,e^{-(\sigma\kappa /2)^2}.
\end{equation}
In this paper we apply it to a ``short energy interval'' Fourier transform.
Figures~\ref{fig:shortTimeFourierTransformResultHONIL}~to~%
\ref{fig:shortTimeFourierTransformResultWS20}
show $\hat{g}(\tau, \epsilon ; \sigma)$ with $\sigma=\sqrt{2}$,
which gives the same size of window widths
in two complementary variables $\tau$ and $\spe$.
The same nucleus $^{208}$Pb is used for this calculation.
The input level density $\tilde{g}(\spe)$ has been smoothed
with $\gamma=1.2$ and $p=3$ in the right-hand panels
(no smoothing has been done for the left-hand panels).
The location of cutoff due to these smoothings are
$\tau_p\rsu{cut}=2\sqrt{p}\gamma^{-1}=2.9$ for $\gamma=1.2$,
but then the cutoff result is blurred by the convolution with the
window function of width $\sqrt{2}/\sigma=1$
(see Eq.~(\ref{eq:shortTimeFourierTransform})).

\begin{figure*}[htbp] 
\includegraphics[width=150mm]{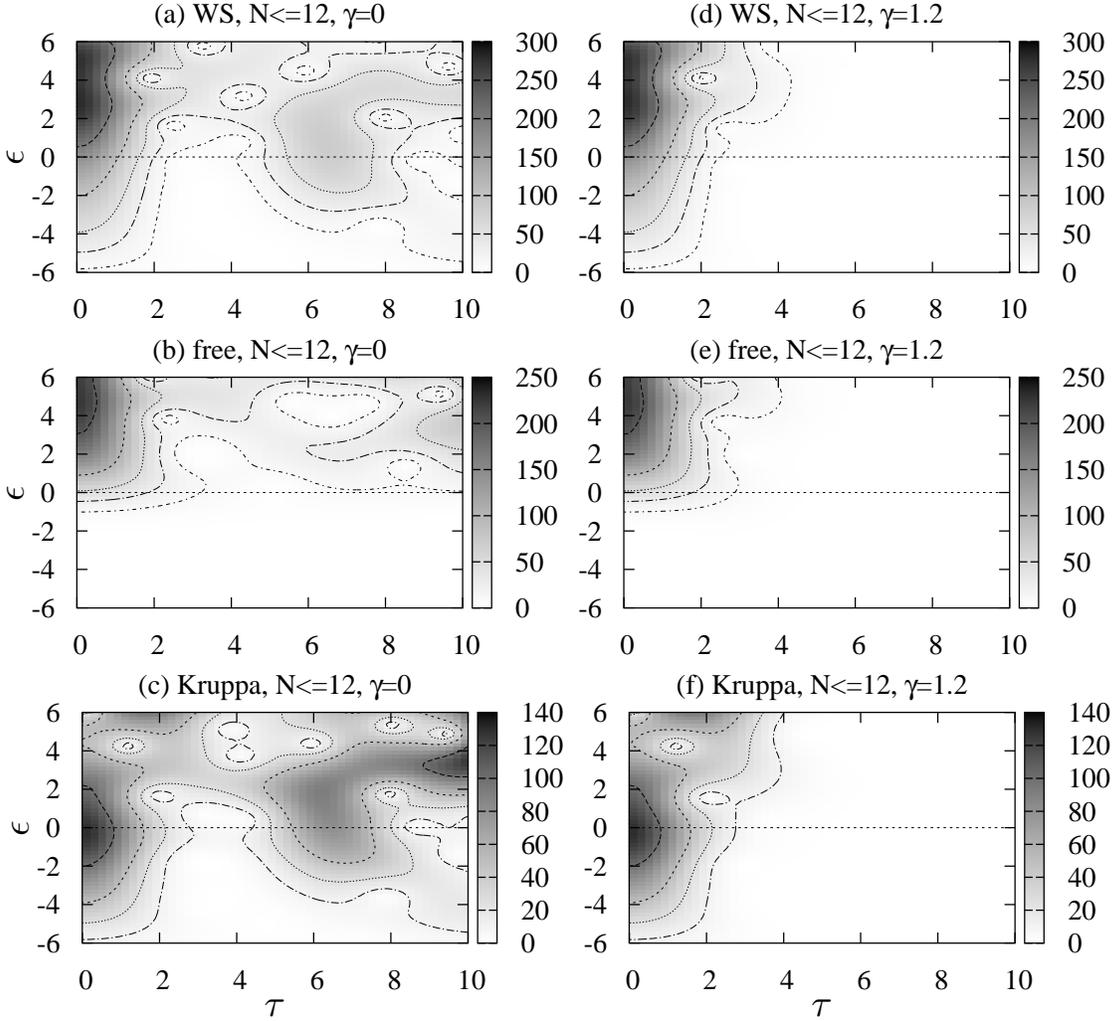}
\vspace*{-5mm}
\caption{
Absolute value of the short-time Fourier transform of the
level density of the Woods-Saxon Hamiltonian calculated with $\noscmax=12$.
The units are $\hbar \omega$ for $\epsilon$
and $(\hbar \omega)^{-1}$ for $\tau$.  See text for explanations.
}
\label{fig:shortTimeFourierTransformResultWS12}
\end{figure*} 

\begin{figure*}[htbp] 
\includegraphics[width=150mm]{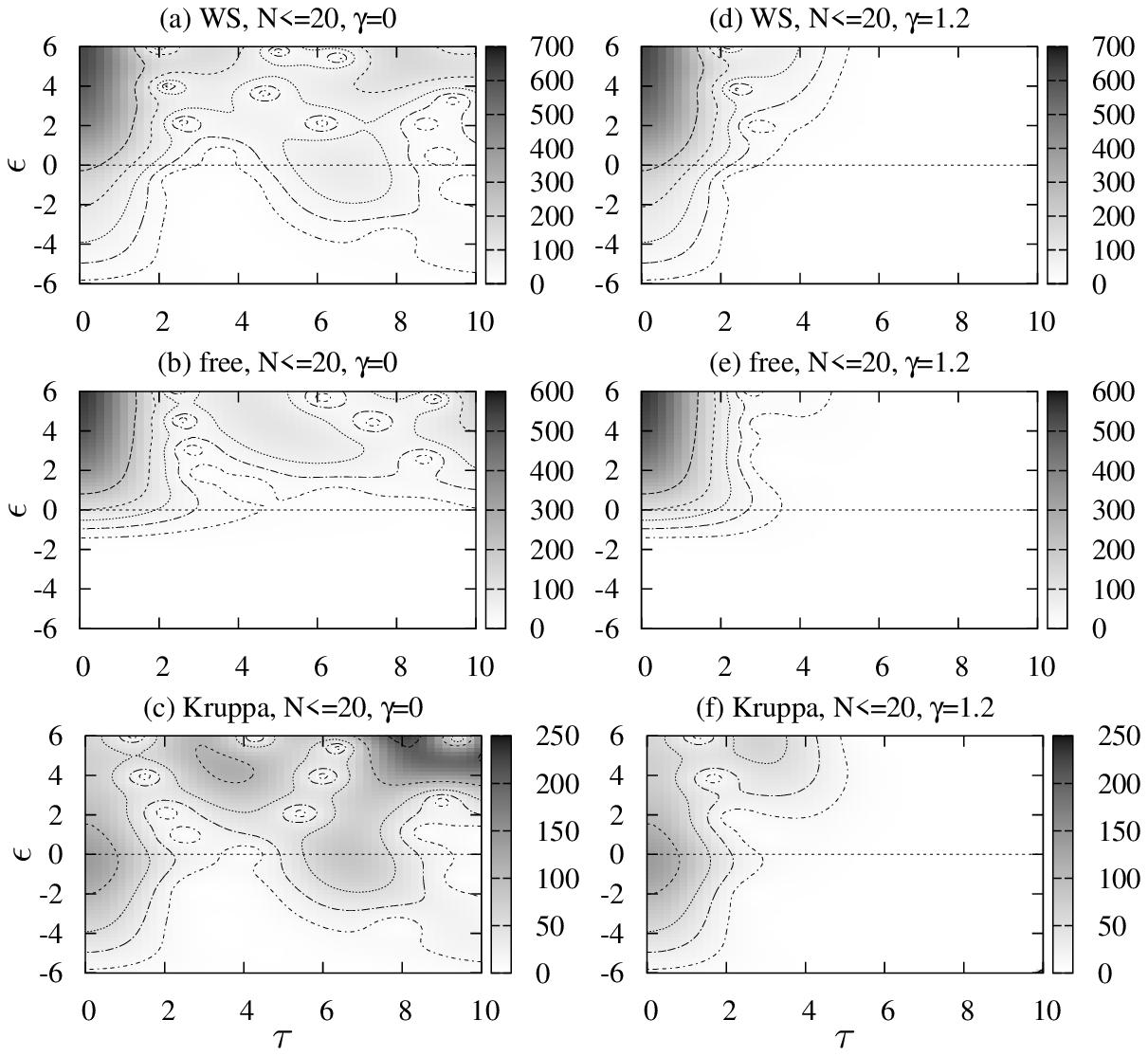}
\vspace*{-5mm}
\caption{The same as in Fig.~\ref{fig:shortTimeFourierTransformResultWS12}
but calculated with $\noscmax=20$.
}
\label{fig:shortTimeFourierTransformResultWS20}
\end{figure*} 

As for the harmonic oscillator spectrum shown in 
Fig.~\ref{fig:shortTimeFourierTransformResultHONIL}~(a),
there persist two main components $\tau=0$ and $\tau=2\pi$
irrespectively of $\epsilon$.
The component $\tau=2\pi$  corresponds to the major shell
spacing ($\hbar\omega$). 
There is a large low-amplitude region between the two hills
along lines $\tau=0$ and $\tau=2\pi$.
The cutoff for the standard smoothing parameters $\gamma=1.2$ and $p=3$ is
$\tau_p\rsu{cut}=2.9$, which is almost at the center of this region.
The result of this standard smoothing is shown in
Fig.~\ref{fig:shortTimeFourierTransformResultHONIL} (b),
in which the hill at $\tau \sim 2\pi$ is removed completely while
that at $\tau \sim 0$ is left almost intact.
This explains the existence of a perfect plateau.

In the case of the Nilsson spectrum
shown in Fig.~\ref{fig:shortTimeFourierTransformResultHONIL}~(c),
the hill at $\tau \sim 2\pi$ becomes distorted and lowered,
which reflects the disturbance that the spin-orbit and the $l^2$ terms of the Nilsson potential
bring to the periodicity with the major shell spacing.
However, the hill at $\tau \sim 0$ and
the low-amplitude region between the two hills
are almost the same as in the harmonic oscillator case.
This clearly explains that the similar good plateau
can be expected in the Nilsson potential.

For the Woods-Saxon spectrum,
we show the absolute values of the short-time Fourier transforms of the full ((a),(d))
and the free ((b),(e)) spectra
as well as the absolute value of their difference (the Kruppa's level density) ((c),(f))
in Fig.~\ref{fig:shortTimeFourierTransformResultWS12} for $\noscmax=12$ and
in Fig.~\ref{fig:shortTimeFourierTransformResultWS20} for $\noscmax=20$.
One can still see the valley between the two hills in the full (a) and the
Kruppa (c) spectra at negative $\epsilon$. 
However, it is much more filled than for the Nilsson spectrum.
At positive $\epsilon$, the landscape is too complicated to be regarded as a
single valley.
In the results of the standard smoothing in panels~(d), (e), and (f),
the contours are much more irregular than those in harmonic oscillator
and Nilsson cases.
These irregularities indicate the existence of nonvanishing structures grown 
in the valley.
Because their contributions change sensitively by small shifts in the cutoff
from the standard value, the plateau can be destroyed completely.

In the meantime,
comparing the Kruppa spectrum with $\noscmax=12$ and $\noscmax=20$,
one sees that the spectrum does not change at negative energies
but continues to change at positive energies versus $\noscmax$.
The changes at positive energies originate in both the full
and the free spectrum.
These time structures at positive energies
are most likely to be the remnant of the property of the diagonalization basis.

A related fact is that
there are no clear changes of the principal time component
of the full Hamiltonian
between positive and negative values of the single-particle energy $\epsilon$.
One can see only obscure and $\noscmax$ dependent changes.
Such clear changes would occur
if the major shell interval were changed altogether at $\epsilon=0$. 
Indeed, in Appendix C of Ref.~\cite{MAF06}, 
Magner et al.\ seem to have obtained such a clear
change in the major shell interval between negative and positive energies
that they could remove the difference through a transformation of the energy
to obtain a plateau behavior.
The difference of the results between them and us
seems to be originated mainly in the difference between
solutions in an infinite wall and those in an oscillator-basis expansion.
In the latter case, the shell structure at positive energies is thought to be
strongly connected with the basis.

\section{Improvements to the shell correction method}
\label{sec:improvements}
\subsection{Reference density method}
\label{sec:refdens}

Although the dependence of the results 
on the smoothing width is unremovable completely,
it is still preferable to make it as small as possible.
In the Kruppa method, this dependence comes
principally from the diffusion of the peak of the level density
at threshold energy ($\epsilon \approx 0$).
This peak is so sharp that it is inevitably more diffused by larger widths.
Since this peak exists already in the (oscillator-basis) Thomas-Fermi approximation,
it should not be diffused but be kept unchanged.

We now propose a prescription to prevent this diffusion, which
we call the reference density (Strutinsky) method. In the method,
one applies the Strutinsky smoothing procedure
not directly to the original discrete spectrum
but to its deviation from some continuous reference level density $g\rs{ref}$,
i.e., to $g(\epsilon)-g\rs{ref}(\epsilon)$.
Note that our concern is the Kruppa level density
$g^{\rm K}(\epsilon)$ of Eq.~(\ref{eq:KruppaDensity}),
which we write $g(\epsilon)$ in this section.

By designating the Strutinsky smoothing procedure
of Eq.~(\ref{eq:smoothedLevelDensity}) with $S$
(do not confuse with the $S$-matrix that does not appear in the followings),
we write $\tilde{g}(\epsilon)$ as $S[g](\epsilon)$.
We do not write $S[g(\epsilon)]$ since $S$ is not a function but a functional.
Using this $S$ operation,
we define the result of the application of the reference density Strutinsky 
method to $g(\epsilon)$ by
\begin{equation} \label{eq:refdnsstrut}
S\rs{ref}[g](\epsilon) = S[g-g\rs{ref}](\epsilon) + g\rs{ref}(\epsilon).
\end{equation}
Owing to the linearity of the Strutinsky smoothing procedure,
the right-hand side of Eq.~(\ref{eq:refdnsstrut}) can be rewritten as
\begin{equation} \label{eq:refdnsstrutRewritten}
S\rs{ref}[g](\epsilon) = S[g](\epsilon) - S[g\rs{ref}](\epsilon) + g\rs{ref}(\epsilon),
\end{equation}
which means that $S\rs{ref}[g]$ and $S[g]$ differ
only where $S[g\rs{ref}](\epsilon)$ $\not=$ $g\rs{ref}(\epsilon)$, i.e.,
where $g\rs{ref}(\epsilon)$ cannot be approximated very well by a polynomial of order $2p$
over an interval of a few $\gamma$ width.
If one defines $g\rs{ref}$ by oversmoothing $g$
(e.g., $g\rs{ref}=S[g]$ with large $\gamma$),
the reference density method and the original Strutinsky smoothing 
give very close results (i.e., almost $S\rs{ref}[g] = S[g]$).
If one superimposes the peak at energy zero to this reference density,
one will obtain $S\rs{ref}[g]$ which is almost equal to $g\rs{ref}$ near energy zero and is very close to $S[g]$ anywhere else.  

It should be mentioned that the new smoothing procedure is generally
more time-consuming than the original one because the integration
of $g\rs{ref}(\epsilon)$ with respect to the single-particle energy
necessary to calculate $S[g\rs{ref}](\epsilon)$ cannot be done
analytically in general.
In practice, we sample $g\rs{ref}(\spe)$ 
at an interval of $0.15 \hbar \omega$
and use a polynomial interpolation between the sampling points.

\subsection{Construction of the reference density}
\label{sec:refdensConstruction}

The remaining problem is how to determine the shape of the 
peak at energy zero, which is the only important part of
the reference density $g\rs{ref}(\epsilon)$.
First we present our best method. Second, we
discuss shortcomings of some other methods which we have tried.

\begin{figure}[htb] 
\includegraphics[width=90mm]{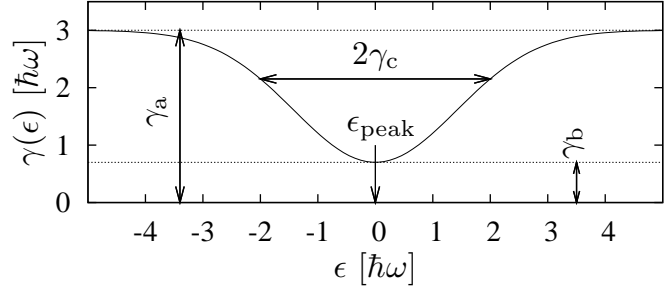}
\vspace*{-5mm}
\caption{$\gamma(\epsilon)$ defined by Eq.~(\ref{eq:gammaOfEpsilon})
to be used to construct the reference density.
}
\label{fig:gammaOfEpsilon}
\end{figure} 

The best method is to define the reference level density as
\begin{equation} \label{eq:refdns}
 g\rs{ref}(\epsilon) = \int_{-\infty}^{\infty} g(\epsilon')
 \frac{1}{\gamma(\epsilon)} 
 f_{0} \left( \frac{\epsilon-\epsilon'}{\gamma(\epsilon)} \right) d\epsilon'
\; , \;\;
 f_{0}(x)=\frac{1}{\sqrt{\pi}} e^{-x^2}.
\end{equation}
This is the same as the Strutinsky smoothing 
without the curvature correction polynomial ($p=0$) except that
$\gamma$ is a function of energy $\epsilon$, for which we assume
\begin{equation} \label{eq:gammaOfEpsilon}
\gamma(\epsilon) = \gamma_{\rm a} +(\gamma_{\rm b} - \gamma_{\rm a})
 \,{\exp} \left[ 
- \left( \frac{\epsilon - \epsilon_{\rm peak} }{\gamma_{\rm c}}
  \right)^2 \right],
\end{equation}
with parameters 
$\gamma_{\rm a} = 3$,
$\gamma_{\rm b} = 0.7$,
$\gamma_{\rm c} = 2$,
and $\epsilon_{\rm peak}=0$ in units of $\hbar \omega$.
Eq.~(\ref{eq:gammaOfEpsilon}) is shown graphically
in Fig.~\ref{fig:gammaOfEpsilon} to elucidate the roles of each parameter.

The quantity $\gamma\rs{a}$ is chosen to be large enough
so that it holds $S[g\rs{ref}] \simeq g\rs{ref}$ and thus
$S\rs{ref}[g]=S[g]$ at energies distant from $\epsilon\rs{peak}$.
The quantity $\gamma\rs{b}$ is chosen to be small enough
so that it holds $S[g\rs{ref}] \simeq S[g]$ and thus
$S\rs{ref}[g]=g\rs{ref}$ at energies near $\epsilon\rs{peak}$,
but not so small as the peak is split into more than two peaks.
The quantity $\gamma\rs{c}$ is determined empirically to obtain smooth results.
The energy $\epsilon\rs{peak}$ is taken as zero for neutrons,
while it should be around the Coulomb-barrier-top energy for protons.
In this paper, we employ the reference density method only to treat
the neutron spectrum,
since for protons the standard Strutinsky method works rather well
due to the fact that the peak energy is considerably larger
than the proton Fermi energy even near the proton drip line.

\begin{figure*}[htb] 
\includegraphics[width=90mm]{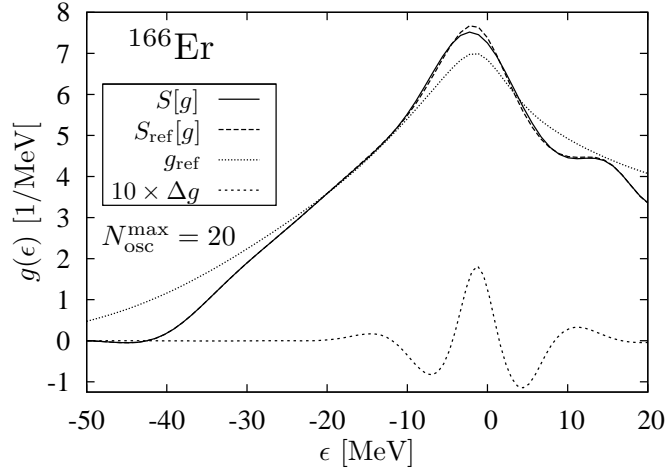}
\vspace*{-5mm}
\caption{
The Kruppa's level density for a neutron in the ground state of
$^{166}$Er smoothed in various ways.
The basis is specified by $\noscmax=20$.
Deformation parameters are $\beta_2=0.280$ and $\beta_4=0.005$.
In the Strutinsky smoothing, $\gamma=1.2\,\hbar \omega$ and $p=3$ are used.
See text for explanations.
}
\label{fig:refdnsDensity}
\end{figure*} 

In Fig.~\ref{fig:refdnsDensity}, smoothed level densities are
shown for the neutron spectrum of $^{166}$Er.
The result with the reference density method ($S\rs{ref}[g]$, long-dash line)
has slightly sharper peak at around $-1$ MeV
than the result without using it ($S[g]$, solid line).
Their difference $\Delta g = S\rs{ref}[g] - S[g]$ looks small but turns out
to play an important role in improving the plateau
in Sec.~\ref{sec:improvedplateau}.
The difference multiplied by ten ($10\times\Delta g$)
is shown with a short-dash line.
From Eq.~(\ref{eq:refdnsstrutRewritten}), 
$\Delta g = g\rs{ref}-S[g\rs{ref}]$, where
$g\rs{ref}$ is shown with a dot line.
From the shape of $\Delta g$ one can see that
the Strutinsky smoothing smears the peak of $g\rs{ref}$ at $-1$ MeV 
by moving the density near the top to its hillsides around $-7$ MeV and $4$ MeV
and that the reference density method cancels out this 
movement by using this $\Delta g$ as a correction term.

Incidentally,
if one makes $\gamma$ a function of $\epsilon$, not
of $\epsilon'$, in Eq.~(\ref{eq:refdns}),
one finds fake dips in both sides of the peak, 
as well as an enhancement of the peak.
(Changing $\gamma$ as a function of $\epsilon$ in the ordinary
Strutinsky method also leads to similar fake dips and bumps.)
Our choice does not suffer from this problem.
One should also note that, although
the total number of levels of the reference level density
is not exactly equal to that of the original discrete spectrum,
\begin{equation}
\int_{-\infty}^{\infty} g\rs{ref} (\epsilon) d\epsilon
\not= \int_{-\infty}^{\infty} g (\epsilon) d\epsilon,
\end{equation}
it still holds
\begin{equation}
\int_{-\infty}^{\infty} S\rs{ref}[g](\epsilon) d\epsilon
= \int_{-\infty}^{\infty} g (\epsilon) d\epsilon.
\end{equation}

We have also examined the possibility of least-square fittings
of trial functions.
For deformed nuclei,
we could determine clearly the center and the width of the peak.
For spherical nuclei, however, we could not because the
levels are multiple degenerated and thus are very sparse.

An alternative method to determine the reference density $g\rs{ref}$
is the semiclassical estimation.
We have tried the level density obtained with the OBTF,
only to find much stronger dependence on $\gamma$
than that of the standard Strutinsky method for a test calculation 
without spin-orbit force.
The OBTF approximation does not seem to be sufficiently precise to calculate the
shell correction energy.
Indeed, it is known that one should include higher order approximation
than the Thomas-Fermi approximation
in the semiclassical Wigner-Kirkwood expansion~\cite{Jen73,JBB75}.
For this purpose, however,
one must extend it to the case of truncated oscillator basis expansion.
In addition, semiclassical approaches are more difficult to 
use for deformed nuclei. They also
have some problems near the particle threshold (drip lines)~\cite{VKL98}.

\subsection{Improvement of plateau condition}
\label{sec:improvedplateau}

\begin{figure*}[htb] 
\includegraphics[width=150mm]{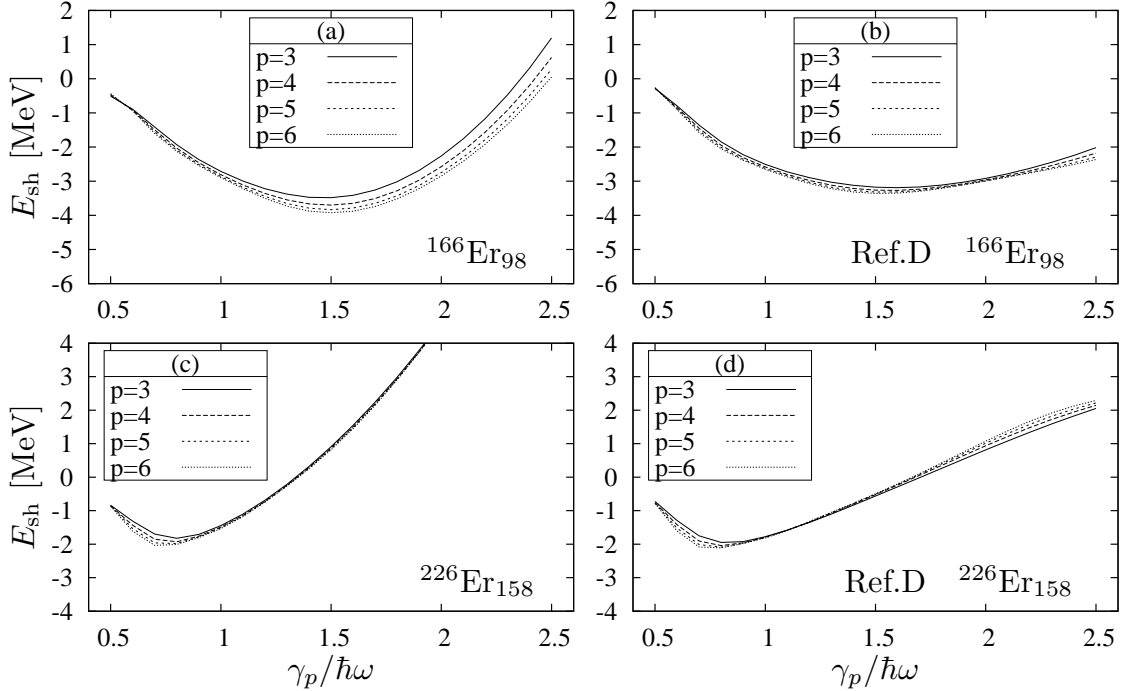}
\vspace*{-5mm}
\caption{
The neutron shell correction energies for $^{166}$Er and $^{226}$Er
calculated with the Kruppa method (left), and
the Kruppa + reference density (right) Strutinsky method as functions
of the scaled smoothing width parameter $\gamma_p$~(\ref{eq:scaledgamma})
in unit of $\hbar\omega$ with different choices for the order 
of the polynomial $p=3-6$.
As for the basis size $\noscmax=30$ is used.
}
\label{fig:refdnsEnergy}
\end{figure*} 

\begin{figure*}[htb] 
\includegraphics[width=150mm]{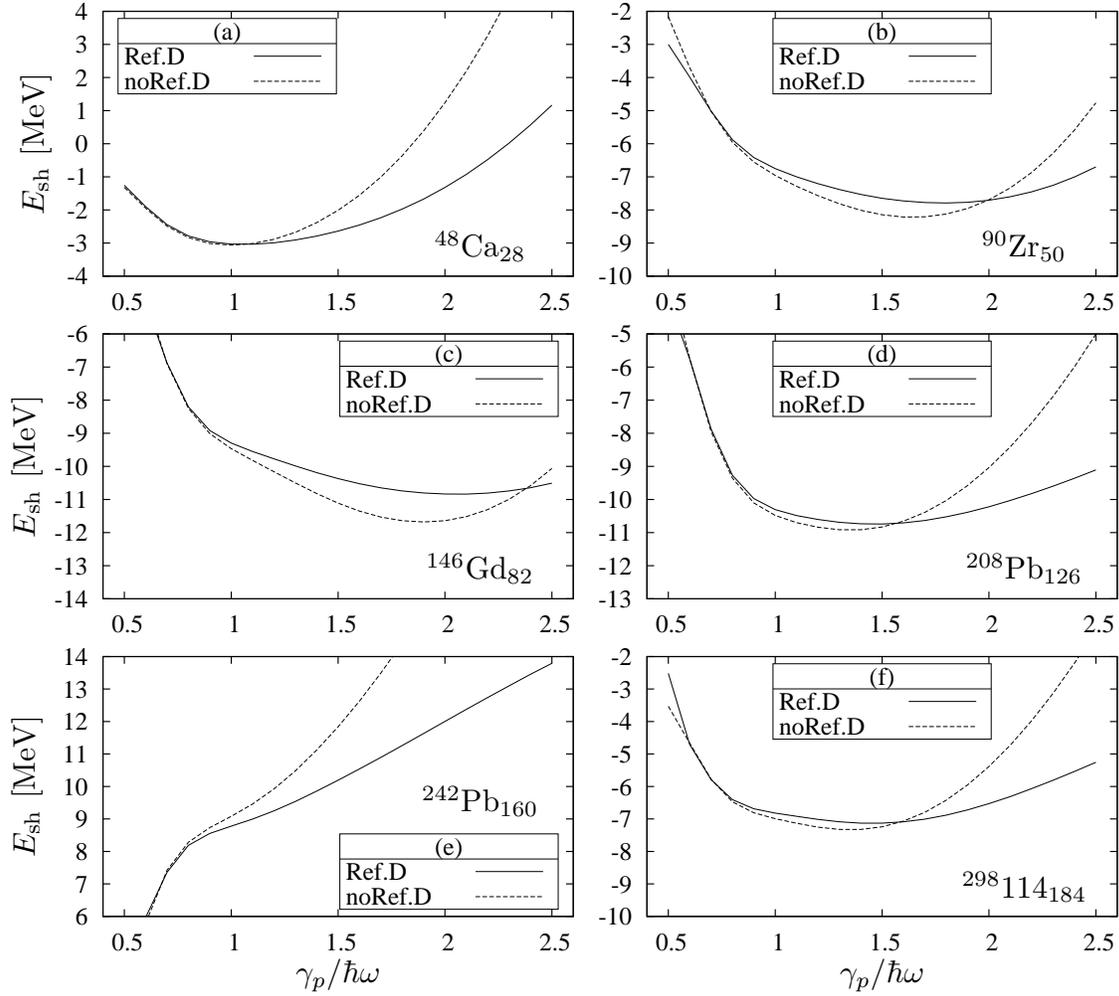}
\vspace*{-5mm}
\caption{
Comparison of the neutron shell correction energies
as functions of the scaled smoothing parameter $\gamma_p$
calculated with the reference density smoothing (solid curves)
and with the ordinary Strutinsky smoothing (dashed curves)
for $^{48}$Ca, $^{90}$Zr, $^{146}$Gd, $^{208}$Pb, $^{242}$Pb,
and the superheavy nucleus $^{298}$114, all of which
are spherical nuclei ($\bm{\beta}=\bm{0}$).
The order of smoothing function is $p=3$ and $\noscmax=30$ is used.
}
\label{fig:refdnsCmp}
\end{figure*} 

We show how the reference density method
improves the plateau condition
in Figs.~\ref{fig:refdnsEnergy} and~\ref{fig:refdnsCmp},
where the Kruppa's prescription is used throughout.
Figure~\ref{fig:refdnsEnergy} depicts the dependence of the shell correction
energies on the scaled smoothing width parameter
$\gamma_p$~(\ref{eq:scaledgamma})
for the $\beta$ stable and very neutron rich nuclei,
$^{166}{\rm Er}$ and $^{226}{\rm Er}$,
considered as examples in Sec.~\ref{sec:plateau}.
Comparing the results with the Kruppa + reference density method (right panels)
to those with the Kruppa method (left panels),
the dependence on the smoothing width is weakened remarkably,
although it is not completely satisfactory in the
case of $^{226}{\rm Er}$.  The dependence
on the order $p$ of the smoothing function is also greatly reduced and
the difference of the shell correction energies between $p=3$ and 6
is typically within a hundred keV.
Since this is a general tendency, we show only the results with $p=3$
in the followings.

Combined with the reference density method, a better stability
against the width and the order of the smoothing function is obtained:
In most cases, the shell correction energy has a minimum
as a function of the smoothing width parameter in the range,
$\gamma_p=(1-2)\hbar\omega$, and around the minimum we often find
a plateau-like quite flat landscape.
According to Ref.~\cite{BP73}, 
the shell correction energy at this minimum
should be adopted even in the case with no pronounced plateau
(the local plateau condition).
We show examples for several test cases from light to heavy spherical nuclei
in Fig.~\ref{fig:refdnsCmp},
where, in each panel, the solid curve is the result
with the reference density method and the dash curve is the one without it.
Comparing two curves, 
one can see that the plateau is always improved by the reference density method.
The improvements are remarkable especially
for $^{48}{\rm Ca}$ and $^{146}{\rm Gd}$.
It should be emphasized that these improvements do not change
when increasing the size of the model space.
However, there are some exceptions as is shown for $^{242}{\rm Pb}$
in Fig.~\ref{fig:refdnsCmp},
where no minimum but a inflection point appears.
Although the dependence is reduced, it is not enough to obtain
plateau-like behavior.
Note that the parameters of the reference density method
are fixed to be the same for all nuclei in this paper.
There is still some room for their further improvement or optimization.

\subsection{Kruppa-BCS equation} \label{sec:KruppaBCS}

Whether the pairing correlation is enhanced or not in nuclei 
near the neutron drip line is still an open question. 
On one hand,
high level density near and above the neutron threshold 
is expected to enhance the pairing~\cite{DobNaz98,DobNaz02}. 
Thicker neutron skin is also likely to make the pairing interaction stronger.
On the other hand,
the radial expansion of the single-particle wavefunctions near the Fermi level
will weaken the pair-scattering matrix elements.
There can be a competition between a spatially expanded normal state 
and a compact super state~\cite{Taj04,Taj05}
(the latter is an manifestation of the pairing anti-halo effect~\cite{BDP00}).

To take into account all of the above effects,
one needs to employ at least mean field models in the
Hartree-Fock-Bogoliubov (HFB) formalism~\cite{DFT84,DNW96}.
To mimic them in the shell correction method, 
one has to extend the method in many aspects by, e.g.,
calculating the pairing matrix elements using the wavefunctions, 
replacing the BCS gap equation with the HFB equation,
making the radius and diffuseness parameters ($R$ and $a$)
not constants but variables to be optimized like deformation parameters
($\beta_2$ and $\beta_4$ in this paper), etc. 
Instead of trying to consider everything,
we aim at only one thing, i.e., 
the usage of the Kruppa's level density in the BCS calculation.
We call this method (or the resulting equation)
the Kruppa-BCS method (equation).

Near the critical point of the transition between
the normal and superfluid phases, it is necessary to go beyond
the mean field treatment by, e.g., the number projection~\cite{RS80}
and/or its approximate version, the so-called
Lipkin-Nogami method~\cite{Lip60,Nog64,PN73},
or the random phase approximation (RPA) method~\cite{SGB89}.
Generalizations of the Kruppa prescription
to such treatments are a quite interesting subject.
We restrict, however, to the simplest BCS treatment in the present work.

Although some generalizations are possible, e.g., to the state-dependent
pairing interaction~\cite{Ton79}, we consider the simplest
seniority-type pairing force for the BCS calculation in this paper.
The following matrix elements are assumed
for the pairing interaction $V\rs{pair}$, 
\begin{equation} \label{eq_vp}
  \langle i \bar{i}' \vert V\rs{pair} \vert
  j \bar{j}' \rangle = - G \,\delta_{ii'}\delta_{jj'}\,
  \cutoff (\epsilon_i) \cutoff (\epsilon_j).
\end{equation}
In the left-hand side,
$\bar{k}$ (for $k=i',j'$)
represents the label for the time reversal partner of
the $k$th eigenstate of the full
or the free single-particle Hamiltonian.
It holds that $\epsilon_{\bar{k}}=\epsilon_k$.
When $k$ is a label for a free-particle state,
$\epsilon_k$ should be read as $\efree{k}$.
Scatterings from a pair of full-Hamiltonian states into a pair of
free-Hamiltonian states, and the reverse processes,
also appear in the Kruppa-BCS equation to be presented later.
In the right-hand side, $G$ is a constant while $\cutoff(\epsilon)$
is a cutoff factor~\cite{BFH85}, for which we use a different form
from that of Ref.~\cite{BFH85},
\begin{equation} \label{eq:fe}
\cutoff(\epsilon) = 
\frac{1}{2} \left[1+
{\rm erf}\left(\frac{\epsilon-\tilde{\lambda}+\Lambda\rs{l}}{d\rs{cut}}\right)
\right]^{1/2}
\left[1+
{\rm erf}\left(\frac{-\epsilon+\tilde{\lambda}+\Lambda\rs{u}}{d\rs{cut}}\right)
\right]^{1/2},
\end{equation}
where the error function is defined by
${\displaystyle
{\rm erf}(x)=\frac{2}{\sqrt{\pi}} \int_{0}^{x}e^{-t^2}dt
}$.  
We use the cutoff parameters of the pairing model space,
$\Lambda\rs{u}=\Lambda\rs{l}=1.2\,\hbar \omega$ and
$d\rs{cut}=0.2\,\hbar \omega$.
$\tilde{\lambda}$ is the smoothed Fermi level
defined by Eq.~(\ref{eq:particleNumberEquation}).
Incidentally, if one uses $\lambda\rs{BCS}$ 
(to be defined in Eqs.~(\ref{eq:KruppaBCSGapEq}) and (\ref{eq:KruppaBCSNumberEq}))
instead of $\tilde{\lambda}$ in Eq.~(\ref{eq:fe}),
one sometimes encounters an instability caused by a positive feedback
from  $\lambda\rs{BCS}$, a part of the solution,
to the equation through $\cutoff(\spe)$.

The energy of the BCS model for a separable interaction like Eq.~(\ref{eq_vp})
can be expressed as~\cite{RS80},
\begin{equation} \label{eq:KruppaBCSEnergy}
  E\rs{BCS}=
  \int_{-\infty}^{\infty}
  \epsilon\, v^2(\epsilon) g(\epsilon) d\epsilon -\frac{\Delta^2}{G},
\end{equation}
where the pairing gap $\Delta$ is given by,
\begin{equation} \label{eq:KruppaBCSGap}
  \Delta = \frac{G}{2} 
  \int_{-\infty}^{\infty} 
  \cutoff (\epsilon) u(\epsilon) v(\epsilon) g(\epsilon) d\epsilon.
\end{equation}
(The pairing gap for the state $i$ is state-dependent,
$\cutoff(\spe_i)\Delta$, because of the cutoff function.)
In Eq.~(\ref{eq:KruppaBCSEnergy}) and in the followings,
the exchange contribution
of the pairing interaction to the particle-hole channel is neglected.
The constraint on the expectation value of the number of particles is expressed as,
\begin{equation} \label{eq:KruppaBCSNumber}
  N=\int_{-\infty}^{\infty} v^2(\epsilon) g(\epsilon) d\epsilon.
\end{equation}
In the above formulation using integrals, 
one regards the BCS $u$ and $v$ factors as continuous functions of
the single-particle energy. 
It should be reminded that these equations are not well defined
due to the divergence of the level density
if the continuum states are included.
One has to replace the level density $g(\epsilon)$
by that of Kruppa in Sec.~\ref{sec:KruppaMethod}.
Thus, we naturally define the Kruppa-BCS model as a model obtained 
by replacing the ordinary level density to
the Kruppa's one~(\ref{eq:KruppaDensity}),
$g(\epsilon)\Rightarrow g\rsu{K}(\epsilon)$, in 
Eqs.~(\ref{eq:KruppaBCSEnergy})~to~(\ref{eq:KruppaBCSNumber}).

In fact the use of the Kruppa level density
makes the gap equation~(\ref{eq:KruppaBCSGap}) convergent
without the energy cutoff function $\cutoff(\spe)$.
This is because the integral diverges logarithmically
as $\spe\rightarrow\infty$ if the level density is constant,
while it is shown in Sec.~\ref{sec:OBTF} that
the Kruppa level density $g\rsu{K}(\epsilon)\propto \spe^{-1/2}$,
see Eq.~(\ref{eq:cobk}).
However, the convergence is slow and it is dangerous to rely on it,
so that we use the cutoff function~(\ref{eq:fe}) in the following calculations.

Considering that only values at discrete points
($\epsilon=\epsilon_i$ and $\efree{i}$) contribute in the Kruppa
prescription, the equation to be satisfied by the minimum-energy state is 
just the standard gap equation~\cite{RS80} and the constraint on the number,
but with the additional (negative) contributions from the free spectra:
\begin{eqnarray}
\label{eq:KruppaBCSGapEq}
 \frac{2}{G} &= & \frac{1}{2}\sum_{i=1}^{M}
\left[
 \frac{\cutoff (\epsilon_i)^2}{\sqrt{(\epsilon_i - \lambda\rs{BCS})^2
 + \cutoff (\epsilon_i)^2 \Delta^2}} 
 - \frac{\cutoff (\efree{i})^2}{\sqrt{(\efree{i} - \lambda\rs{BCS})^2
 + \cutoff (\efree{i})^2 \Delta^2}}
\right], \\
\label{eq:KruppaBCSNumberEq}
 N &=& \frac{1}{2} \sum_{i=1}^{M}
\left[
 - \frac{\epsilon_i-\lambda\rs{BCS}}{\sqrt{(\epsilon_i - \lambda\rs{BCS})^2
 + \cutoff (\epsilon_i)^2 \Delta^2}} 
 + \frac{\efree{i}-\lambda\rs{BCS}}{\sqrt{(\efree{i} - \lambda\rs{BCS})^2
 + \cutoff (\efree{i})^2 \Delta^2}}
\right].
\end{eqnarray}
As in the case of the usual BCS equation, the pairing gap and
the chemical potential $(\Delta,\lambda\rs{BCS})$ are determined
by these two coupled equations for given force strength $G$.
Note that $M$ is two times the number of the pairs of time-reversal states,
namely the degeneracy is explicitly counted in the level density.
One should assume
$\epsilon_1$ = $\epsilon_2$ $\le$ $\epsilon_3$ = $\epsilon_4$ $\le$ $\cdots$
$\le$ $\epsilon_{M-1}$ = $\epsilon_{M}$
and
$\efree{1}$ = $\efree{2}$ $\le$ $\efree{3}$ = $\efree{4}$ $\le$ $\cdots$
$\le$ $\efree{M-1}$ = $\efree{M}$
to understand the reason of appearing a factor $\frac{1}{2}$ in many of the equations
in this paper, e.g., in the right hand side
of Eqs.~(\ref{eq:KruppaBCSGapEq}) and~(\ref{eq:KruppaBCSNumberEq}).
In the case of odd particle number the blocking BCS calculation
should be done~\cite{RS80};
i.e., the single-particle level occupied by the last odd particle,
for example $i=N$, should be eliminated from the pairing model space,
and the resultant BCS equation with number $N-1$ is the same as
in the case of even particle number.

It is known that the BCS equation does not necessarily have finite
pairing gap solutions if $\spe_N < \spe_{N+1}$~\cite{RS80}; namely
the system is not in the superfluid phase but in the normal phase.
The critical force strength $G\rs{crit}$ ($\Delta=0$ if $G\le G\rs{crit}$)
is given by
\begin{equation} \label{eq:Gcrit}
 \frac{2}{G\rs{crit}} =  \mathop{\mbox{min}}_{\spe_N < \lambda' <\spe_{N+1}}
\left[
 \frac{1}{2}\sum_{i=1}^{M}
\left(
 \frac{\cutoff (\epsilon_i)^2}{|\epsilon_i - \lambda'|}
 - \frac{\cutoff (\efree{i})^2}{|\efree{i} - \lambda'|}
\right) \right],
\end{equation}
where the minimum value of the right hand side is searched
with respect to $\lambda'$
(in the case of odd particle number, it is always $\spe_N = \spe_{N+1}$,
and for the blocked level $i=N$ the minimum value in Eq.~(\ref{eq:Gcrit})
should be searched for $\spe_{N-1} < \lambda' <\spe_{N+1}$).
Although the Fermi energy $\lambda$ in the normal phase
is arbitrary within $\spe_N < \lambda <\spe_{N+1}$,
it is desirable to define the Fermi energy uniquely
in the later discussion (see Sec.\ref{sec:potentialdepth}).
Therefore, we define $\lambda\rs{BCS}$ when $\Delta=0$ as the $\lambda'$
that gives $G\rs{crit}$ in Eq.(\ref{eq:Gcrit}).

In selfconsistent methods, one only needs to deal with particle-bound nuclei
with negative Fermi energies.
In shell correction approaches, however, one needs some reasonable solution
for positive energy Fermi levels.
This is because negative Fermi levels of the microscopic part
do not always mean positive separation energies
calculated from the total energies of the shell correction method,
see Sec.~\ref{sec:potentialdepth}.

One must be careful in applying the Kruppa-BCS method
to the particle-unbound cases.
More precisely, if the Fermi energy is higher than the lowest energy
of the free spectra $\{\efree{i};i=1,...,M\}$.
For example, if $\spe_N < \efree{1} < \spe_{N+1}$
the right hand side of Eq.~(\ref{eq:Gcrit})
has no minimum because of the negative contribution of the free spectra,
and $G\rs{crit}$ cannot be defined.
As an another peculiar feature of the Kruppa-BCS equation,
the solution is not always unique for a positive Fermi level.
This nonuniqueness is easy to explain for the normal states ($\Delta=0$).
There are more than one ways to fill the spectrum as normal states,
i.e., to choose the Fermi level $\lambda$ such that
$\epsilon_i \le \lambda < \epsilon_{i+1}$,
$\efree{j} \le \lambda < \efree{j+1}$, and $i-j=N$.
For example, in a case
$\efree{4} < \epsilon_{N+4} < \efree{6} < \epsilon_{N+6}$,
both $\lambda=\epsilon_{N+4}$ and $\lambda=\epsilon_{N+6}$
have correct number of particles.
One has to pay attention to choose the physically most reasonable solution;
e.g., the one which gives more continuous total energy with respect to
the change of deformation parameters.

\subsection{Extension of the Kruppa's prescription to other observables}
\label{sec:ExtKruppa}

Subtraction of the free contributions in the Kruppa level density
in Eq.~(\ref{eq:KruppaDensity}) reminds us of the counter term
in the renormalization procedure; both contributions diverge but
the difference remains finite being independent of the cutoff.
Therefore, it may be natural to extend this idea to other observables:
\begin{equation} \label{eq:OKruppa}
\langle O \rangle \quad\Rightarrow \quad
\langle O \rangle\rsu{K}=
\langle O \rangle - \langle O \rangle_0,
\end{equation}
where the first term is the expectation value with respect to the wave
function calculated by the diagonalization of the Woods-Saxon potential
and the second term is that of the free Hamiltonian
(or the repulsive Coulomb Hamiltonian for protons).
We consider only one-body observables in the mean field
approximation for the many-body wave function.
In the simple independent particle approximation,
e.g., the Hartree-Fock theory,
the second (free) term does not contribute as long as
the Fermi energy is below the particle threshold.
If the residual interaction is included, however,
the occupation probabilities of unoccupied states become non-zero,
and then the free-spectrum terms do contribute, which is exactly the situation
in the case of the BCS theory for the pairing correlation.

The Kruppa-BCS gap and the number equations, Eqs.~(\ref{eq:KruppaBCSGapEq})
and~(\ref{eq:KruppaBCSNumberEq}), can be regarded as examples of
the above extended procedure~(\ref{eq:OKruppa})
because they are derived from
\begin{equation}
 \Delta=G\langle \hat{P}^\dagger \rangle\quad\mbox{and}\quad
 N=\langle \hat{N} \rangle,
\label{eq:BCSeqFromMean}
\end{equation}
where $\hat{P}^\dagger$ is the pair transfer operator, whose matrix elements
are $\langle i\bar{i'} | \hat{P}^\dagger|0\rangle
=\delta_{ii'}f\rs{c}(\epsilon_i)$,
and $\hat{N}$ is the nucleon number operator.

Note that the simple BCS calculation of observables
composed of the spatial coordinate $\bm{r}$ diverges 
as the basis size is increased for nuclei near the particle threshold.
This is because the continuum states have finite occupation probabilities;
the so-called the ``neutron gas'' problem~\cite{DFT84,DNW96}.
Therefore, it is impossible to obtain a reliable estimate for,
e.g., the root mean square radii~\cite{DNW96a} or the quadrupole moments.
It can be shown, however, that with the prescription~(\ref{eq:OKruppa})
such observables also converge as $\noscmax\rightarrow \infty$,
by employing the oscillator-basis Thomas-Fermi approximation
in Sec.~\ref{sec:OBTF} as in the same way as for the level density.
Thus, the Kruppa method relieves the conventional BCS method
from the failure of the neutron gas problem in nuclei near the drip line.
The results will be reported elsewhere~\cite{OS10a}.

\subsection{Determination of the strength of the pairing interaction}
\label{sec:pairingForceStrength}

The strength of the pairing interaction $G$ is often determined
so as to reproduce the empirical smooth trend of the pairing gap
in the continuous spectrum approximation, in which
the smoothed level density is used in the BCS
calculation~\cite{Str68,FunnyHills,TTO96,Taj01a,TTO98}.
This method is almost indispensable in order to treat,
say, all the nuclei in the nuclear chart on a single footing.
A consistent usage of the Kruppa's level density 
also applies to this procedure.

However, in most of the existing shell correction calculations,
e.g., Ref.~\cite{MNM95,MN90},
this procedure is not followed rigorously: The so-called uniform
level density approximation~\cite{FunnyHills,RS80} is additionally employed,
i.e., the energy dependence of the level density is neglected
and it is replaced by a constant value at the Fermi energy,
$\tilde{g}(\tilde{\lambda})$.

As is discussed in Sec.~\ref{sec:KruppaMethod}, the Kruppa's level density
has a peak near the particle threshold.  We have found that this peak
strongly affects the pairing correlation in nuclei near the drip line.
Therefore, the usual method to approximate the level density
as a constant, $\tilde{g}(\tilde{\lambda})$, over the entire 
energy interval where the pairing is active is inadequate.
Instead, the energy dependence of the level density should be evaluated exactly.
We solve the following continuous version of
the gap equation and the constraint on the number,
\begin{eqnarray}
\label{eq:CSAgap}
\frac{2}{G}&=&\frac{1}{2}
\int_{-\infty}^{\infty} \frac{\cutoff (\epsilon)^2}
{\sqrt{(\epsilon-\tilde{\lambda}\rs{BCS})^2 + \cutoff (\epsilon)^2
\tilde{\Delta}^2}} \, \tilde{g}(\epsilon) d \epsilon,
\\
\label{eq:CSApartnum}
N &=& \frac{1}{2} \int_{-\infty}^{\infty} \left[
1 - \frac{\epsilon - \tilde{\lambda}\rs{BCS}}
{\sqrt{(\epsilon-\tilde{\lambda}\rs{BCS})^2 + \cutoff (\epsilon)^2
\tilde{\Delta}^2}}
\right] \tilde{g}(\epsilon) d \epsilon.
\end{eqnarray}
Substituting $\tilde{\Delta}$ with a value from some empirical formula for
the pairing gap, one can determine the Fermi level $\tilde{\lambda}\rs{BCS}$
from Eq.~(\ref{eq:CSApartnum}) and then
the force strength $G$ from Eq.~(\ref{eq:CSAgap}),
and obtain the smoothed BCS energy,
\begin{equation} \label{eq:CSAEBCS}
\tilde{E}\rs{BCS}=
\frac{1}{2} \int_{-\infty}^{\infty} \left[
1 - \frac{\epsilon - \tilde{\lambda}\rs{BCS}}
{\sqrt{(\epsilon-\tilde{\lambda}\rs{BCS})^2 + \cutoff (\epsilon)^2
\tilde{\Delta}^2}}
\right] \epsilon\,\tilde{g}(\epsilon) d \epsilon
- \frac{\tilde{\Delta}^2}{G}.
\end{equation}
This completes the formula for the total energy in Eq.~(\ref{eq:EtotB}).
The force strength $G$ determined by Eq.~(\ref{eq:CSAgap}) is used
in the Kruppa-BCS (or the usual BCS) method in the previous section.
Needless to say, the level density should be replaced,
$g(\epsilon)\Rightarrow g\rsu{K}(\epsilon)$,
in Eqs.~(\ref{eq:CSAgap})~to~(\ref{eq:CSAEBCS})
for the Kruppa-BCS calculation.
Although various kinds of input $\tilde{\Delta}$ can be presumed~\cite{MN90},
We use a standard choice $\tilde{\Delta}$ = $13/\sqrt{A}$ MeV in this paper.

In the uniform level density approximation, the energy integral
in Eqs.~(\ref{eq:CSAgap})~to~(\ref{eq:CSAEBCS}) can be performed
analytically~\cite{RS80} with a sharp cutoff of a pairing model space,
$\tilde{\lambda}-\Lambda_{\rm l} < \epsilon <\tilde{\lambda}+\Lambda_{\rm u}$,
($d\rs{cut}\rightarrow 0$ in Eq.~(\ref{eq:fe})),
and the pairing strength can be calculated by
\begin{eqnarray} \label{eq:Gsmoothap}
{\displaystyle \frac{2}{G}} &=&
{\displaystyle
\frac{1}{2}\,\tilde{g}(\tilde{\lambda})
\log\left[\left(
 \sqrt{\left(\frac{\Lambda\rs{l}}{\tilde{\Delta}}+1\right)^2}
 +\frac{\Lambda\rs{l}}{\tilde{\Delta}}\right)
\left(
 \sqrt{\left(\frac{\Lambda\rs{u}}{\tilde{\Delta}}+1\right)^2}
 +\frac{\Lambda\rs{u}}{\tilde{\Delta}}\right)
 \right] }
 \vspace*{1mm} \\
&\approx&
{\displaystyle
\frac{1}{2}\,\tilde{g}(\tilde{\lambda})
\log\left(
 \frac{\Lambda\rs{l}\Lambda\rs{u}}{{\tilde{\Delta}}^{2}} \right) }.\nonumber
\end{eqnarray}
Moreover, the smooth pairing energy is replaced by
the corresponding approximate expression;
\begin{eqnarray} \label{eq:Epairsmoothap}
 \tilde{E}\rs{BCS} - \tilde{E}\rs{s.p.}
 &\Rightarrow&
 -{\displaystyle
 \frac{1}{4}\,\tilde{\Delta}^2\, \tilde{g}(\tilde{\lambda})
 \left[1-
 \left(
 \sqrt{\left(\frac{\Lambda_{\rm u}}{\tilde{\Delta}}+1\right)^2}
 -\frac{\Lambda_{\rm u}}{\tilde{\Delta}}
 \right)
 \left(
 \sqrt{\left(\frac{\Lambda_{\rm l}}{\tilde{\Delta}}+1\right)^2}
 -\frac{\Lambda_{\rm l}}{\tilde{\Delta}}
 \right)\right]} \\
 & \approx &
 -{\displaystyle
 \frac{1}{4}\,\tilde{\Delta}^2\, \tilde{g}(\tilde{\lambda}) }.\nonumber
\end{eqnarray}
In Ref.~\cite{JD73}, the results of the continuous BCS equation,
Eqs.~(\ref{eq:CSAgap})~to~(\ref{eq:CSAEBCS}),
and of its uniform level density approximation,
Eqs.~(\ref{eq:Gsmoothap}) and (\ref{eq:Epairsmoothap}),
were compared.  The difference in the smooth pairing energy,
$\tilde{E}\rs{BCS} - \tilde{E}\rs{s.p.}$, was found to be
smaller
than a few hundred keV in most cases.
In our calculations,
the difference is even smaller, less than a hundred keV 
both for neutrons and protons. 

According to the sharp cutoff in the uniform level density approximation,
the number of levels included in the conventional BCS calculation
(corresponding to Eqs.~(\ref{eq:KruppaBCSGap}) and (\ref{eq:KruppaBCSNumber}))
is often restricted from $i=N\rs{l}$ up to $N\rs{u}$
in the following way~\cite{MN90};
\begin{equation} \label{eq:ConstLevNum}
N\rs{u}=N+\tilde{g}(\tilde{\lambda})\Lambda\rs{u},\quad
N\rs{l}=\left\{\begin{array}{lll}
 N-\tilde{g}(\tilde{\lambda})\Lambda\rs{l}+1 &\mbox{for}&
    N > \tilde{g}(\tilde{\lambda})\Lambda\rs{l},\cr
 1&\mbox{for} &N \le \tilde{g}(\tilde{\lambda})\Lambda\rs{l}.
 \end{array}\right.
\end{equation}
Namely, the actual cutoff is often done not for the single-particle energy
but for the number of levels.
In contrast, in the Kruppa-BCS method,
the cutoff must be done in terms of energy.
We use the same cutoff parameters
as used in the smooth energy cutoff factor~(\ref{eq:fe})
($\Lambda\rs{l}=\Lambda\rs{u}=1.2\,\hbar\omega$)
when we test the sharp cutoffs in Sec.~\ref{sec:resultKruppaBCS}.

  
\subsection{Results of the Kruppa-BCS calculations}
\label{sec:resultKruppaBCS}

\begin{figure*}[htb] 
\includegraphics[width=150mm]{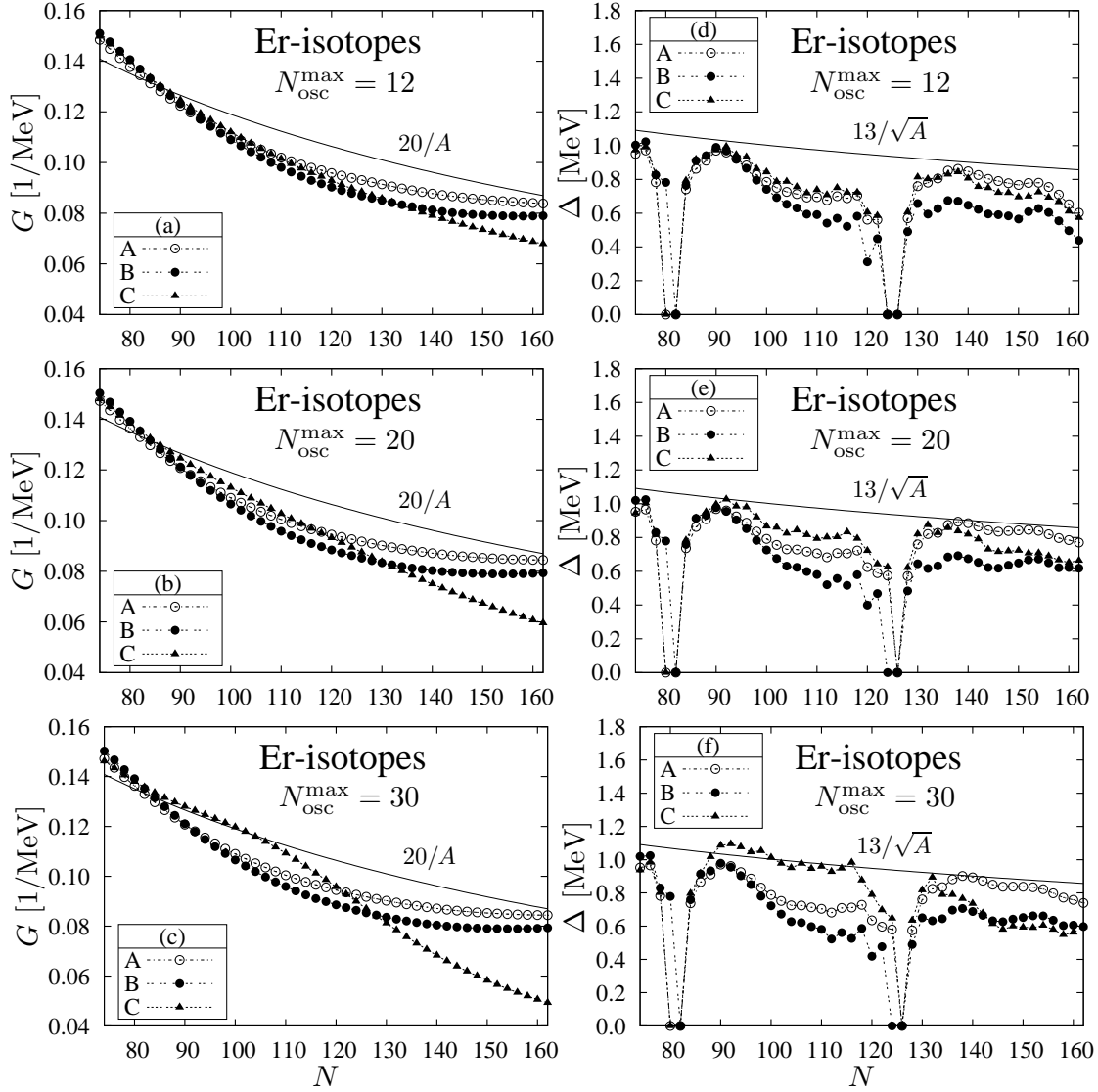}
\vspace*{-5mm}
\caption{Neutron's pairing force strength $G\rs{n}$ (left)
and the pairing gap $\Delta\rs{n}$ (right)
as functions of the neutron number $N$ for $Z=68$ (Er) isotope chain.
The top, middle, bottom panels are for $\noscmax=12,20,30$, respectively.
The results with three different variants of BCS calculations
are included;
A: Kruppa-BCS + continuous $G$ with $\tilde{g}\rsu{K}$,
B: Kruppa-BCS + uniform $G$ with $\tilde{g}\rsu{K}$,
and
C: BCS with number cutoff + uniform $G$ with $\tilde{g}$
(see text for a detailed explanation).
The solid curves, $G=20/A$ [1/MeV] and
$\tilde{\Delta}\rs{n}=13/\sqrt{A}$ [MeV], are also included.
}
\label{fig:pairingGapForIsotopeChain}
\end{figure*}

\begin{figure*}[htb] 
\includegraphics[width=75mm]{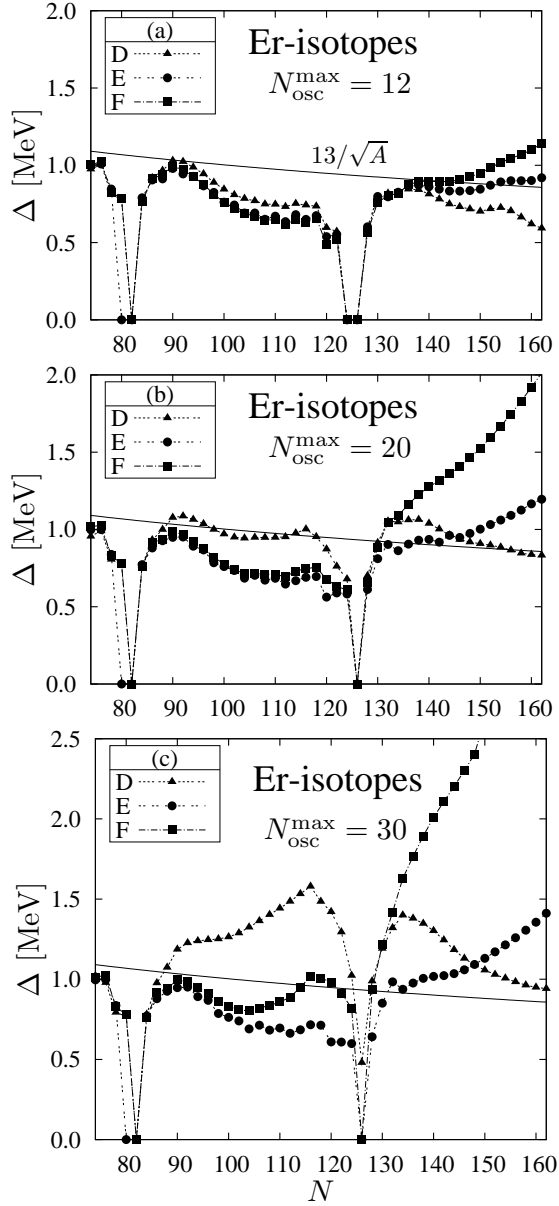}
\vspace*{-5mm}
\caption{
Same as in the right panels of Fig.~\ref{fig:pairingGapForIsotopeChain} but with
non-conventional choices of the BCS calculation;
D: BCS with energy cutoff + uniform $G$ with $\tilde{g}$,
E: BCS with number cutoff + uniform $G$ with $\tilde{g}\rsu{K}$,
and
F: BCS with energy cutoff + uniform $G$ with $\tilde{g}\rsu{K}$
(see text for a detailed explanation).
}
\label{fig:pairingGapForIsotopeChainC}
\end{figure*}

In this subsection we compare the results of several variants
of the BCS calculations.  We make a combined use of
four kinds of classifications to specify those variants.
The first kind of classification is
the Kruppa-BCS method or the ordinary BCS method.
The second one is the smooth-energy cutoff or the level-number cutoff.
Because the level-number cutoff is not applicable to the Kruppa-BCS method,
there are three possible combinations of the first
and the second classifications, 
which we call ``Kruppa-BCS''(with energy cutoff),
``BCS with energy cutoff'', and ``BCS with number cutoff''.
The third classification is whether one determines the strength $G$
by the continuous gap equation~(\ref{eq:CSAgap})
or by its uniform level density approximation~(\ref{eq:Gsmoothap}).
We call the former ``continuous $G$'' and the latter ``uniform $G$''.
The fourth is whether one uses
the smoothed Kruppa level density $\tilde{g}\rsu{K}(\epsilon)$
or the usual one $\tilde{g}(\epsilon)$ in determining $G$.
There are four possible combinations
of the third and the fourth classifications,
which we call ``continuous $G$ with $\tilde{g}\rsu{K}$'',
``uniform $G$ with $\tilde{g}$'', etc.
In total there are twelve possible variants, among which
we choose the most reasonable three to show
in Fig.~\ref{fig:pairingGapForIsotopeChain}
and three unconventional variants to show
in Fig.~\ref{fig:pairingGapForIsotopeChainC}.

In Fig.~\ref{fig:pairingGapForIsotopeChain} we show the
calculated strength $G$ and pairing gap $\Delta$
for neutrons in the $Z=68$ (Er) isotope chain
covering a few more numbers beyond the proton and neutron drip lines.
The deformation parameters $\beta_2$ and $\beta_4$ are determined to
minimize the total energy~(\ref{eq:Etot}) for each nucleus.
The basis size is changed as $\noscmax$=12, 20 and 30.
The figure includes the results of three variants of the BCS calculation,
A: Kruppa-BCS + continuous $G$ with $\tilde{g}\rsu{K}$,
B: Kruppa-BCS + uniform $G$ with $\tilde{g}\rsu{K}$,
and
C: BCS with number cutoff + uniform $G$ with $\tilde{g}$.
The choices A and B are based on the Kruppa's prescription and are
new variants we introduce in this paper.
The choice C is the conventional
one employed in, e.g., Refs.~\cite{MN90,MNM95}.

\begin{figure}[htb] 
\includegraphics[width=95mm]{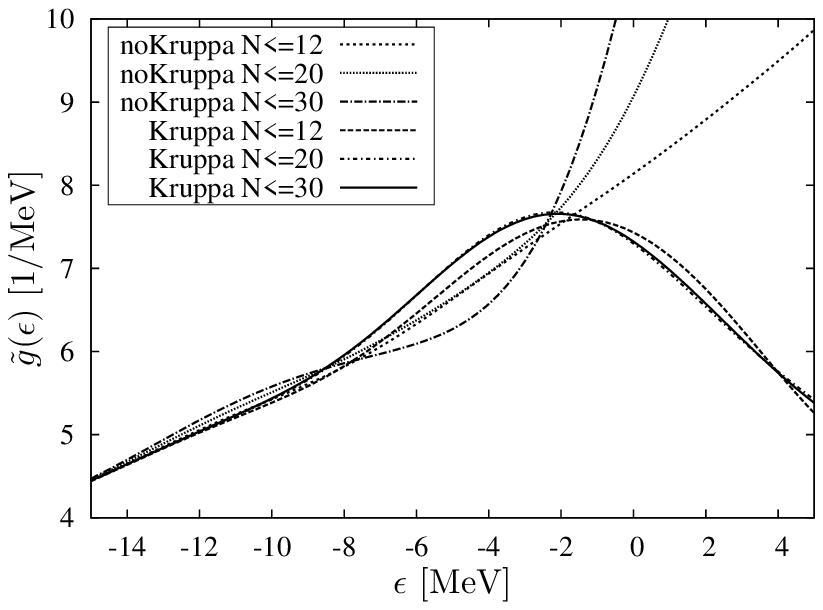}
\vspace*{-5mm}
\caption{
Enlargement of the smoothed level densities
with and without Kruppa prescription shown in Fig.~\ref{fig:KruppaDensity}
in the energy range of the most influential part for BCS calculations.
The Kruppa's densities calculated with $\noscmax=20$ and $\noscmax=30$
are almost indistinguishable.
}
\label{fig:KruppaDensFine}
\end{figure} 

The results of the Kruppa-BCS method
(A and B in Fig.~\ref{fig:pairingGapForIsotopeChain}~(a)~to~(f))
are, first of all, stable against the increase of $\noscmax$
as the shell correction energy $E\rs{sh}$ is.
In this way, the microscopic quantity $E\rs{pair}$, as well as $E\rs{sh}$,
can be calculated to any desired accuracy by increasing the basis size.
The whole procedure is consistent and unambiguous, which is
the first and most important purpose of the present paper.
Second, in Fig.~\ref{fig:pairingGapForIsotopeChain}~(a)~to~(c),
one sees that the continuous strength $G$ (A) is systematically larger
than the uniform strength $G$ (B) on the neutron-rich side.
This difference can be traced back to the behavior of
the Kruppa's level density in Fig.~\ref{fig:KruppaDensity},
enlargement of which in the energy range of
the most influential part to BCS calculations
is shown in Fig.~\ref{fig:KruppaDensFine}.
The Kruppa level density decreases as the single-particle energy
exceeds the threshold, which leads to the increase
of the pairing strength compared to the case of uniform level density
for nuclei near the drip line
(note that the strength is inversely proportional to the level density
at the Fermi level).
This means that the uniform level density approximation is inappropriate
when the peak at energy zero is close to the Fermi level.
Third, in Fig.~\ref{fig:pairingGapForIsotopeChain}~(d)~to~(f),
one sees that the pairing gap of the calculation A is closer to
the input value $\tilde{\Delta}=13/\sqrt{A}$ MeV than that of B,
which means that the continuous $G$ choice is
preferable to the uniform $G$ choice.
We propose the method A, i.e., the Kruppa-BCS method with
the strength $G$ calculated by the continuous gap equation
with the Kruppa's level density,
as the best method for reliable calculations of, e.g., nuclear masses.

Let us examine also the results of the conventional BCS calculation.
When the basis is as small as $\noscmax=12$,
the pairing gap of the conventional BCS calculation 
(C in Fig.~\ref{fig:pairingGapForIsotopeChain}~(d))
agrees very well with that of 
the Kruppa-BCS with continuous $G$ (A).  This agreement is, again, accidental 
since increasing the basis size changes the results considerably
(compare the calculations C
in Fig.~\ref{fig:pairingGapForIsotopeChain}~(d)~to~(f)).
Even when the basis size is as large as $\noscmax=30$,
the pairing gap of the conventional treatment
(C in Fig.~\ref{fig:pairingGapForIsotopeChain}~(f))
does not look totally wrong~\cite{NWD94}, e.g.,
being closest to the input values for $N \le 132$.
However, the force strength $G$ in this case 
(C in Fig.~\ref{fig:pairingGapForIsotopeChain}~(c))
behaves rather unnaturally.
It is quite different from
those of A and B in Fig.~\ref{fig:pairingGapForIsotopeChain}~(c)
as well as from frequently used simple expressions like $G=20/A$ [1/MeV].
It is large in the stable region
but decreases dramatically toward the neutron drip line.
This peculiar behavior
is caused by a spurious effect caused by including
more and more continuum states coming into the pairing model space
when increasing the basis size in the conventional treatment
with $\tilde{g}$ (not with $\tilde{g}\rsu{K}$),
as is clearly seen in Fig.~\ref{fig:KruppaDensFine}.

This strong reduction in $G$ combined with the level number cutoff
is helpful to prevent the pairing gaps from becoming extremely large
as is often the case for the other unreasonable variants
(see Fig.~\ref{fig:pairingGapForIsotopeChainC}).
However, this result is also unphysical as it is inspected from
the results of non-conventional calculations
shown in Fig.~\ref{fig:pairingGapForIsotopeChainC}.
Here we show three non-conventional choices of the BCS method,
D: BCS with energy cutoff + uniform $G$ with $\tilde{g}$,
E: BCS with number cutoff + uniform $G$ with $\tilde{g}\rsu{K}$,
and
F: BCS with energy cutoff + uniform $G$ with $\tilde{g}\rsu{K}$.
The cases E and F are included to see what happens if reasonable
values of the strength $G$
(calculated by $\tilde{g}\rsu{K}$ instead of $\tilde{g}$) are used.
For sufficiently large basis sizes, the resultant pairing gaps are too large,
if the energy cutoff
(D in Fig.~\ref{fig:pairingGapForIsotopeChainC}~(c))
is used in place of the level number cutoff
(C in Fig.~\ref{fig:pairingGapForIsotopeChain}~(f)).
Other non-conventional choices
(E and F in Fig.~\ref{fig:pairingGapForIsotopeChainC})
uses more reasonable pairing force strength $G$
(the same as B in the left panels in Fig.~\ref{fig:pairingGapForIsotopeChain}),
and give reasonable values of $\Delta$ in stable nuclei ($N<120$),
but the pairing gaps diverges when approaching to the drip line
(see Fig.~\ref{fig:pairingGapForIsotopeChainC}~(c)).
Only with relatively small basis sizes such as $\noscmax=12$,
reasonable pairing gaps are obtained; actually all the six variants,
A to F in Figs.~\ref{fig:pairingGapForIsotopeChain}
and~\ref{fig:pairingGapForIsotopeChainC},
gives almost the same results for stable nuclei with $\noscmax=12$.

A lesson of these test calculations is that
the conventional BCS calculation is very dangerous
if one uses the continuum states obtained by
the diagonalization with a large basis.
The subtraction procedure of the free contributions,
i.e., the Kruppa prescription, is indispensable to treat the pairing
correlation in nuclei far from the stability.

\subsection{Readjustment of the potential depth for the Fermi level consistency}
\label{sec:potentialdepth}

In Strutinsky calculations,
the neutron and proton Fermi levels of the single-particle potentials
are not equal to the derivatives of the total energy,
$\partial E / \partial N$ or $\partial E / \partial Z$,
in general unlike in mean field models.
As it is explained in Sec.\ref{sec:shellCorrectionMethod},
the total energy is divided into the macroscopic and microscopic parts
and they are calculated separately:  The Fermi energies
corresponding to them are generally different.
Since most of nuclei are in the superfluid phase and
the microscopic energy is calculated by the BCS method
(see Eq.~(\ref{eq:EtotB})), we define the Fermi energies,
\begin{equation} \label{eq:FermEmac}
\fermiMacNeu = \frac{\partial E\rs{mac}(N,Z)}{\partial N},\qquad
\fermiMacPro = \frac{\partial E\rs{mac}(N,Z)}{\partial Z},
\end{equation}
for the macroscopic part, and for the microscopic part,
\begin{equation} \label{eq:FermEmic}
\fermiMicNeu \equiv\lambda\rs{BCS}\rsu{(n)}
= \frac{\partial E\rs{BCS}\rsu{(n)}(N)}{\partial N},\qquad
\fermiMicPro \equiv\lambda\rs{BCS}\rsu{(p)}
= \frac{\partial E\rs{BCS}\rsu{(p)}(Z)}{\partial Z}.
\end{equation}
The ``total'' Fermi energy is directly related to the two-particle
separation energy, $-S\rs{2n}/2$ or $-S\rs{2p}/2$,
with the definition of the separation energies,
\begin{equation} \label{eq:separationE}
\begin{array}{l}
S\rs{2n}(N,Z) \equiv E(N-2,Z)-E(N,Z),\\
S\rs{2p}(N,Z) \equiv E(N,Z-2)-E(N,Z).\end{array}
\end{equation}
Although the physically meaningful quantity is only the total one,
it is desirable that all the macroscopic, microscopic, and total
Fermi energies coincide with each other.
As it is shown in the following, however,
most of the existing Woods-Saxon parameter sets lead to
$\fermiMicNeu>0$ and $\fermiMicPro>0$,
i.e., particle-unbound, at the drip lines,
which is not only inconsistent conceptually but also can be problematic
for the Kruppa-BCS calculation as is mentioned in Sec.~\ref{sec:KruppaBCS}.
Therefore, we consider how to avoid this problem.

\begin{figure*}[htb] 
\includegraphics[width=95mm]{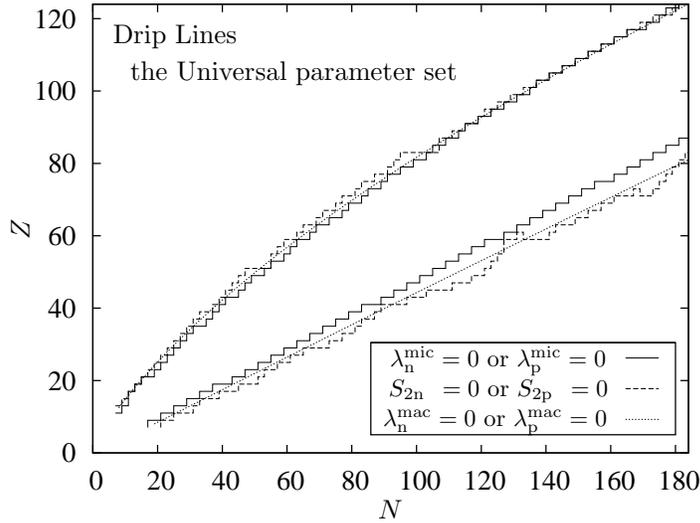}
\vspace*{-5mm}
\caption{
Two-nucleon drip lines calculated with the universal parameter set.
Solid lines are two-nucleon drip lines, which pass between adjacent
bound and unbound even-even nuclei.  The dash lines are the boundary
between nuclei having the positive and negative microscopic Fermi levels.
The dot lines are the drip lines of the macroscopic part of the model.
$\noscmax=20$ is used.
}
\label{fig:dripLinesUnvPrmOrgV}
\end{figure*} 

\begin{figure*}[htb] 
\includegraphics[width=95mm]{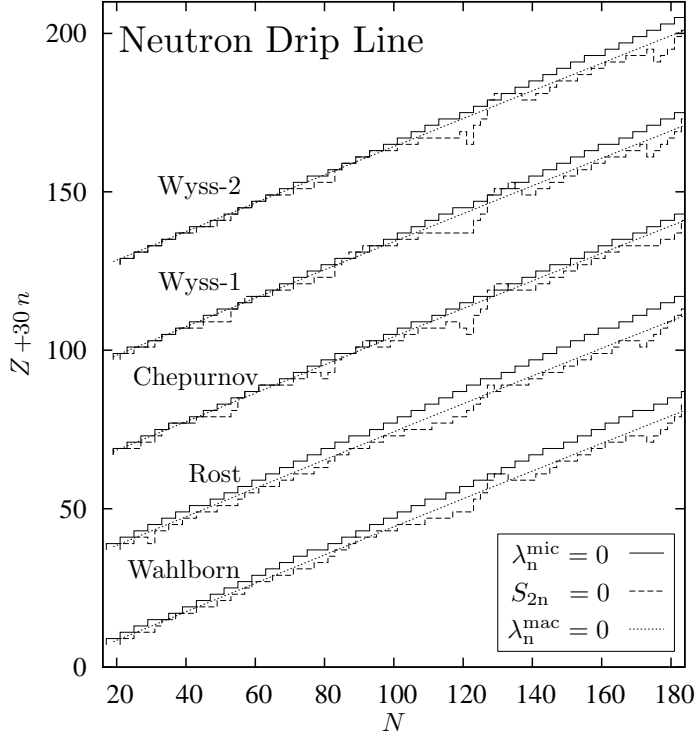}
\vspace*{-5mm}
\caption{
Same as in Fig.~\ref{fig:dripLinesUnvPrmOrgV} but
for other parameter sets. Only the neutron drip lines are shown.
The ordinate values are shifted upward by $30 n$ where
$n=0$ for the parameter set Wahlborn,
$n=1$ for Rost,
$n=2$ for Chepurnov,
$n=3$ for Wyss-1, and 
$n=4$ for Wyss-2.
}
\label{fig:neuDripLineOrgV}
\end{figure*} 

As a test calculation of the improved microscopic-macroscopic method
developed in the present work, we have done global mass calculations for
even-even nuclei with $8\le N \le 184$ and $8\le Z \le 126$.
The basis size specified by $\noscmax=20$,
the Strutinsky smoothing parameters $\gamma=1.2\,\hbar\omega$ and $p=3$,
and the smoothed pairing gap $\tilde{\Delta}=13/\sqrt{A}$ MeV
are used for these mass calculations.
In Fig.~\ref{fig:dripLinesUnvPrmOrgV},
we show the two-neutron drip line ($S_{\rm 2n}=0$)
and the two-proton drip line  ($S_{\rm 2p}=0$) 
calculated with the universal parameter set~\cite{CDN87,DSW81} for the
Woods-Saxon potential.
The drip lines tend to fluctuate outside
the drip lines of the macroscopic part of the model,
defined by equations $\fermiMacNeu =0$ or $\fermiMacPro=0$,
due to the shell effect.
On the other hand, the line in which neutron's Fermi level is zero
($\fermiMicNeu = 0$) is located by 7~(12) neutrons inside the line
$\fermiMacNeu =0$ at $Z$=40~(80): The microscopic Fermi energy $\fermiMicNeu$
is positive and non-negligible at the neutron drip line.

Neutron drip lines for five other potentials are shown 
in Fig.~\ref{fig:neuDripLineOrgV}. 
For potentials of Wahlborn \cite{BW60} and Rost \cite{Ros68}, the 
displacement of the line  $S_{\rm 2n}=0$ from the line $\fermiMicNeu =0$ 
is as large as $\Delta N$=6 to 9~(13) at $Z$=40~(80).
For potentials Wyss-1 \cite{BVCprep09} and Wyss-2 \cite{WysPrivate05}, 
$\Delta N$=2 (9) at $Z$=40 (80).
(The values of the parameters for the potential Wyss-2 can be found
in Table I of Ref.~\cite{SS09}.
The numberings for Wyss's two potentials are tentative.)
The smallest displacement is obtained for Chepurnov's potential \cite{Che67} 
for which $\Delta N=$2~(5) at $Z$=40~(80).
These displacements clearly show that the linear dependence of the depth of the
potential on $N-Z$ as in Eq.~(\ref{eq:WSform}) is oversimplified 
for nuclei far from stability.

As for protons, on the other hand, the situation is much better;
three lines almost coincide in Fig.~\ref{fig:dripLinesUnvPrmOrgV}
as well as for the other potentials (not shown),
which will be mainly because the proton drip line
is closer to experimentally known nuclei than the neutron drip line is.
However, again, the microscopic Fermi energy $\fermiMicPro>0$ at some places
on the proton drip line.

The reliability of
shell correction energies for nuclei in the area between lines
$\fermiMicNeu=0$ and $S\rs{2n}=0$ is not very high
because $E\rs{s.p.}$ is affected by basis-dependent
discretized continuum levels directly (not via smoothing).
Thus, it is preferable to modify the potential parameters
in such a way that the Fermi levels are consistent with the liquid-drop part
of the model.
Among the parameters of the central potential in Eq.~(\ref{eq:centralPotential}),
the radius $R\rs{0CE}$ and the surface diffuseness $a\rs{CE}$ are related
directly to other observables than energy. 
Hence we choose the depth,
\begin{equation} \label{eq:Vdepth}
 V\rs{depth}=-V\rs{0CE}\left[1\pm\kappa\rs{CE}\frac{N-Z}{A}\right],
\end{equation}
(see Eq.~(\ref{eq:WSform}))
to modify.

Our procedure is as follows.
We consider only spherical shape (we use the same depth
for deformed shapes), and we neglect the spin-orbit potential.
The single-particle Hamiltonian is then given by Eq.~(\ref{eq:sphamiltonian}).
The local number density of neutrons or protons at the energy $\epsilon$
is represented by $\rhotf(r,\epsilon)$ of Eq.~(\ref{eq:rhore})
in the Thomas-Fermi approximation.
Given a set of neutron and proton numbers ($N,Z$),
we substitute $\epsilon$ with $\fermiMacNeu$ in Eq.~(\ref{eq:FermEmac})
and readjust $V\rs{depth}$ so as to fulfill
\begin{equation} \label{eq:numberofnucleons}
N = 4 \pi \int_0^{\infty} \rhotf(r, \fermiMacNeu) r^2 dr .
\end{equation}
For protons, $N$ and $\fermiMacNeu$ are replaced with $Z$ and $\fermiMacPro$,
respectively, and the integral is only inside the Coulomb barrier
for $\fermiMacPro >0$.
It should be noted that, in our method,
the depths of the central potentials are determined
by the other parameters than $V\rs{0CE}$ and $\kappa\rs{CE}$.
We use the original parameter value of $V\rs{0CE}$
(multiplied with $\lambda\rs{SO}$) only
to determine the depths of the spin-orbit potentials.
We do not use $\kappa\rs{CE}$ anywhere.

\begin{figure*}[htb] 
\includegraphics[width=105mm]{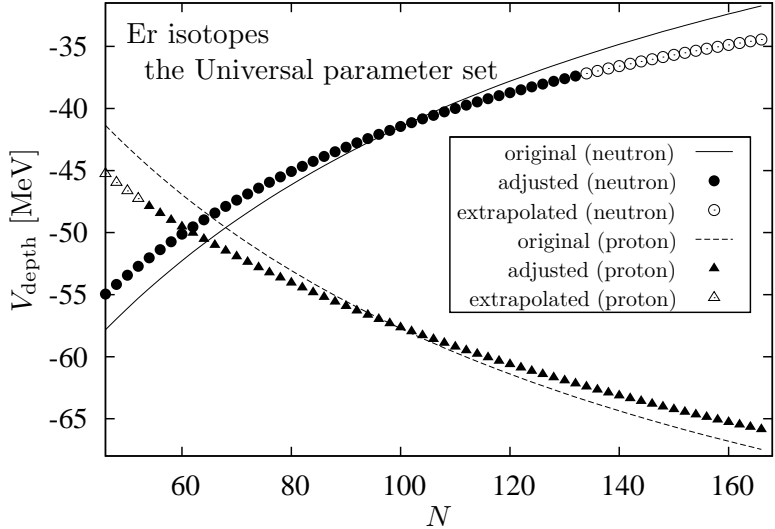}
\vspace*{-5mm}
\caption{
The depths defined in Eq.~(\ref{eq:Vdepth})
of the central potentials of the universal parameter set
(solid and dash lines) and
those readjusted in the Thomas-Fermi approximation
(filled markers) for Er isotopes.
For nuclei near or beyond the
particle thresholds (or the Coulomb barrier top for protons) extrapolated
values are plotted (empty markers).
$\noscmax=30$ is used.
}
\label{fig:ErPotentialDepth}
\end{figure*} 

If the Fermi level of the liquid-drop model
is close to zero, the readjusted depth of the potential becomes 
significantly shallower than
the smooth continuation from the results for more bound nuclei.
It is due to the tail of the Woods-Saxon potential.
This deviation from the smooth trend does not seem to
be physically meaningful. 
Thus we switch to an extrapolation of the smooth
trend if the Fermi level of the liquid-drop model is higher than
some predefined energy: We take this energy as $-2$ MeV in this paper.
For neutrons (protons), we use a polynomial of second degree in
$N$ ($Z$) determined by three heaviest even-$N$ isotopes (even-$Z$ isotones)
not matching the above condition.

This extrapolation is also indispensable to
determine the potential depth for nuclei which are outside the drip lines
of the liquid-drop model. One has to calculate such nuclei because
they may be bound since the shell effect can shift the drip lines.
It is also because nuclei just beyond the drip lines are necessary 
to determine the drip lines themselves.

\begin{figure*}[htb] 
\includegraphics[width=105mm]{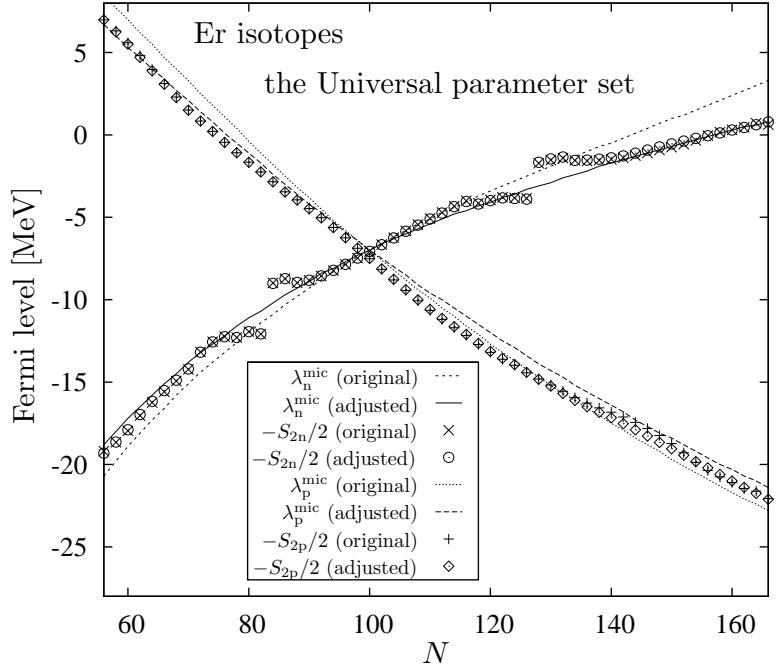}
\vspace*{-5mm}
\caption{
Proton and neutron Fermi levels
of Er isotopes calculated using
the universal parameter set with and without the readjustment of
the depth of the central potentials (four kinds of lines).
Two nucleon separation energies divided by two with the sign inverted
are also shown (four kinds of markers).
$\noscmax=30$ is used.
}
\label{fig:ErFermiLevel}
\end{figure*} 

The resulting readjusted potential depths
are shown in Fig.~\ref{fig:ErPotentialDepth} for Er isotopes.
The original parameter set is the universal one.
While $V\rs{depth}$ is readjusted,
the other parameters are kept unchanged.
For both proton and neutron, the changes due to the readjustment are almost
vanishing for stable nuclei at $N \sim 100$:
This is totally non-trivial because the original parameter
is determined by completely different requirements.
In the neutron (proton) drip line at $N=156$ ($N=76$), 
the readjustment of the neutron (proton) potential has 
non-negligible size 
$\Delta V\rs{depth}=-2.3$ ($-1.3$) MeV.
This means that the original parameter set is quite reasonable 
near the $\beta$-stability line but not
sufficiently accurate to be applicable to the driplines.

Our readjusted depths look very smooth 
functions of $N$ and $Z$, which seem to be fitted nicely
with simple functions having only a few parameters. 
Such a fitting ought to be done
when we will publish an optimized parameter set in future.
Indeed, Nazarewicz et al.\ introduced an extra $(N,Z)$ dependence
to potential parameters for the same purpose \cite{NWD94}.
At present we make a direct use of the depths determined in the 
Thomas-Fermi approximation.

\begin{figure*}[htb] 
\includegraphics[width=95mm]{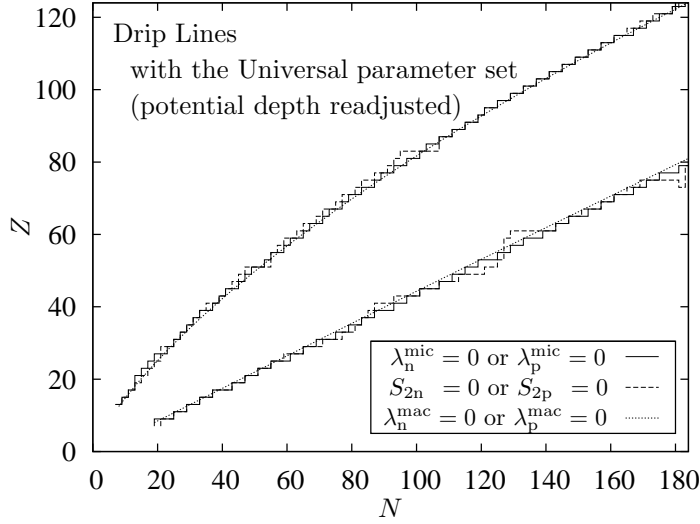}
\vspace*{-5mm}
\caption{
Same as in Fig.~\ref{fig:dripLinesUnvPrmOrgV} but
with readjusted depths for the central potentials.
}
\label{fig:dripLinesUnvPrmTFV}
\end{figure*} 

\begin{figure*}[htb] 
\includegraphics[width=95mm]{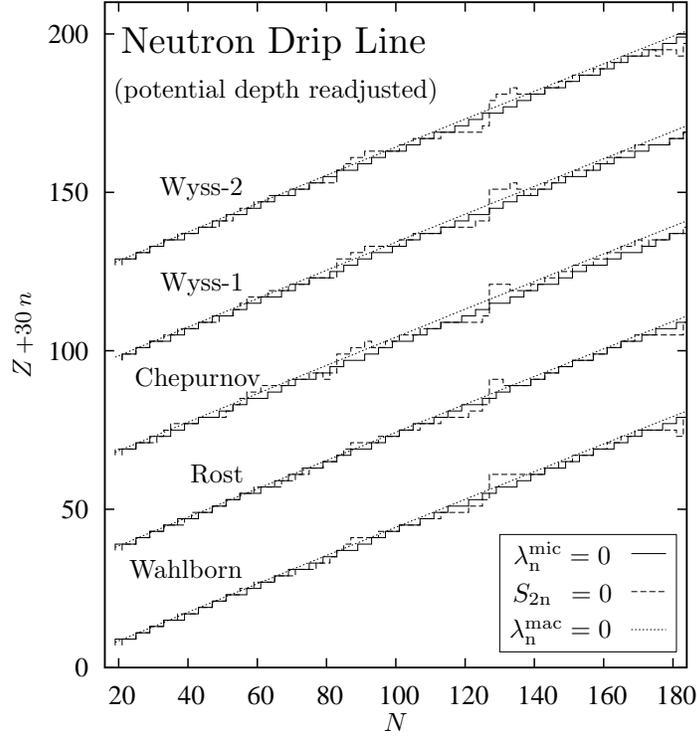}
\vspace*{-5mm}
\caption{
Same as in Fig.~\ref{fig:neuDripLineOrgV} but
with readjusted depths for the central potentials.
}
\label{fig:neuDripLineTFV}
\end{figure*} 

In Fig.~\ref{fig:ErFermiLevel} we show the various kinds
of Fermi levels for Er isotopes.  One can see near drip lines that
Fermi levels calculated with the readjusted potential
depths are in good agreement with separation energies.
This fact, as well as the fact that the original and readjusted depths
are almost equal for stable nuclei,
confirm the soundness of our prescription,
although it may change slightly the single-particle spectrum
from the optimized one of the original potentials.

The drip lines with the readjusted potential depths are shown
in Fig.~\ref{fig:dripLinesUnvPrmTFV} for the universal parameter set 
and in Fig.~\ref{fig:neuDripLineTFV} for the other parameter sets.
One can see that the three lines are now quite close to each other.
With this readjustment method, one can
treat more reliably nuclei near the driplines.
Furthermore, it is worth stressing that
one can control the drip lines by changing the macroscopic part while keeping
automatically the consistency with the microscopic part.

Incidentally, the relation between macroscopic part and the microscopic part
has been payed attention to by Myers~\cite{Myers70} already in 1970.
He determined the droplet model parameters in terms of
a Thomas-Fermi statistical model
with a phenomenological velocity-dependent force
applied to infinite and semiinfinite nuclear matter.
However, this approach seems rather distant from what is proposed in the present paper.

\section{CONCLUSION}
\label{sec:conclusion}

We have examined the Fourier transform of the smoothing function
of Strutinsky to find that it is nothing but a low-pass filter, which
passes only short-time components.  The polynomial part of the filter
is simply a truncated Taylor expansion of $e^{(k/2)^2}$ where $k$ is the
time multiplied by the smoothing width $\gamma$ and divided by
$\hbar$. It may be redefined as a polynomial to minimize the
distortion of the filter near $k=0$, which seems simpler than the
original definition as a curvature correction.  We have also derived a
relation between $\gamma$ and the order of the polynomial part $p$
that changing $\gamma$ proportionally to $\sqrt{p}$ leaves the results
of smoothing almost unchanged.
This picture of the Strutinsky smoothing as a low-pass filter
is general and will be useful for investigating the other smoothing
functions, e.g., those of Ref.~\cite{SKVprep10}, than
the standard one considered in the present work.

From this point of view, we have given a negative perspective to the
problem of the plateau for the Woods-Saxon spectrum, a problem
concerning the shell correction energy in the microscopic-macroscopic
method.  It has been known that the shell correction energy for the
Nilsson spectrum behaves like a long flat plateau as a function of
$\gamma$ while that for the Woods-Saxon spectrum does not show such a
magnificent plateau in general.  We have noticed that the Fourier
transform of the Nilsson spectrum has an interval of time components
where the amplitude is almost vanishing, while that of the Woods-Saxon
spectrum does not have such an interval.  A plateau appears
when the cutoff of the filter is in such an interval.

Instead, we have proposed a new method to weaken the dependence on
the smoothing width $\gamma$.
We call it the reference density method, in which the
smoothing is applied only to the deviation from a reference level
density, which was prepared in such a way that the peak around
energy zero of the Kruppa's level density, which is another principal
subject of this paper, is not washed away.  
We have demonstrated that the method works well in the desired direction.

To apply the Woods-Saxon-Strutinsky method (the
microscopic-macroscopic method with finite-depth potentials) to nuclei
near the nucleon drip lines, it seems necessary to employ the Kruppa's
prescription for positive energy levels, in which the spectrum is defined
as the Woods-Saxon spectrum subtracted by the free-nucleon
spectrum, both of which are obtained through diagonalizations in the
same oscillator basis.  We have discussed the ground for this
prescription as well as its relation to the continuum level density.

We have also proposed the oscillator-basis Thomas-Fermi approximation,
with which one can describe spectra obtained from diagonalization in
truncated oscillator bases.  We have demonstrated that this
approximation can reproduce average behaviors of the Woods-Saxon, free,
and Kruppa's level densities.  One can also use this approximation to
show analytically the convergence of the results
with Kruppa's prescription versus the size of the oscillator basis.

We have also introduced the Kruppa-BCS method, in which we modified
the BCS equation for the pairing correlation so that it can be applied
to the Kruppa's level density by taking into account negative
contributions from the free-nucleon spectrum.  The Kruppa-BCS method
is applicable to any cases, while the ordinary BCS method becomes very
faulty especially when the diagonalization basis is not small and the
nucleus is very neutron-rich.

We have also studied how to determine the strength of the pairing
interaction to be used in the Kruppa-BCS method.  An important
conclusion is that, in adjusting the strength to reproduce the
empirical smooth trend of the pairing gap with the smoothed Kruppa's
level density in the gap equation, one should carry out the energy
integral without using the uniform level density approximation.

The inconsistency between the macroscopic and the microscopic parts
(i.e., the liquid drop model and the single-particle potential) is
another problem to the application of the method to nuclei near the
drip lines.  Calculating masses in the whole nuclear chart
with several parameter sets for the Woods-Saxon potential
using the methods developed in this paper,
we have found that the neutron drip line of the microscopic
part is located typically more than ten neutrons inside the dripline
of the total energy.  We have proposed a method to readjust the depths
of the central potentials to achieve the consistency within the
Thomas-Fermi approximation. Although the method contains two
simplifications, assuming the spherical symmetry and neglecting
the spin-orbit potential, the method has worked very well
to shift the microscopic dripline close to the total dripline.

We are going to apply the methods presented in this paper to extend
our studies \cite{TS01,TSS02} on the origin of the prolate-dominance of
the atomic nucleus from the Nilsson potential to the Woods-Saxon
potential.

\section{ACKNOWLEDGEMENTS}

The authors would like to thank
Dr.~N.~Onishi and Dr.~K.~Arita for discussions,
and Dr. R.~Wyss for providing an unpublished parameter set
for the Woods-Saxon potential.
This work is supported in part by the JSPS Core-to-Core Program,
International Research Network for Exotic Femto Systems (EFES),
and by Grant-in-Aid for Scientific Research~(C) 
No.~18540258 from Japan Society for the Promotion of Science.
A part of the formula manipulations were carried out on the computer system
at YITP in Kyoto University.


\end{document}